\documentclass[oldversion,longauth]{aa}
\def\simlt{\lower.5ex\hbox{$\; \buildrel < \over \sim \;$}}
\def\simgt{\lower.5ex\hbox{$\; \buildrel > \over \sim \;$}}
\def\Mstar {${\cal M}$}
\def\MstarL {${\cal M}/L$}
\def\Mstars {${\cal M}^*$}
\def\logM {$\log ({\cal M}/{{\cal M}_{\odot}})$} 
\def\logMtot {$\log ({\cal M}_{\rm tot}/{{\cal M}_{\odot}})$} 
\def\logSSFR {$\log({\it SSFR}$/Gyr$^{-1}$)} 
\def\Mlim {${\cal M}_{\rm lim}$}
\def\Mbias {${\cal M}_{\rm min}$}
\def\Ms {${\cal M}_{\odot}$}
\def\Msun {${\cal M}_{\odot}$}
\def\Mcross {${\cal M}_{\rm cross}$}
\def\SFR {${\it SFR}$} 
\def\SFRs {${\it SFRs}$} 
\def\sSFR {${\it SSFR}$} 
\def\SSFR {${\it SSFR}$} 
\def\sSFRs {${\it SSFR}s$} 
\def\SSFRs {${\it SSFR}s$} 
\def\PT {${\cal PT}$}
\def\ZT {${\cal ZT}$}
\def\MT {${\cal MT}$}
\def\ST {${\cal ST}$}
\def\IMF {${\it IMF}$}
\def\GSMF {${\it GSMF}$}
\def\GBMF {${\it GBMF}$}
\def\MF {${\it GSMF}$}
\def\MFs {${\it GSMF}$s}
\def\dNdt {$\dot \rho_N$}
\def\dMdt {$\dot \rho_{\cal M}$}
\def\zbuilding {$z_{\rm building}$}
\def\zassembly {$z_{\rm assembly}$}
\def\zform {$z_{\rm form}$}
\usepackage{graphicx}

\newcommand{\Vmax}{$1/V_{\rm max}$}
\newcommand{\D}{{\rm d}}
\newcommand{\mass}{{\cal M}}
\begin{document}
\title{zCOSMOS - 10k-bright spectroscopic sample
         \thanks{based on data obtained with the European Southern Observatory 
Very Large Telescope, Paranal, Chile, program 175.A-0839
}
}
\subtitle{The bimodality in the galaxy stellar mass function: exploring
its evolution with redshift
}

\author{
     L.~Pozzetti    \inst{1}  
\and M.~Bolzonella  \inst{1}  
\and E.~Zucca \inst{1} 
\and G.~Zamorani \inst{1}  
\and S.~Lilly \inst{2} 
\and A.~Renzini\inst{3} 
\and M.~Moresco \inst{4} 
\and M.~Mignoli \inst{1} 
\and P.~Cassata \inst{5,6} 
\and L.~Tasca \inst{7,6} 
\and F.~Lamareille \inst{8} 
\and C.~Maier\inst{2} 
\and B.~Meneux\inst{9,10} 
\and C.Halliday\inst{11} 
\and P.~Oesch\inst{2} 
\and D.~Vergani \inst{1} 
\and K.~Caputi\inst{2} 
\and K.~Kova\v{c}\inst{2} 
\and A.~Cimatti\inst{4} 
\and O.~Cucciati\inst{6,12} 
\and A.~Iovino\inst{12} 
\and Y.~Peng \inst{2} 
\and M.~Carollo \inst{2} 
\and T.~Contini\inst{8} 
\and J.-P.~Kneib\inst{6} 
\and O.~Le~F\'evre\inst{6} 
\and V.~Mainieri\inst{9} 
\and M.~Scodeggio\inst{7} 
\and S.~Bardelli\inst{1} 
\and A.~Bongiorno\inst{9} 
\and G.~Coppa\inst{1} 
\and S.~de~la~Torre\inst{6} 
\and L.~de~Ravel\inst{6} 
\and P.~Franzetti\inst{7} 
\and B.~Garilli\inst{7} 
\and P.~Kampczyk\inst{2} 
\and C.~Knobel\inst{2} 
\and J.-F.~Le~Borgne\inst{8} 
\and V.~Le~Brun\inst{6} 
\and R.~Pell\`o\inst{8} 
\and E.~Perez~Montero\inst{8} 
\and E.~Ricciardelli\inst{3} 
\and J.D.~Silverman\inst{2} 
\and M.~Tanaka\inst{9} 
\and L.~Tresse\inst{6} 
\and U.~Abbas\inst{6} 
\and D.~Bottini\inst{7} 
\and A.~Cappi\inst{1} 
\and L.~Guzzo\inst{12} 
\and A.M.~Koekemoer\inst{13} 
\and A.~Leauthaud\inst{6} 
\and D.~Maccagni\inst{7} 
\and C.~Marinoni\inst{14} 
\and H.J.~McCracken\inst{15} 
\and P.~Memeo\inst{7} 
\and C.~Porciani\inst{2} 
\and R.~Scaramella\inst{16} 
\and C.~Scarlata\inst{17} 
\and N.~Scoville\inst{18} 
}
   \offprints{Lucia Pozzetti \email{lucia.pozzetti@oabo.inaf.it} }

\institute{
INAF -- Osservatorio Astronomico di Bologna, via Ranzani 1, I-40127, Bologna, Italy
\and 
Institute of Astronomy, Swiss Federal Institute of Technology (ETH H\"onggerberg), CH-8093, Z\"urich, Switzerland
\and 
Dipartimento di Astronomia, Universit\'a di Padova, Padova, Italy
\and 
Dipartimento di Astronomia, Universit\'a degli Studi di Bologna, via Ranzani 1, I-40127, Bologna, Italy
\and 
Department of Astronomy, University of Massachusetts, 710 North Pleasant Street, Amherst, MA 01003, USA 
\and 
{Laboratoire d'Astrophysique de Marseille, Universit\'{e} d'Aix-Marseille, CNRS, 38 rue Frederic Joliot-Curie, F-13388 Marseille Cedex 13, France}
\and 
INAF -- Istituto di Astrofisica Spaziale e Fisica Cosmica di Milano, via Bassini 15, I-20133 Milano, Italy;
\and 
Laboratoire d'Astrophysique de Toulouse-Tarbes, Universit\'{e} de Toulouse, CNRS, 14 avenue Edouard Belin, F-31400 Toulouse, France
\and 
Max Planck Institut f\"ur Extraterrestrische Physik,  Giessenbachstrasse, D-84571 Garching, Germany
\and 
Universitats Sternwarte, Scheinerstrasse 1, D-81679 Muenchen 
\and 
{INAF -- Osservatorio Astrofisico di Arcetri, Largo Enrico Fermi 5, I-50125 Firenze, Italy }
\and 
INAF -- Osservatorio Astronomico di Brera, via Brera 28, I-20121  Milano, Italy
\and 
{Space Telescope Science Institute, 3700 San Martin Drive, Baltimore, MD 21218, USA}
\and 
Centre de Physique Theorique, Marseille, Marseille, France
\and 
Institut d'Astrophysique de Paris, Universit\'e Pierre \& Marie Curie, Paris, France
\and 
{INAF -- Osservatorio Astronomico di Roma, via di Frascati 33, I-00040 Monteporzio Catone, Italy}
\and 
Spitzer Science Center, Pasadena, CA, USA 
\and 
{California Institute of Technology, MC 105-24, 1200 East California Boulevard, Pasadena, CA 91125, USA}
}

\authorrunning {L.Pozzetti, et al.}

\titlerunning {zCOSMOS: Galaxy bimodality in the stellar mass function}

\date{Received -- -- ----; accepted -- -- ----}

\abstract{ 
We present the galaxy stellar mass function (\GSMF) to redshift $z\simeq 1$, 
based on the analysis of about 
8500 galaxies with $I<22.5$ (AB mag) over 1.4 deg$^2$,
which are part of the zCOSMOS-bright 10k spectroscopic sample.
We investigate the total \MF, as well as the contributions of 
early- and late-type galaxies (ETGs and LTGs, respectively), defined
by different criteria (broad-band spectral energy distribution,  
morphology, spectral properties, or star formation activities). 
We unveil a galaxy bimodality in the global \MF, whose
shape is more accurately represented by 2 Schechter functions, one linked
to the ETG and the other to the LTG populations.
For the global population, we confirm a mass-dependent evolution 
(``mass-assembly downsizing"),
i.e., galaxy number density increases with cosmic time 
by a factor of two between $z=1$ and $z=0$ for intermediate-to-low mass 
(\logM$\sim10.5$) galaxies but less than 15\% for \logM$>11$. We find that
the \MF\ evolution at intermediate-to-low values of \Mstar\ (\logM$<10.6$) is mostly 
explained by the growth in stellar mass driven by smoothly decreasing star formation 
activities, despite the redder colours predicted in particular at low redshift.
The low residual evolution is consistent, on average, with $\sim 0.16$ merger 
per galaxy per Gyr (of which fewer than $0.1$ are major), 
with a hint of a decrease with cosmic time but not a clear dependence on the mass.
From the analysis of different galaxy types,  we find that ETGs, 
regardless of the classification method, increase in number density with cosmic 
time more rapidly with decreasing \Mstar, i.e., follow a top-down building history,
with a median ``building redshift" increasing with mass ($z>1$ for \logM$>11$), 
in contrast to hierarchical model predictions.  
For LTGs, we find that the number density of blue or spiral galaxies with \logM$>10$
remains almost constant with cosmic time from $z\sim1$.
Instead, the most extreme population of star-forming galaxies (with high specific 
star formation), at intermediate/high-mass, rapidly decreases in number density 
with cosmic time. Our data can be interpreted as a combination of different effects. 
Firstly, we suggest a transformation, driven mainly by SFH, from blue, active, 
spiral galaxies of intermediate mass to blue quiescent and subsequently (1-2 Gyr after) 
red, passive types of low specific star formation. 
We find an indication that the complete morphological transformation, probably 
driven by dynamical processes, into red spheroidal galaxies, occurred on longer 
timescales or followed after 1-2 Gyr. A continuous replacement of blue galaxies is 
expected to be accomplished by low-mass active spirals increasing their stellar mass.
We estimate the growth rate in number and mass density of the red galaxies at 
different redshifts and masses. The corresponding fraction of blue galaxies that, 
at any given time, is transforming into red galaxies per Gyr,
due to the quenching of their \SFR, is on average $\sim 25$\% for \logM$<11$. 
We conclude that the build-up of galaxies and in particular of ETGs follows the 
same downsizing trend with mass (i.e. occurs earlier for high-mass galaxies) as the 
formation of their stars and follows the converse of the trend predicted by current SAMs.
In this scenario, we expect there to be a negligible evolution of the galaxy baryonic 
mass function (\GBMF) for the global population at all masses and a decrease with cosmic 
time in the  \GBMF\ for the blue galaxy population at intermediate-high masses.

         }

\maketitle

\section{Introduction}\label{sec:intro}

Tracing the history of both galaxy star formation (Lilly et al. 1996; Madau et al. 1996; 
Madau, Pozzetti, Dickinson 1998) and stellar mass assembly (Dickinson et al. 2003) 
over cosmic time represent major challenges in modern cosmology. In particular,
it is still uncertain how the bimodality seen in the local Universe in terms of galaxy 
properties (e.g., colour, morphology, star formation, and spectral features; 
Kauffmann et al. 2003, Baldry et al. 2004, Brinchmann et al. 2004) evolves and when it 
was created. Galaxies in the local Universe exhibit distinctive bimodal colour 
distributions (Strateva et al. 2001, Hogg et al. 2002, Blanton et al. 2003),
which suggests that different evolution histories exist for galaxies lying on the 
two sequences (Menci et al. 2005; Scarlata et al. 2007b; De Lucia et al. 2007). 
Even if colour bimodality has already been observed and studied 
at higher redshift (Bell et al. 2004, Weiner et al. 2005, up to 
$z\sim 1$; Franzetti et al. 2007, Cirasuolo et al. 2007, up to $z\sim 1.5$; 
Giallongo et al. 2005, Cassata et al. 2008, Williams et al. 2009, up to $z\sim 2$), 
no study until now 
has fully evaluated how the number densities of the two populations compare and 
evolve with cosmic time.

It is still unclear, indeed, how the galaxy stellar mass function (\MF) is linked to the galaxy bimodality.
Baldry et al. (2004, 2006, 2008) noted a bimodal shape in the local \MF\ in the SDSS 
with an upturn at masses 
lower than $10^9$ \Ms, related to the two different galaxy populations.
It is, therefore, extremely interesting to explore how this shape evolves with redshift
and which mechanisms contribute to its appearance. 
From previous surveys, it is well established that there has been an
increase/decrease in the fraction of red/blue or early/late-type galaxies with 
cosmic time at intermediate masses since $z\sim1$
(Fontana et al. 2004; Bundy et al. 2006; Arnouts et al. 2007; Scarlata et al. 2007b; Vergani et al. 2008),
while the precise evolution in absolute number densities of the two populations
remains controversial, in particular for intermediate/massive galaxies. 
A knowledge of this evolution may help to constrain the 
quenching mechanisms responsible for downsizing.

In agreement with the first formulation introduced by Cowie
et al. (1996), a downsizing scenario in both age and star formation history
has been proposed by several observational studies,
i.e., more massive galaxies form their stars earlier and more rapidly than lower mass ones
(hereafter age downsizing: Brinchmann \& Ellis 2000; Gavazzi \& Scodeggio 1996; 
Fontana et al. 2004, Kodama et al. 2004; Bauer et al. 2005; Feulner et al. 2005a,b; 
Juneau et al. 2005; Borch et al. 2006; Cucciati et al. 2006; Vergani et al. 2008).
Age downsizing occurs 
separately for each of the two populations causing galaxy bimodality, i.e., a red peak
(Thomas et al. 2005, Fontana et al. 2004, Thomas et al. 2009)
and a blue one (Noeske et al. 2007a,b). 
However, it remains unclear whether
the age downsizing is coupled with a mass-assembly downsizing scenario for galaxy evolution and formation
(Fontana et al. 2004, 2006; Pozzetti et al. 2003, 2007; Cimatti et al. 2006; Bundy et al. 2006),
i.e., if the more massive galaxies assembled their mass earlier than lower mass ones. 
Furthermore, do low-mass galaxies contain younger stars and assemble later  
even within the same spectral type?
The hierarchical model of De Lucia et al. (2006) predicts, for example, 
a bottom-up assembly history for elliptical galaxies (also called ``upsizing")
following the hierarchical growth  of dark matter haloes, 
in contrast to a top-down, downsizing scenario for the formation of their stars.  

A fundamental role towards answering these questions is played by deep surveys, 
which sample thousands of galaxies across large portions of the sky.
Very deep surveys have been exploited to describe the 
shape of the stellar mass function at high redshift (Fontana et al. 
2006; Drory et al. 2005; Gwyn \& Hartwick 2005; Bundy et al. 2006; Pozzetti et al. 2007), 
but a clear picture about stellar mass assembly and how it depends on mass 
(mass-assembly downsizing) and galaxy type has not yet emerged.
Previous studies  have explored the evolution of different galaxy types 
in deep near-IR surveys, such as K20, by means of the $K$-band luminosity function 
(Pozzetti et al. 2003) and the galaxy stellar mass function (Fontana et al. 2004),  
using spectral classification (i.e., absorption-line galaxies versus emission-line galaxies), 
and in larger optical and near-IR surveys
such as VVDS, COMBO17, and DEEP2, using colours or spectra to define galaxy types (Bell et al. 2004; 
Cimatti et al. 2006; Faber et al. 2007; Zucca et al. 2006; Arnouts et al. 2007; Vergani et al. 2008). 
Even if the results of  these surveys remain disputed (compare for example 
Bell et al. 2004 and Cimatti et al. 2006 for the same  dataset), most of these studies agree
that luminous and rather massive old galaxies were
already quite common at $z\sim1$ and that their number density declines rapidly at yet higher 
redshift. This suggests that merger events are ruled out
as the major mechanism behind their assembly history below $z\simeq1$. However, 
observational results about major merging and dry merging
are still contradictory (see Bell et al. 2006; van Dokkum 2005; Lin et al. 2004;
de Ravel et al. 2009; and Renzini 2007 for a summary).
It is also known that the
less luminous/massive ETGs decline in number density steadily with 
redshift (Cimatti et al. 2006). Scarlata et al. (2007b) show similar results for the 
photometric survey COSMOS, using both morphologically and photometrically selected 
subsamples of early-type galaxies. The evolution in number density of the most massive 
late-type star-forming galaxies continues, however, to be unclear.

Another open question is
whether the evolution of the observed \GSMF\ with cosmic time is driven mainly by either merging events at any 
given mass, as predicted by hierarchical galaxy formation models (Cole et al. 2000;
Menci et al. 2005; Bower et al. 2006; De Lucia et al. 2006; Monaco et al. 2006)
or star formation histories (SFHs, 
see Vergani et al. 2008; Walcher et al. 2008). 
It is widely believed that galaxies are assembled by hierarchical mergers between 
massive cold dark matter haloes, in which baryonic star-forming matter is embedded.
However, most of the hierarchical galaxy assembly models
are unable to completely account for the observed \GSMF\ and its evolution
(see Fontana et al. 2004, 2006, Monaco et al. 2006, Caputi et al. 2006, Marchesini et al. 2009, 
Fontanot et al. 2009 for a detailed comparison with models). For instance,
some models tend to underpredict the high-mass tail and
overpredict the rate of its evolution (Fontana et al. 2004, 2006), 
even in the extreme case of evolution driven purely by mergers
(Monaco et al. 2006). Only the assumption that a significant fraction, 
$\sim 30$ \%, of stars are scattered within the diffuse stellar component at each merger event 
leads to significant suppression of the predicted evolution rate, 
in closer agreement with observational constraints (Monaco et al. 2006).
On the other hand,
most of the models overpredict the number density of relatively low-mass galaxies 
(see Fontana et al. 2006, Kitzbichler \& White 2007, Marchesini et al. 2009, Fontanot et al. 2009).
Furthermore, the models do not reproduce the downsizing trend in stellar mass observed 
for elliptical galazies (Cimatti et al. 2006).
According to the latest observational results, the current  galaxy formation 
and evolutionary scenario is becoming one 
in which a smoother evolution in mass growth and star formation (due to cold gas accretion) plays a major,  
if  not dominant,  role compared to dark  matter (major)  merging events.

In this paper, we use the zCOSMOS spectroscopic survey (Lilly et al. 2007, 2009)
to complete a comprehensive analysis of the \GSMF.  Compared to previous spectroscopic surveys, 
the larger area and number of spectroscopic redshifts of zCOSMOS allows higher precision, 
higher quality statistics, and lower cosmic variance
(Lilly et al. 2009) to be achieved in our analysis, in particular for the massive end of the \MFs.
We derive the \GSMF\ and its evolution with cosmic time since $z=1$, 
as well as the contribution of different galaxy populations, using numerous classification methods 
defined in terms of their colours, morphologies, star formation activities, or spectroscopic classifications.
By using spectroscopic redshifts, we are able to study the shape of the global \GSMF\ to high precision, 
its evolution, and how it is related to the bimodalities in galaxy properties 
(colours, morphologies, spectral properties) observed to high redshifts.
We explore the different roles of merging and SFHs as a function of \Mstar\ and cosmic time 
to explain the observed \MF\ evolution.
We also explore the evolution with cosmic time of the \GSMF\ for different galaxy types and propose 
an evolutionary scenario.

Throughout the paper, we adopt the cosmology $\Omega_m = 0.25$ and
$\Omega_\Lambda = 0.75$, with $h_{70} = H_0 / (70$ km s$^{-1}$ Mpc$^{-1}$). 
Magnitudes are given in the AB system.

\section{COSMOS and the bright 10k zCOSMOS spectroscopic sample}

The Cosmic Evolution Survey (COSMOS, Scoville et al. 2007) is the largest HST survey (640 orbits) 
ever undertaken, imaging a field of $\sim 2$ deg$^2$ with the Advanced Camera for Surveys (ACS)
with F814W (Koekemoer et al. 2007).
This survey was designed 
to probe galaxy evolution and the effects of environment to high redshift. 
COSMOS observations include good coverage of the field with 
multiband photometry from the UV (with GALEX, Zamojski et al. 2007), 
optical (with Subaru and CFHT, Taniguchi et al. 2007, Capak et al. 2007), 
NIR (with CTIO, KPNO, Capak et al. 2007, and CFHT, McCracken et al. 2010), to MIR and 
FIR (S-COSMOS with Spitzer, Sanders et al. 2007),
in combination with a multiwavelength dataset from radio (with VLA, Schinnerer et al. 2007), 
millimetre (with MAMBO-2 at the IRAM telescope, Bertoldi et al. 2007), 
to X-rays (with XMM, Hasinger et al. 2007, and Chandra, Elvis et al. 2009).

The zCOSMOS spectroscopic survey (Lilly et al. 2007) is an ongoing ESO Large Programme 
($\sim 600$ hours of observations) 
aiming to map the COSMOS field with
the VIsible Multi-Object Spectrograph (VIMOS, Le~F\`evre et al. \cite{vimos}), mounted on the
ESO Very Large Telescope (VLT).
The zCOSMOS survey consists of a {\it bright} part, 
with spectroscopy limited to objects in the magnitude range $15.0<I<22.5$, 
and of a {\it deep} part, which measures redshifts of $B<25.25$ of galaxies 
colour-selected to be in the range $1.4<z<3$, within the central 1 deg$^2$. 
The bright part has already produced redshifts and spectra for about $10~000$ 
galaxies over $\sim 1.4$ deg$^2$, the so-called 10k-bright spectroscopic 
sample (Lilly et al. 2009), with an average sampling rate of about $33\%$.
For more details about the zCOSMOS 10k-bright sample, we refer to Lilly et al. (2009). Here we recall that
the VIMOS spectroscopic observations were completed using the red medium resolution $R \sim 600$ MR grism 
(5550-9650 \AA), the 
spectra were reduced using the VIMOS Interactive Pipeline
Graphical Interface software (VIPGI, Scodeggio et al. 2005)
and redshift measurements were visually determined after a first estimate had been provided by 
an automatic package (EZ, Garilli et al. in preparation).
A quality flag was assigned to each redshift measurement. This flag ranges
from 0 (failed measurement) to 4 (100\% confidence level), and
flag 9 indicates spectra with a single emission line, for which
multiple redshift solutions are possible. 
Additional details about the spectroscopic quality flags and their probability 
of being robust are given in Lilly et al. (2009; see their Table 1).
In addition to the confidence 
classes described in Lilly et al. (2007), a decimal place (from 5 to 1, see Table 3 in 
Lilly et al. 2009) in the class is added to indicate the level of consistency
between the spectroscopic and photometric redshifts obtained 
by the Zurich Extragalactic Bayesian Redshift Analyzer (ZEBRA; Feldmann et al. 2006),
using the optical to infrared SED.


\subsection {The selected galaxy spectroscopic sample}\label{sec:sample}

The analysis presented in this paper is based on the zCOSMOS-10k bright sample (Lilly et al. 2009).
From the total sample of $10~644$ objects observed spectroscopically,
we used the objects within the statistical sample defined in the magnitude range $15 <I< 22.5$,
and removed the spectroscopically confirmed stars, broad-line AGNs, and 
the galaxies with low quality redshift flag (flag$<$ 1.5, i.e., with a verification rate $<90\%$ 
and a spectroscopic redshift inconsistent with the photometric redshift). To achieve 
reliable SED fitting to multi-band photometric data, we excluded objects with apparent magnitudes 
measured in fewer than 5 bands ($\sim 1.7$\%) and objects for which the
ground photometry can be affected by the blending of sources, as
inferred from the number of ACS sources brighter than $I = 22.5$ within
$0.6''$ ($\sim 0.5$\%).  
This resulted in the selection of 8450 galaxy spectra with secure spectroscopic measurements ($7936$ in the redshift range
where the following analysis is carried out, $z=0.1-1$) over 5045.80 arcmin$^2$.
Objects with redshift flags $<1.5$ are taken into account
statistically (see Sect. \ref{sec:GSMF}, and Bolzonella et al. 2009  and Zucca et
al. 2009 for details). 

The final selected sample of $8450$ galaxies, which have
highly reliable spectroscopic redshifts, have complete multi-band photometric coverage
(from UV to IRAC) and morphological classifications, along with morphological parameters
 from  both the Zurich Estimator of Structural Types (ZEST; 
Scarlata et al. 2007a) and estimates obtained in Marseille 
(Cassata et al. 2008, Tasca et al. 2009, MRS hereafter).

\subsection{Photometric data }\label{sec:photo}

The COSMOS field has been covered by multiband photometry over a wide range of wavelengths.
In this paper, we used the observed magnitudes in 10 photometric bands
(CFHT $u^*$ and $K_s$, Subaru $B_J$, $V_J$, $g^+$, $r^+$, $i^+$, and $z^+$, 
and Spitzer IRAC at 3.6 $\mu$m and 4.5 $\mu$m). 
Descriptions of the photometric catalogs
are given in Capak et al. (2007), Sanders et al. (2007), and McCracken et al. (2010).
Following the same approach as Capak et al. (2007; see their Table 13),
the photometry was adjusted by applying fixed zeropoint offsets 
to the observed magnitudes in each band
to statistically reduce the differences between 
observed and reference magnitudes computed from a set of template SEDs,
finding in general for each band very similar offsets.

\section{Estimate of the stellar masses}\label{sec:mass}

We used the stellar masses (\Mstar) estimated from
a fit to the multicolour spectral energy distribution (SED), using 
the observed magnitudes in 10 photometric bands from $u^*$ to 4.5 $\mu$m 
(see Sect. \ref{sec:photo}), and following the method described in Pozzetti et al. (2007).
In a parallel paper, Bolzonella et al. (2009) describe the different methods used to compute stellar masses, 
based on different assumptions about the population synthesis models 
(Bruzual \& Charlot 2003, BC03; Maraston 2005, M05; Charlot \& Bruzual 2007, CB07) 
and the star-formation histories (SFHs), such as smooth exponentially decreasing or 
complex SFHs with the addition of secondary bursts. In Bolzonella et al., the associated 
uncertainties and degeneracies are also discussed 
(see also Fontana et al. 2004, Pozzetti et al. 2007, Marchesini et al. 2009).
Here we just recall that the accuracy of the photometric stellar masses
is satisfactory overall,   with typical dispersions caused by statistical uncertainties and
 degeneracies of the order of $0.2$ dex.
We note that the addition of secondary bursts to
a continuous star-formation history produces systematically higher (up to 40\% on average) stellar masses
(Pozzetti et al. 2007), and in particular Fontana et al. (2004) demostrated that the effect is larger 
when the fraction of mass produced in the burst is small. 
On the other hand,
population synthesis models with a TP-AGB phase (Maraston 2005, Charlot \& Bruzual 2007) produce
shifts of up to $\sim 0.2$ dex towards lower \Mstar. 
Finally, the uncertainty in the absolute value of the \Mstar\ 
related to the assumptions about the {\it initial mass function} (\IMF) is within 
a factor of 2 for the typical \IMF s usually adopted in the literature.

In this paper, we adopt stellar masses derived with {\it Hyperzmass} (see Pozzetti et al. 2007),
a modified version of the photometric redshift code {\it Hyperz} (Bolzonella et al. 2000). We used
smooth exponentially decreasing SFHs
(\SFR$(t) \propto \exp(-t/\tau)$ with timescale $\tau= [0.1, \infty ]$ and age $t= [0.1, 20]$ Gyr
constrained to be shorter than a Hubble time at each redshift). 
This parametrisation may be 
inappropriate for actively star-forming galaxies (especially 
at high redshifts, Renzini 2009), because it assumes that all galaxies are 
observed while having their minimum SFR. However, for the redshift range explored ($z<1$) 
and for passively evolving galaxies this assumption does not appreciably affect the
mass determinations. 
We checked, for example, that the inclusion of delayed exponential SFHs 
(\SFR$(t) \propto t/\tau^2 \times \exp(-t/\tau)$ do not 
significantly affect the stellar masses estimates. As we noted previously, 
small secondary bursts may cause a systematic increase in the mass estimate (Fontana et al. 2004).
For this study, we adopted the 
Calzetti et al. (2000) extinction law, solar metallicity (see also Table 1 in Pozzetti et al. 2007),
Bruzual \& Charlot (2003) population synthesis models, and a Chabrier \IMF\ (Chabrier 2003)
with lower and upper cutoffs of 0.1 and 100 $M_{\odot}$,
to which we collectively refer hereafter as the default parameters, if not specified otherwise.
All \MFs\ were computed by using in addition M05 and CB07 population synthesis models.
In particular, we are aware that the estimate of ETG stellar masses at high-$z$ are 
particularly sensitive to the TP-AGB phase of the stellar population around 1-2 Gyr, 
producing a shift to lower masses of up to 0.2 dex (Maraston et al. 2006). 
In the redshift range sampled here ($0.1<z<1.0$), ETGs have on average relatively old 
stellar populations (Thomas et al. 2005) that are not dominated by the TP-AGB phase, 
in particular for massive objects.
At higher redshifts ($z>1-1.5$), the effect of the TP-AGB phase in younger (1-2 Gyrs) ETGs 
becomes very important and should not be neglected.
In our data (limited to $z<1$), we find that by using M05 models instead our main results and 
conclusions remain almost unchanged and we discuss the differences later in the text. 


\section{Galaxy classification}\label{sec:types}

We used 5 different methods to classify each galaxy
as either early- or late-type (ETGs and LTGs, hereafter),
according to its colour, star formation activity, morphology, or spectroscopy, as follows:

\begin{enumerate}

\item {\it Photometric classification (red and blue galaxies):}

We derived galaxy {\it photometric types} (\PT)
from the best-fit to the multi-band photometry 
(from $u^*$ to $K_s$ band, see previous section).
Following Zucca et al. (2006),
we used the empirical set of (62) SEDs described in Ilbert et al. (2006).
These SEDs were derived by interpolating between the four local observed spectra 
of Coleman et al. (1980, ranging from that of the old stellar population of both M31 and 
M81 to Sbc, Scd, and Im SEDs)
and two starburst SEDs from Kinney et al. (1996).
These templates were also linearly extrapolated to the
ultraviolet ($\lambda < 2000$ \AA) and near-infrared wavelengths using the 
GISSEL synthetic models (Bruzual \& Charlot 2003).
Galaxies were divided into four types, according to their spectral energy distribution 
between the UV and near-IR. These types correspond to a red E/Sa
template (\PT=1), an early spiral template (\PT=2), a late spiral template
(\PT=3), and an irregular or a starburst template (\PT=4).
Using the data from the VVDS spectroscopic survey, Zucca et al. (2006) 
verified statistically the consistency between
the photometric classification and the average galaxy spectral properties.
Using stacked spectra of all galaxies of each of the four types, they detected an 
increasingly bluer continuum with stronger emission lines from \PT=1 to \PT=4,
 confirming the robustness of this classification scheme.

\begin{figure} 
\centering
\includegraphics[width=0.49\hsize]{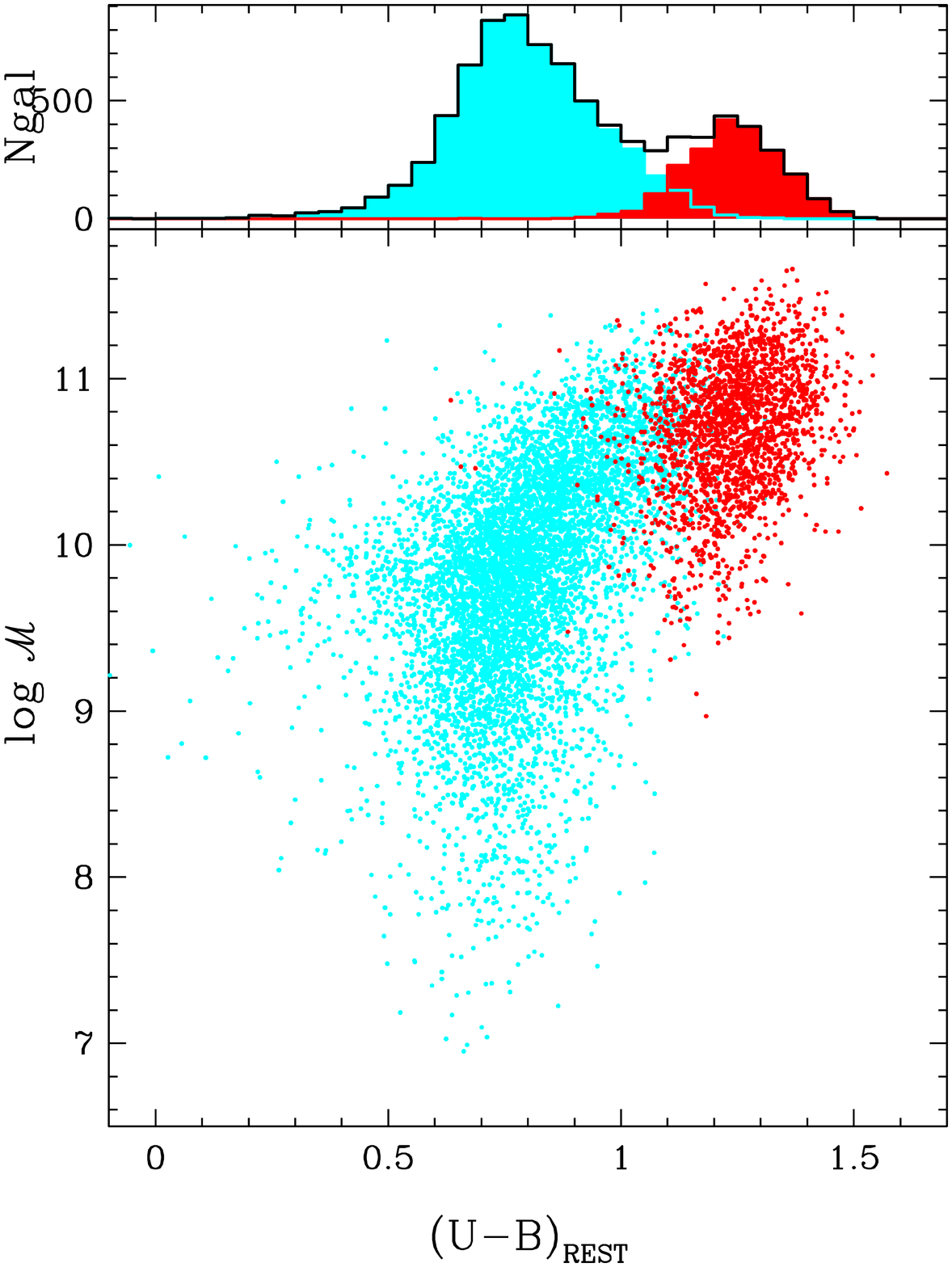}
\includegraphics[width=0.49\hsize]{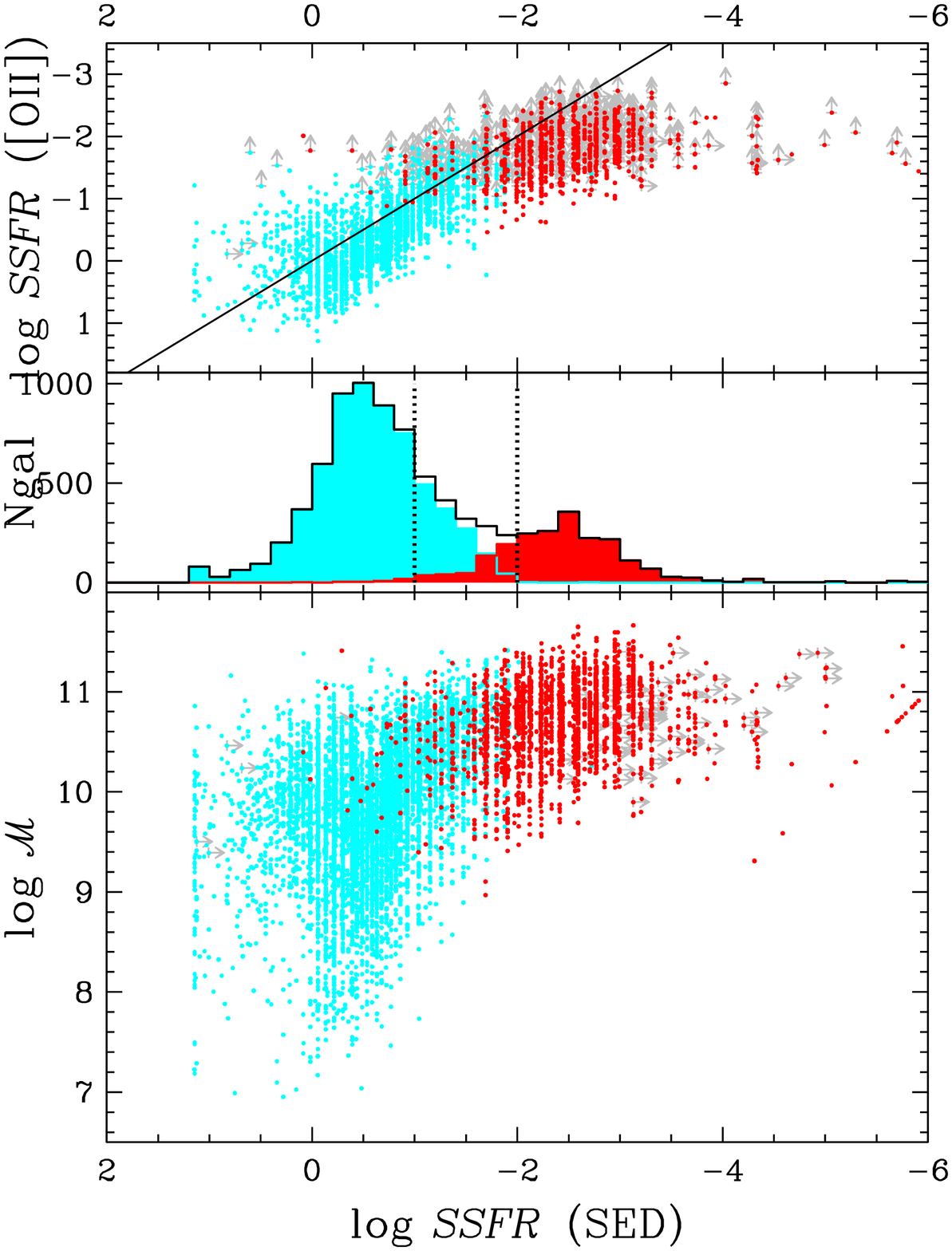}
\caption{ 
{\it Left figure:} 
The upper panel shows the $U-B$ rest-frame colour distribution of galaxies with different 
{\it photometric types}.
The bottom panel shows the colour-mass diagram.
{\it Right figure:} \SSFR\ of galaxies with different 
{\it photometric types}.
The upper panel shows the comparison between estimates from the SED fitting and from [OII] for galaxies 
in the range $0.5<z<0.9$. The middle panel shows the \SSFR\ distribution of galaxies, 
along with the two dotted lines used to separate active from quiescient (\logSSFR$=-1$) 
or from passive galaxies (\logSSFR$=-2$). The bottom panel shows the \logSSFR--mass 
diagram. Red dark and light cyan colours represent SED-ETGs (\PT=1) and SED-LTGs 
(\PT=2, 3, 4), respectively.
}
\label{fig:UB_Mass_SSFR}
\end{figure}

In Fig. \ref{fig:UB_Mass_SSFR} (left panel), we present the $U-B$ rest-frame colour distributions
of galaxies in our sample divided according to the {\it photometric types}. From this
figure, it is evident that the colour bimodality present in the sample
can be described well by
dividing the sample according to their {\it photometric type}: red galaxies that have 
\PT=1 (2 103 objects, also called SED-ETGs hereafter) and blue galaxies that have \PT= 2, 3, 4 
(6 347 objects, SED-LTGs), respectively.
In Fig. \ref{fig:UB_Mass_SSFR} (left panel), we also show the colour-mass diagram, which confirms 
that when our {\it photometric classification} is applied 
to the 10k-bright zCOSMOS sample it is consistent
with previous selections of red-sequence galaxies based on colour-luminosity diagrams 
(Bell et al. 2004 and Cimatti et al. 2006) and has the
advantage of using the entire multiband photometric coverage rather than only two bands. 
In principle, given the good multiband coverage, the SED fitting is able to break
the degeneracy between a dust-extinguished star-forming galaxy with a red but smooth SED and 
a passive, old galaxy with a strong $4000$ \AA\ break in its SED (Pozzetti \& Mannucci 2000).

\item {\it Star formation activity classification (active, quiescent, and passive galaxies):}

We divided the sample according to the galaxy star formation rate (\SFR) activity. 
We used the \SFR, and also the specific star formation rate (\sSFR=\SFR/\Mstar) 
derived from the SED fitting (see previous section) finding, in general, a good correlation 
between these and the \SSFR\ derived from [OII] (Maier et al. 2009) for galaxies in the 
redshift range $0.5<z<0.9$ (shown in Fig. \ref{fig:UB_Mass_SSFR}, right upper panel).
We therefore divided the sample into active and quiescent galaxies depending on whether 
\logSSFR\ is above or below $-1$, i.e., galaxies that would take less or more than 10 Gyr to 
double their \Mstar\ at the present \SFR, respectively.
In our sample, we find 5051 active galaxies and 3403 quiescent galaxies.
With this definition, we note that the number of galaxies in the high \sSFR\ sample is $\sim 80$\% of the blue 
galaxies defined on the basis of the SED fitting (see Fig. \ref{fig:UB_Mass_SSFR}, right middle panel).
In addition, we investigate in this paper the population of ``passive" galaxies, i.e.,  1612 galaxies 
with \logSSFR$<-2$. This dividing value separates well the two parts of the bimodal distribution of the \sSFR\
(Fig. \ref{fig:UB_Mass_SSFR}, right middle panel).

\begin{figure}[h!]
\centering
\includegraphics[width=0.49\hsize]{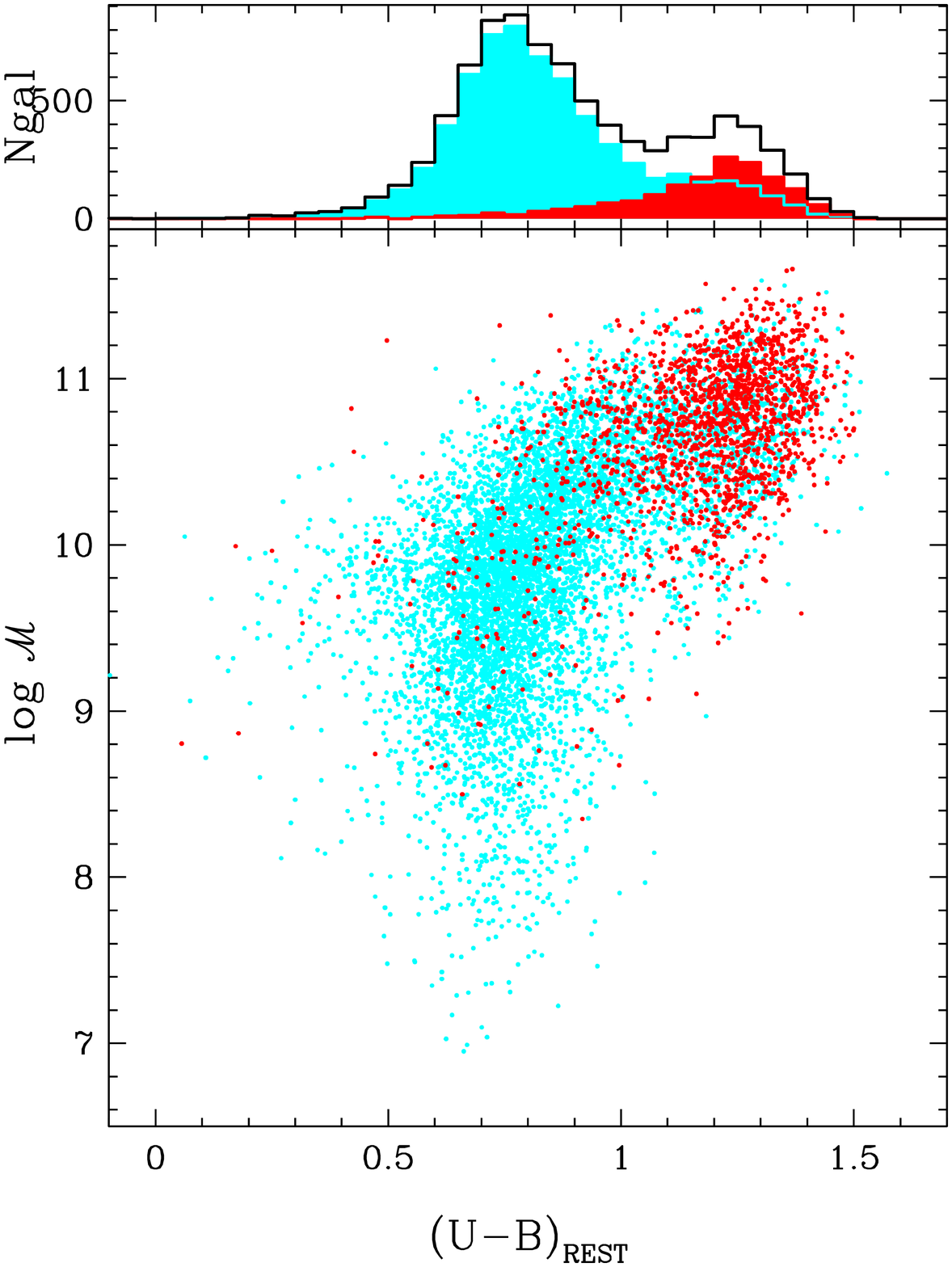}
\includegraphics[width=0.49\hsize]{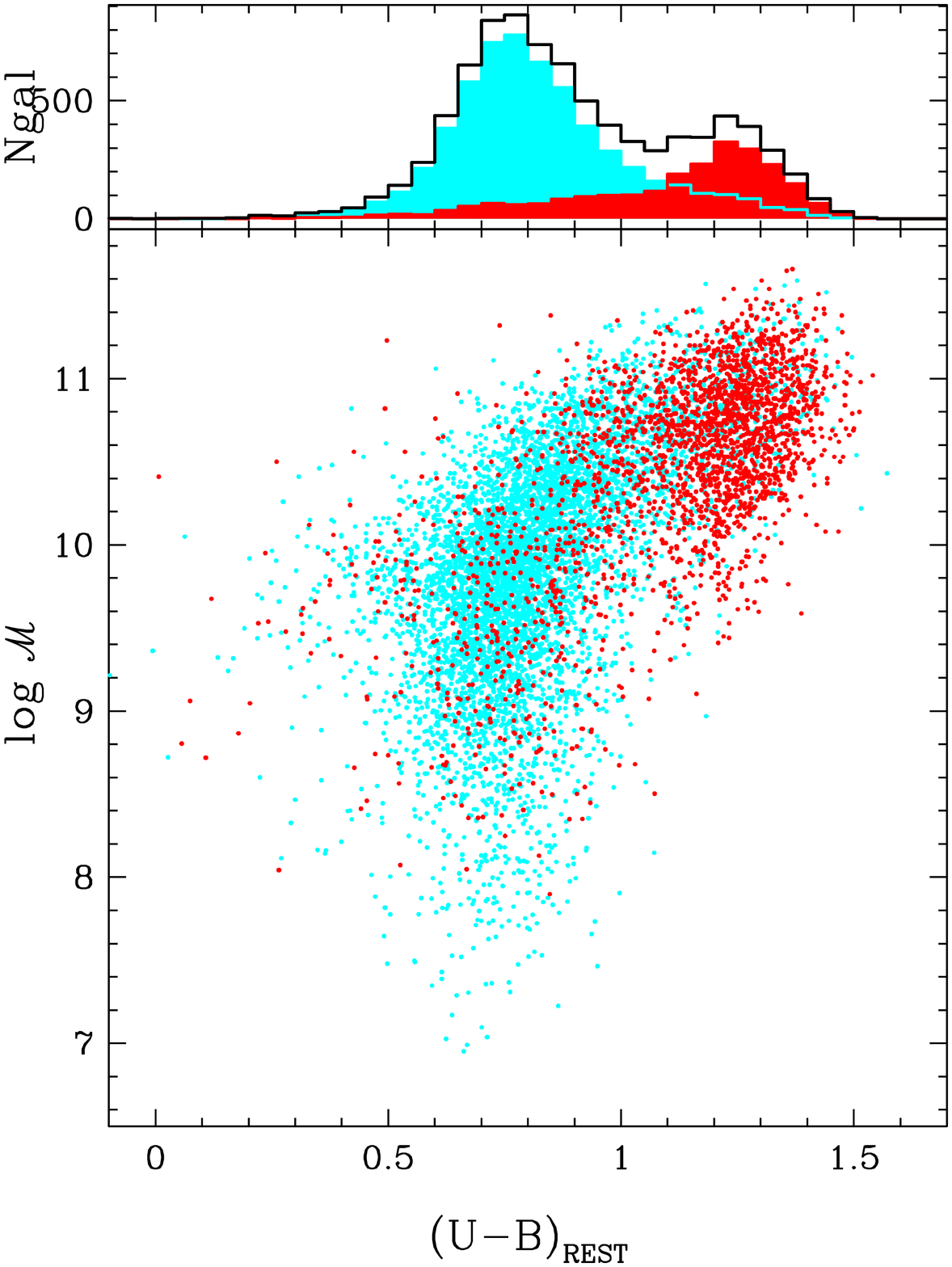}
\caption{$U-B$ rest-frame colours of galaxies with different morphological classes 
using ZEST (left panel) and MRS (right panel) classifications.
The colour-mass diagrams are shown in the bottom panels: red dark and light cyan colours represent 
morpho-ETGs (spheroids) and morpho-LTGs (disc+irregulars), respectively.
}
\label{fig:UBmorph}
\end{figure}

\item {\it Morphological classification (spheroidal and disc + irregular galaxies):}

Taking advantage of the COSMOS HST+ACS images (Koekemoer et al. 2007), we also divided the sample 
according to {\it morphological class}.
To take into account uncertainties in the 
morphological classification,  we used two of the available estimates, 
i.e., the Zurich Estimator of Structural Types (ZEST; Scarlata et al. 2007a) and 
a non-parametric estimate derived by our collaborators in Marseille (MRS hereafter, Cassata et al. 2007, 2008, 
Tasca et al. 2009). Only a small fraction ($\sim$ 3-4\%)
of objects do not have a morphological classification.

ZEST quantitatively describes the galaxy structure by performing a principal component 
analysis (PCA) in the five-dimensional parameter space of asymmetry (A), concentration (C), 
Gini coefficient (G), the second-order moment of the brightest 20\% galaxy pixels (M20; 
Abraham et al. 2003), and the ellipticity of the light distribution ($\epsilon$).  
The PCA indicates that the first three PC variables account for more than 90\% of the 
variance in the original data set, thus almost completely describing the galaxy structure.
The ZEST classification associates each PC value with a type (ZEST type, \ZT, =1 for 
early-type galaxies; =2 for disc galaxies; 
and =3 for irregular galaxies) and a ``bulgeness"
parameter according to the median value of the distribution of  S\'ersic indices $n$ (Sargent et al. 2007) 
of all galaxies brighter than
$I=22.5$ (bulgeness=0,1,2,3, from $n>2.5$ being bulge-dominated to $n<0.75$ being disc-dominated galaxies). 
In this study, the total sample was divided into morpho-ETGs and morpho-LTGs, with
the morpho-ETGs sample including 1680 galaxies classified by ZEST 
as elliptical (\ZT=1, 759) or bulge-dominated (\ZT=2.0) galaxies, while the morpho-LTGs subsample
includes 6413  galaxies (with \ZT\ greater than 2.0, i.e., from not-bulge-dominated to disc-dominated and 
irregular galaxies). In addition, we distinguish ellipticals (\ZT=1) 
from bulge-dominated galaxies (\ZT=2.0).

For comparison, we also used MRS morphological estimates. 
This classification scheme separates galaxies on the basis of
their position in the multi-dimensional parameter space of the four
non-parametric diagnostics of galaxy structures (C,
A, G,  and axial ratio). First, the structural parameters are
measured for all galaxies in the sample. Then, a randomly
extracted subset of 500 galaxies is visually classified (LT, PC) as
ellipticals, spirals, or irregulars. This reference catalogue is used to
explore the multi-dimensional parameter space. For each new
galaxy that needs to be classified, the distance in the parameter
space to the 500 reference galaxies is measured, and the 11 closest
selected. Finally, each galaxy is assigned to the most
frequent visual class among these 11 nearest reference galaxies.  This
procedure allows us to convert structural parameters into morphological
classes: spheroidals (MRS type, \MT=1), spirals (\MT=2), and irregulars
(\MT=3). 
For this paper, we also therefore used this morphological classification 
to divide the sample into morpho-ETGs and 
morpho-LTGs, as in the case of ZEST morphologies,
yielding 2383 spheroids (\MT=1) and 
5831 disc+irregular (\MT=2+3) galaxies, respectively.

In Fig. \ref{fig:UBmorph}, we show the $U-B$ rest-frame colour distributions and colour-mass diagrams of 
these different morphological classifications. For both morphological classifications, there is a non-negligible 
number of blue spheroidal-type galaxies ($\sim 37$\% in MRS and $\sim 29$\% in ZEST)
(see Sect. \ref{sec:ETGs}) as well as red disc+irregular galaxies ($\sim 13$\% in ZEST and $\sim 9$\% 
in MRS).

\item {\it Spectroscopic classification (emission- and absorption-line galaxies):} 
 
Using measurements of spectral features (Lamareille et al. in preparation), we divided the sample
into spectral-ETGs and spectral-LTGs following the criteria 
defined in Mignoli et al. (2009) for the zCOSMOS spectroscopic sample
in the $EW_0[OII] - D4000$ plane.
In the redshift range $0.55<z<1.0$, we find 
1079 galaxies without strong emission lines and large $4000$ \AA\ breaks
($EW_0([OII])<5$ \AA\  and $D4000+0.33 \log(EW_0([OII]))>1.5$, 
and spectroscopic type \ST=1, i.e., spectral-ETGs)
and 3304 galaxies with strong emission lines and small $4000$ \AA\ breaks
($EW_0([OII])>5$ \AA\  and $1.50 < D4000 + 0.33 \log (EW_0([OII])) <2.22$, spectroscopic type \ST=2, 
i.e., spectral-LTGs), 
respectively.

\item {\it Combined classification: a clean sample of ``bona-fide ETGs":}

Given the remaining controversy about morphology or colour-selected
early-type galaxies (see Franzetti et al. 2007 for a discussion of the contamination colour-selected samples),
we defined a more conservative sample of ``bona-fide ETGs" by combining different 
criteria related to colours, morphology, and spectral properties.
From the sample of red galaxies (with {\it photometric type}=1), we selected 
only those whose properties are more accurately described by the 4 reddest templates ($\sim 70$\%) 
and we removed galaxies with either strong emission lines 
($EW_0([OII])$ or $EW_0(H_\alpha)>5$ \AA, $\sim 30$\%) or reliably classified disc+irregular morphologies 
(using the intersection between MRS and ZEST estimates, $\sim 20$\%) or strong emission at 24 $\mu$m 
($K-m(24\mu m)>-0.5$, $\sim 5$\%),
obtaining a final sample of 981 red passive spheroids (``bona-fide ETGs"). For further details, 
we refer to Moresco et al. (2010).
\end{enumerate}

\section{The galaxy stellar mass function }\label{sec:GSMF}

\subsection{The method}

To derive the galaxy stellar mass function (\MF), 
we follow traditional techniques used in computing the
luminosity function. Here we apply the classical non-parametric $1/V_{\rm max}$ 
formalism (Schmidt 1968)
and estimate 
the best-fit Schechter (1976) parameters ($\alpha, {\cal M}^*, \phi^*$). In performing the fit, 
we allow ourselves to use up to 2 Schechter function components.

To correct for both the non-targeted sources in spectroscopy and those for which the
spectroscopic redshift measurement failed,
we use a statistical weight associated with each galaxy with a secure redshift
measurement. This weight is the inverse of the product of the {\it target sampling rate} (TSR)
and the {\it spectroscopic success rate} (SSR).
Accurate weights were 
derived by Bolzonella et al. (2009, see also Zucca et al. 2009) for all objects with secure spectroscopic
redshifts, taking into account the magnitude, colour, and redshift dependence of the SSR, 
as well as the objects observed as compulsory
($\sim 2\%$) and secondary objects ($\sim 2\%$) in the slits, which have different TSRs from the 
whole sample.

\begin{figure}
\centering
\includegraphics[width=0.99\hsize]{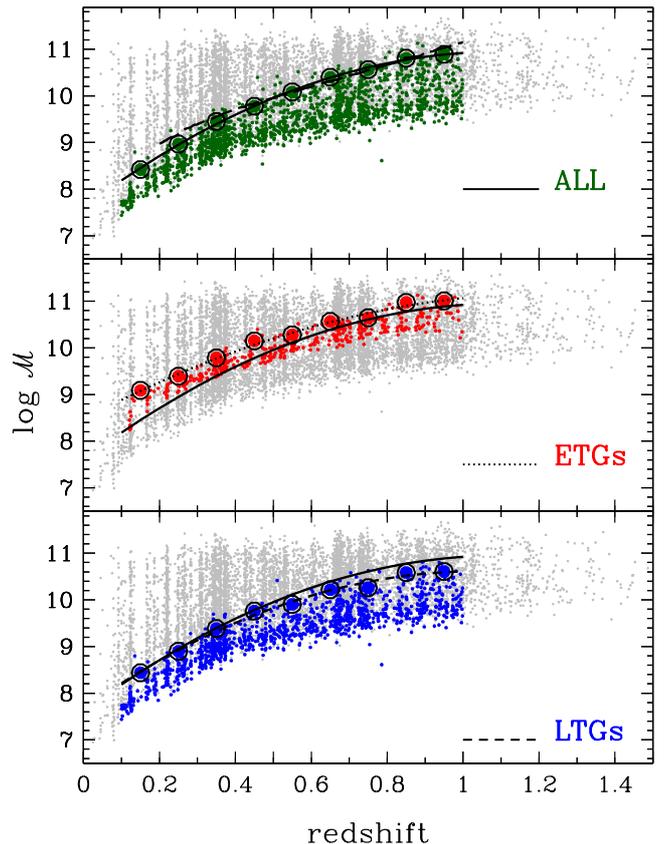}
\caption{Stellar mass as a function of redshift (small grey dots). Also shown
are \Mlim\ (intermediate size dark dots) and \Mbias\ (big black circles and lines) at 95\% of the \Mstar/L 
completeness level (see text): upper panel for the total population, middle and lower panel 
for SED-ETGs and SED-LTGs, respectively,
defined by their {\it photometric type}. Also shown with a dashed black line is the \Mstar\ threshold
(corresponding to 95\% of the completeness level for the global \MF) derived using mock Millenium samples 
by Meneux et al. (2009).
}
\label{fig:masslim}
\end{figure}

\subsection{The limits in mass}

In a magnitude-limited sample, the minimum stellar mass for which observations were completed 
depends on both the redshift and the stellar mass-to-light ratio \MstarL. This latter quantity obviously 
depends on the stellar populations and therefore on galaxy colours.
To account for this limit, we define at each redshift a minimum mass, \Mbias, above which the derived \MF\
is essentially complete because all types of galaxies are potentially observable above this mass.

\begin{figure*}
\centering
\includegraphics[angle=-90,width=0.99\hsize]{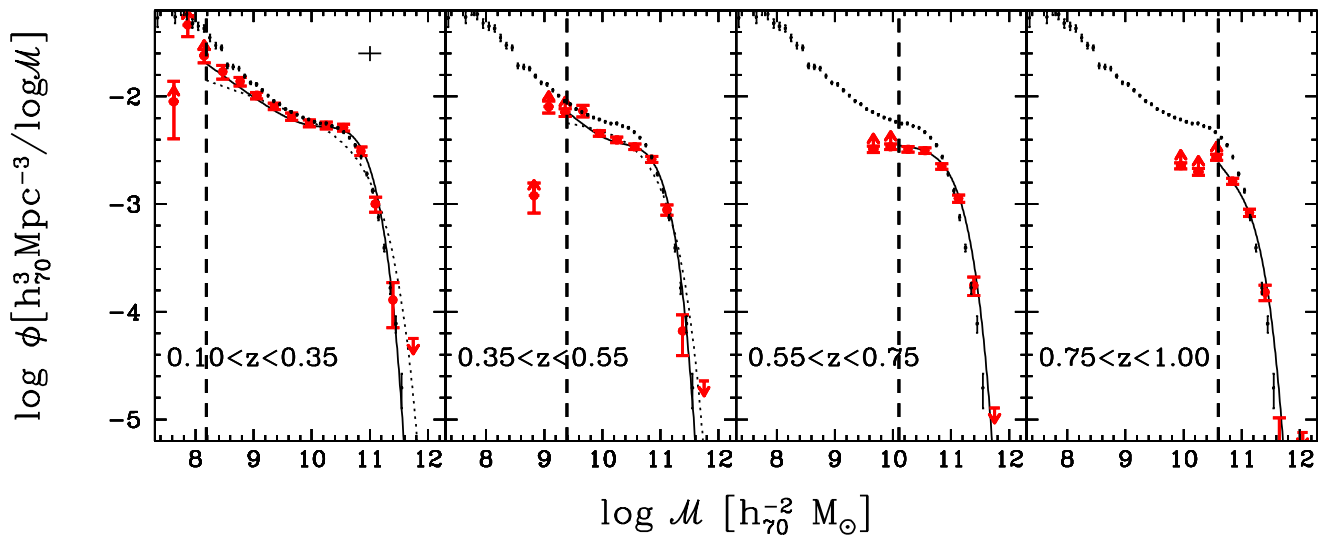}
\caption{Global galaxy stellar mass functions in four different redshift bins.
Red filled points represent \Vmax\ determination and associated Poisson errors, 
while the lines represent the Schechter fits.
The global \MFs\ are not reproduced well by a single Schechter function (dotted lines), but by two Schechter 
functions (continuous lines) up to $z=0.55$.
In the first two redshift bins, the \MFs\ show the inflection points of the bimodality 
around \Mstar $\sim 5 \times 10^{9}$ \Ms.
The dashed vertical lines represent the
mass limit in the corresponding redshift bin (\Mbias). 
Data are plotted as lower limits below \Mbias.
Upper limits at 1$\sigma$ (i.e. 1.84 objects) are shown at the high-mass end.
Small black dots in all panels represent the local \GSMF\ by Baldry et al. (2008).
The cross at the right top of the first panel shows an estimate of the cosmic variance and mass uncertainties.
}
\label{fig:MFsch}
\end{figure*}

To derive \Mbias, we calculate the limiting stellar mass (\Mlim) of each galaxy, 
i.e., the mass it would have, at its spectroscopic redshift,  if its apparent 
magnitude were equal to the limiting magnitude of the survey ($I_{\rm lim}=22.5$) given by
$\log({\cal M}_{\rm lim})=\log($\Mstar$)+0.4(I-I_{\rm lim})$.
The result is a distribution of limiting stellar masses, \Mlim, that reflects the distribution of 
stellar \MstarL\ ratios at each redshift in our sample.
To derive a representative limit for our sample, we use the \Mlim\ of the 20\% faintest 
galaxies at each redshift. This choice takes into account the 
colour-luminosity relation and therefore includes only galaxies with a typical \MstarL\ close to the magnitude 
limit. By doing this, we  avoid the artificial use of a too stringent limit related to the 
brightest and reddest (with the highest \MstarL) 
galaxies, which do not significantly contribute close to the 
magnitude limit of the survey.
Figure \ref{fig:masslim} shows the distribution of stellar masses for all galaxies and of \Mlim\ for 
the 20\% faintest galaxies. We then define \Mbias$(z)$ as the upper envelope of the \Mlim\ distribution 
below which lie 95\% of the \Mlim\ values at each redshift.
This \Mbias\ corresponds to a 95\% completeness limit to the \MstarL\ ratio at each redshift observable by 
the survey, and is taken to be the completeness limit of the \MF. 
Meneux et al. (2009) instead used
mock survey samples to derive a mass limit.
For a given redshift range and mass threshold, 
the completeness was simply defined
as the ratio of the number of galaxies 
brighter than the observed flux limit to those of all fluxes. 
Interestingly, even if this method is model-dependent,
it leads to a similar completeness limit as the previous one.
For example, for the global population we find that our \Mbias\ are in fairly good 
agreement with the 95\% completeness limit (black curve in Fig. \ref{fig:masslim})
derived by Meneux et al. (2009) for zCOSMOS, in most of the redshift range, and in any case
the completeness for \Mstar$>$\Mbias\ is never lower than 85\% at any redshift.

We derived \Mbias\ at each redshift and for each galaxy subsample used. 
For the global population, we note that \Mbias\ at low redshift is close to the limit of 
the  bluest population (see Fig. \ref{fig:masslim}), because blue galaxies dominate the 
zCOSMOS population close to the magnitude limit at that 
redshift, while the global \Mbias\ shifts towards that of the reddest population at high redshift
(at which redshift a significant number of red galaxies is present at the limit of our survey, and
have a higher \MstarL\ and therefore higher mass limit).

The \MFs\ derived using the $1/V_{\rm max}$ technique, which
corrects in addition for volume incompleteness, are formally complete for 
\Mstar$\ge$\Mbias$(z_{\rm inf})$, i.e., \Mbias\ 
 at the lowest redshift of the considered bin.
In the parametric fit, we 
estimate the best-fit Schechter parameters using data above \Mbias$(z_{\rm inf})$ and
plotting as lower limits the data below \Mbias, where the \MFs\ are incomplete.
In addition, we took into account the upper limits that are above the maximum mass 
found in each considered redshift bin, deriving 1$\sigma$ upper limits by following Gehrels (1986) 
in the case of Poisson statistic for $n=0$ events (i.e., $\le 1.84$ at  $1\sigma$).

\section{The bimodality in the zCOSMOS global galaxy stellar mass function}
\label{sec:bimodality}

The resulting \MFs\ of the zCOSMOS galaxy sample were derived 
for the redshift range $0.1<z<1.$ 
We divided this redshift range into 4 redshift bins to obtain an approximately comparable 
number of objects in each redshift bin. This choice also 
dilutes the effect of the most prominent large-scale structures at all redshifts (Lilly et al. 2009;
Kova\v{c} et al. 2010). 
Using the deviations from the median number densities in numerous redshift bins, affected by different 
structures, we estimate that the systematic error in our 
\MFs\ caused by cosmic variance is about 15-20\%.

Figure \ref{fig:MFsch} shows the global \MFs\ in different redshift bins derived using the \Vmax\ 
technique, with the associated Poisson errors.
The most clearly evident results are that at $z<0.55$ the global \MFs, derived 
using the $1/V_{\rm max}$ technique, are bimodal, exhibit an upturn 
with a steep slope below \Mstar $\sim 10^{9.5}$ \Ms, and
are not reproduced well by a single Schechter function. 
At higher redshifts, the narrower mass range due to the increase in \Mbias\ does not allow us to 
explore the upturn in the low-to-intermediate mass regime.
We therefore allowed the fit to include up to two Schechter functions.
We verified that the characteristic mass (\Mstar$^*$) of the second Schechter function, 
which dominates at low mass, is poorly constrained and consistent with 
the characteristic mass of the Schechter function dominating at high mass.
Therefore, following Baldry et al. (2008), we considered hereafter a double Schechter function with a single 
\Mstar$^*$ value, given by
\begin{equation}
  \phi(\mass) \, \D \mass = e^{-\mass/\mass^{*}} \left[ \phi^{*}_1 \left(
  \frac{\mass}{\mass^{*}} \right)^{\alpha_1}+ \phi^{*}_2 \left(
  \frac{\mass}{\mass^{*}} \right)^{\alpha_2} \right] \,
  \frac{ \D \mass }{ \mass^{*} },
\label{eqn:double-schechter}
\end{equation}
where $\phi_\mass \, \D \mass$ is the number density of galaxies with
mass between $\mass$ and $\mass + \D \mass$, and $\alpha_2 <
\alpha_1$, so that the second term dominates at the lowest masses.

\begin{figure}
\centering
\includegraphics[width=0.99\hsize]{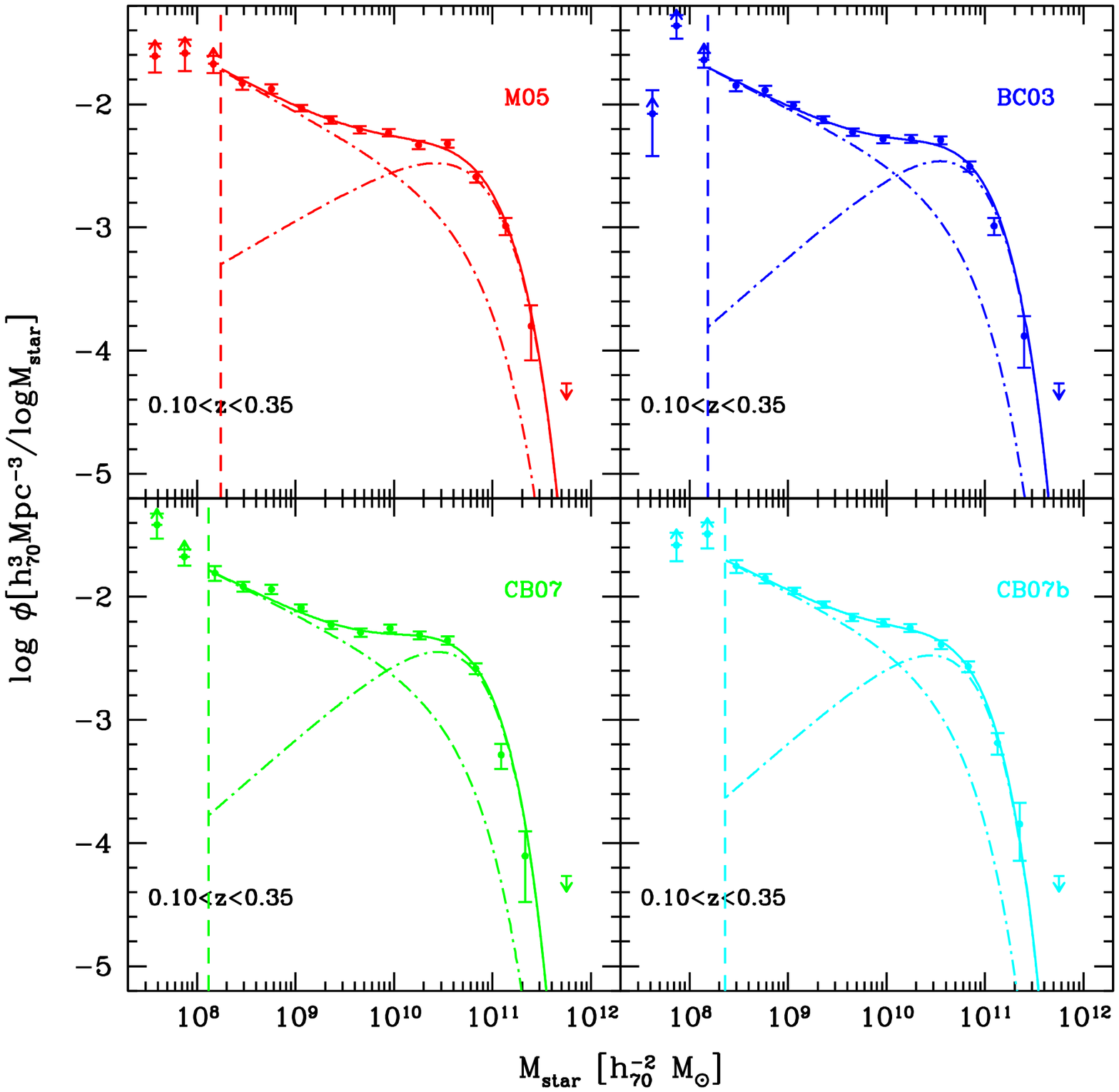}
\caption{Global \MF\ in the first redshift bin using different SFHs and population synthesis models 
to derive the stellar masses
(BC03, M05, CB07, CB07b=CB07+ secondary burst).
Points and lines have the same meaning as in Fig. \ref{fig:MFsch}.
The dot-dashed lines represent the 2 Schechter components:
they intersect around  \Mstar $\simeq 10^{10}$ \Ms.
}
\label{fig:MFmodels}
\end{figure}

Using galaxies with \Mstar$\ge$\Mbias, we find that the slope parameters of the 2 Schechter functions 
differ significantly.
Using the F-test, we find that the fit with two Schechter functions is tighter than the fit with a 
single function at $\ge 3 \sigma$ confidence level in the first two redshift bins.
We checked that our findings for the double Schechter shape are insensitive to the particular SFHs 
and set of population synthesis models used to estimate \Mstar. 
This is shown in Fig. \ref{fig:MFmodels}, where we plot  
the \MF\ derived in the first redshift bin ($0.1<z<0.35$) using different SFHs and population 
synthesis models to obtain the mass estimates.

The strength of the bimodality in the global \MF\ is related to the mutual ratio of the 
two Schechter functions 
that dominate the fit at the high and low-mass end, respectively. 
In Sect. \ref{sec:MFtypes}, we explore  whether
the populations of ETGs and LTGs can explain this bimodality in the \MF.

For many years, there has been evidence of a  `faint-end upturn' in the optical galaxy luminosity functions,
whereby the luminosity function is found to rise steeply about 3--5
magnitudes below the characteristic luminosity ($L^*$), in both clusters
(Driver et al. 1994, Popesso et al. 2006) and the field (Zucca et al. 1997;
Blanton et al. 2005).  
We note, however, that the upturn is not evident for all samples (Norberg et al. 2002).
The existence of a similar upturn is less clearly evident in the \MF.
Baldry et al. (2004, 2006, 2008) fitted a similar shape to the local \MF\ in the SDSS  
with an upturn at masses below $\sim 10^9$ \Ms.  Baldry et al. (2008)
converted the galaxy luminosity function of rich galaxy clusters in the SDSS (Popesso et al. 2006)
to an equivalent mass function and found an even more prominent upturn in their \MFs\ than that found for 
the field \GSMF. This upturn has  
not been clearly investigated until now in the \MFs\ derived from deep surveys, even if
a substantial population of low-mass galaxies ($<10^9 M_\odot$) at low redshift ($z\simeq0.2$),
consisting of faint blue galaxies, has been noticed in deep surveys such as the VVDS (Pozzetti et al. 2007).
We note, {\it a posteriori}, by comparing with other \MFs\ from deep surveys 
(VVDS, Deep2, MUSIC, GOODS, FDF, COMBO17)
that previous surveys (in particular MUSIC, Fontana et al. 2006, and COMBO17, Borch et al. 2006) 
show some  hint of bimodality in the shape of their \MFs, even up to $z\sim1$ 
(see Fig. 10 in Pozzetti et al. 2007).
Ilbert et al. (2010) also found that the \MFs\ derived in COSMOS using photometric redshifts cannot be 
reproduced well with a single Schechter function, 
but the upturn is less evident than in the zCOSMOS sample. This may be caused by
the higher precision of the \Mstar\ determination achieved by using spectroscopic redshifts
instead of photometric ones.

In the zCOSMOS survey, the dependence of the \MF\ shape, 
on the galaxy environment, as seen in the local Universe (Baldry et al. 2006),
is investigated in a parallel paper 
(Bolzonella et al. 2009), in which the bimodality in the global mass 
function is found to be even stronger in high density enviroments.

In the following sections, we study the evolution of the global \GSMF\ and
how its shape is related to the bimodality in the galaxy properties.
We derive the contribution of the different galaxy types to the \MF\ and its bimodal 
shape, and we explore its evolution with cosmic time.
In Sect. \ref{sec:MFz0}, we also explore how the evolution related to
SFHs, which differ between high- and low-mass galaxy populations, changes the shape of the \MF\ bimodality.

\section{The evolution of the galaxy stellar mass function}\label{sec:MFevol}

\begin{figure}[h!]
\centering
\includegraphics[width=0.99\hsize]{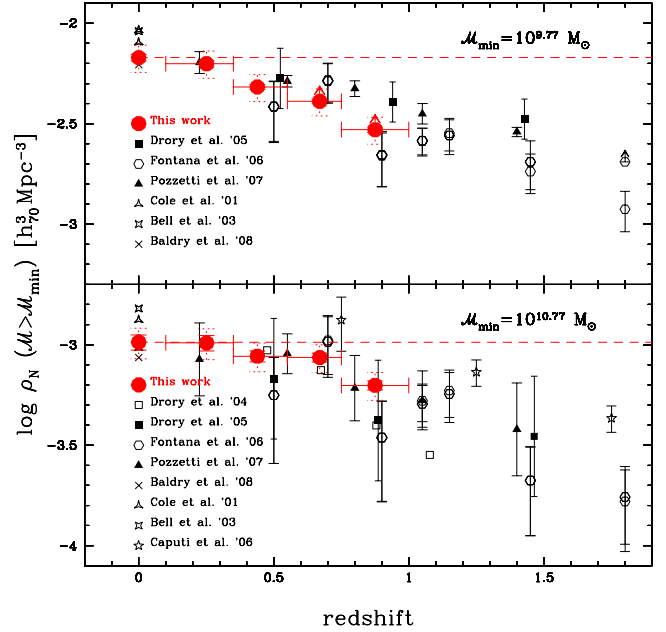}
\caption{Cosmological evolution of the galaxy number density as a function of
redshift, derived using \Vmax\ from the zCOSMOS for 2 different mass thresholds
($>10^{9.77} M_\odot$ and $>10^{10.77} M_\odot$ from top to bottom).
Dotted errors include cosmic variance estimates.
zCOSMOS data at $z=0$ have been derived from the ``SFH-evolved \MFs" (see Sect. \ref{sec:MFevol} and Fig. \ref{fig:MFz0}).
The dashed lines correspond to the no-evolution solution normalized at $z=0$.
Results from previous surveys (small black points)
are also shown.
}
\label{fig:NDliterature}
\end{figure}

We explore  how the evolution of the global \MF\ with cosmic time depends on galaxy mass.
From Fig. \ref{fig:MFsch},
a mass-dependent evolution of the global \MFs\ 
is clearly evident: the massive tail is almost stable up 
to $z=1$, while the number density of less massive galaxies increases continuously with cosmic time, 
in agreement with previous studies  
(Fontana et al. 2006, Pozzetti et al. 2007).
This mass-dependent evolution 
suggests that most of the massive galaxies assembled their mass earlier than lower mass galaxies
(``mass-assembly downsizing").
Figure \ref{fig:NDliterature} shows the number densities ($\rho_N$), derived by the
\Vmax\ method,  
for two different mass limits  of \logM$>9.77, 10.77$ (corresponding to \logM$>10, 11$ for 
a Salpeter (1955) IMF, 
as often used in the literature). We compare our results with literature data
from other high-$z$ deep surveys (VVDS, DEEP2, GOODS-MUSIC, FORS Deep Field, COMBO17;
see Pozzetti et al. 2007 for details and references), 
and for the local Universe (Cole et al. 2001, Bell et al. 2003, Baldry et al. 2008).
The local values for zCOSMOS were derived assuming an evolution from the lowest redshift bin to
$z=0$ (see next section for details).
zCOSMOS data are consistent with most previous observations at $z<1$ 
and are more accurately determined.
We confirm a continuous evolution in number density for \logM$>9.77$, which increases towards that of 
the local Universe,
and a slightly milder evolution for \logM$>10.77$, which is negligible for $z<0.7$ ($<20$\%) and more rapid 
above this redshift ($<60\%$ since $z\sim1$).

\begin{figure}[h!]
\centering
\includegraphics[width=0.99\hsize]{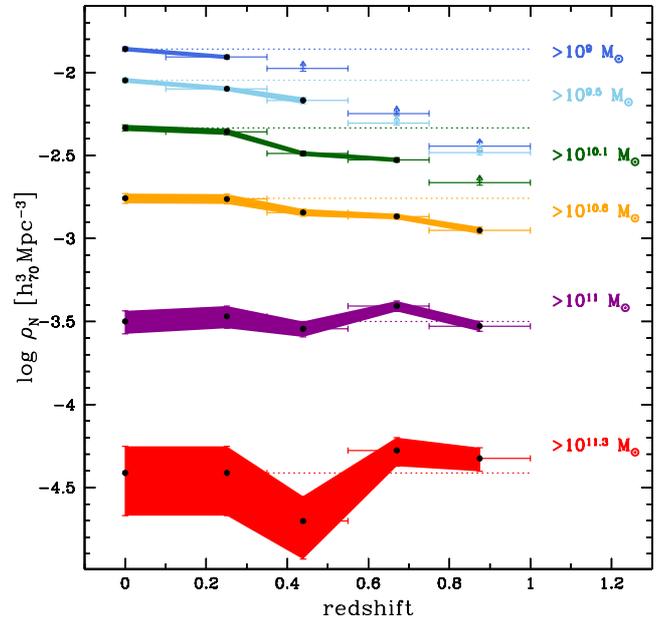}
\caption{Number density evolution using different mass limits for stellar masses.
zCOSMOS data at $z=0$ have been derived from the ``SFH-evolved \MFs" (see Sect. \ref{sec:MFevol} and Fig. \ref{fig:MFz0}).
The dotted lines correspond to the no-evolution solution normalized at $z=0$.
}
\label{fig:NDall}
\end{figure}

\begin{figure*}
\centering
\includegraphics[angle=-90,width=0.99\hsize]{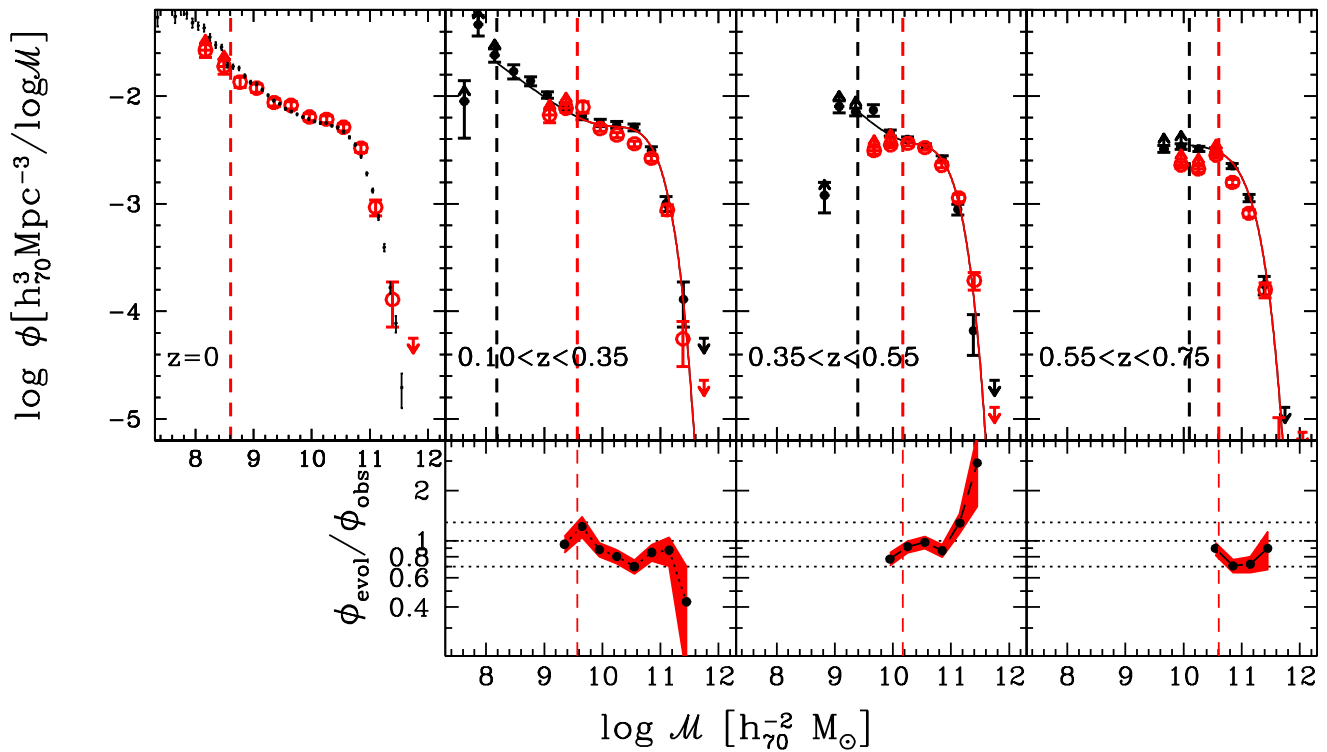}
\caption{{\it Upper panels:} ``SFH-evolved \MF" (empty red points) compared to
observed \MF\ at redshift $z$ (black filled points, from Fig. \ref{fig:MFsch}).
In the first panel, we plot the \MF\ evolved from the first redshift bin to $z=0$ (empty red points),
compared with the local \GSMF\ (Baldry et al. 2008, shown as small dots).
Vertical lines represent the respective mass limits.
{\it Lower panels:} Ratio of ``SFH-evolved \MF" to observed \MF\ (both estimated using \Vmax\ technique).
The dotted lines correspond to $\pm 30\%$ around 1.
}
\label{fig:MFz0}
\end{figure*}

The same trend with mass is evident in 
Fig. \ref{fig:NDall}, which shows the number densities ($\rho_N$) as a function of redshift  using  a more extended range in mass limits 
(from \logM$>9$ to \logM$>11.3$). Over the entire mass range, our data are consistent with a  
faster and steeper increase in the number densities  with cosmic time going from high to low-mass galaxies:
the evolution from $z=0.44$ to $z=0$ is $0.12\pm 0.02$ dex for \logM$>9.5$ and 
$0.09\pm 0.03$ for \logM$>10.6$.
Only galaxies with \logM$>11$ show negligible evolution from $z=1$ to the present time ($0.03\pm0.07$ dex), while
the evolution is $0.19\pm 0.03$ dex for \logM$>10.6$, 
the lowest mass for which we are complete over the whole redshift range.

zCOSMOS data are consistent with a mass-dependent assembly history,
with more massive galaxies evolving earlier than lower mass galaxies (mass-assembly downsizing).
We therefore confirm that mass is an important parameter driving galaxy evolution. 
In the context of the hierarchical scenario,
galaxy formation is predicted to be controlled primarily by their dark matter halo mass, and therefore,
one would expect the evolution of individual galaxies to be affected also by their enviroment.
The relative importance of stellar mass and galaxy environment to the shape and evolution of the \GSMF\
is investigated for the zCOSMOS dataset in a parallel paper (Bolzonella et al. 2009). 
This study found an accelerated trend of  downsizing in overdense regions (see also Iovino et al. 2010).

A detailed comparison of the global \MF\ with hierachical semi-analytical models (SAM) 
is postponed to a future paper. 
In Sect. \ref{sec:SAM}, we compare the predictions of SAMs with observational ETG \MFs.
Here we recall that most hierarchical galaxy assembly models
are unable to fully account for the observed \GSMF\ and its evolution
(see Fontana et al. 2004, 2006;  Caputi et al. 2006;  Kitzbichler \& White 2007; Marchesini et al. 2009, 
Fontanot et al. 2009 for a detailed comparison with models). 
A revision of the physical treatment of the baryonic component, such as its star-formation
history/timescale, of the role of feedback, dust content, and/or AGN feedback
(Menci et al. 2006; Monaco et al. 2007; Bower et al. 2006), and
the introduction of the uncertainties in the mass determination, may help to reduce the disagreement
(Kitzbichler \& White 2007; Cattaneo et al. 2008; Fontanot et al. 2009).
Fontanot et al. (2009) compare a broad compilation of available 
data sets with the predictions of three different semi-analytic models of galaxy formation 
within the $\Lambda CDM$ framework.
When observational errors on stellar mass are taken into account, 
they also find that the models acceptably 
reproduce the observede number density of massive galaxies (\logM$>11$), but that
low-mass galaxies (\logM$=9-10$)
are predicted to form too early in the models and are too passive at late times.
Thus, the models do not correctly reproduce the downsizing trend in stellar mass.
We show in Sect. \ref{sec:SAM} that the typical uncertainties in the mass determination cannot 
fully account for the 
observed excess of massive galaxies relative to some hierarchical model predictions, and, when applied, 
the hierarchical models tend to overestimate the
high-mass galaxy number density at low redshift 
(see also Fig. 1 in Fontanot et al. 2009).

In the following section, we analyse
whether the evolution of the observed \GSMF\ with cosmic time is driven mainly by the SFHs
at any given mass, and quantify the residual importance of merging events.

\subsection{The importance of the SFHs to the evolution at $z<1$}\label{sec:MFz0}

We attempt to interpret the mass-dependent evolution with cosmic time found for the \GSMF, 
i.e., the increase in number density at a given \Mstar\ or the shift in \Mstar\ at a given number density, 
in terms of only a ``pure growth in stellar mass" with cosmic time 
related in turn to the mass-dependent star-formation history of each galaxy 
(Thomas et al. 2005, Noeske et al. 2007a,b).
In this way, we aim to quantify the residual importance of merging events or other processes
after accounting for the role of the 
galaxy star formation activity, which drives the growth in stellar mass with cosmic time. 
A ``SFH driven evolution" was also adopted by Bell et al. (2007), assuming a constant 
specific SFR between two adjacent redshift bins and an instantaneous return fraction of 45\%.

For each galaxy, we derived the \Mstar\ evolved following its SFH (exponentially decreasing) 
derived from the SED fitting. 
Indeed, the best-fit parameters of the stellar population models (such as the star-formation history timescale
$\tau$  and the age) derived from the observed SEDs, allow us to follow the evolution of stellar mass
in cosmic time, based on the assumption that the galaxies evolved in isolation (i.e., without merging) 
and continued to form stars in a smooth way with the same SFH 
derived from the SED fitting. 
For each galaxy at redshift $z$, we derived the \Mstar\ evolved to the mean redshift of
the previous (i.e., lower) redshift bin.
The cosmic time elapsed between the mean redshifts of two adjacent redshift bins ranges between $1.1$ and 
$1.6$ Gyr.
For galaxies in the lowest redshift bin ($z_{\rm mean}=0.25$), we predicted the 
 evolved \Mstar\ to $z=0$,
i.e., after about $3$ Gyr.
We performed simulations to test the reliability of this ``evolved stellar mass" (\Mstar$_{\rm evolved}$): 
using models, we created a simulated multi-band catalogue at the depth of our catalogue 
(following Bolzonella et al. 2000),
from which we were able to recover not only the present \Mstar\ of the input model 
(see also Pozzetti et al. 2007) but also the 
model ``evolved stellar mass" after a given time; the agreement overall is satisfactory,
with no systematic shifts and 
with the typical dispersions caused by statistical uncertainties and
degeneracies being of the order of $0.13-0.21$ dex, depending on the elapsed time.

We find that the mass growth derived from the SED fitting in our sample is on average about a factor 
$1.7$ ($0.24$ dex) for 
\Mstar=$10^{9.5}$ \Msun\ and $1.22$ ($0.09$ dex) for \Mstar=$10^{10}$ \Msun, 
but reaches a factor of $30$ 
for the most extreme star-forming galaxies with \logM$<10$.
In Fig. \ref{fig:MFz0}, we show the \MF\ predicted at the average redshift of each bin 
using the galaxies in the higher redshift bin 
and compare it with the measured \MF\ at the observed redshift.
In the first panel, we also show the \MF\ predicted at $z=0$ using galaxies in the first redshift 
bin ($0.1<z<0.35$).

\begin{figure}
\centering
\includegraphics[width=0.99\hsize]{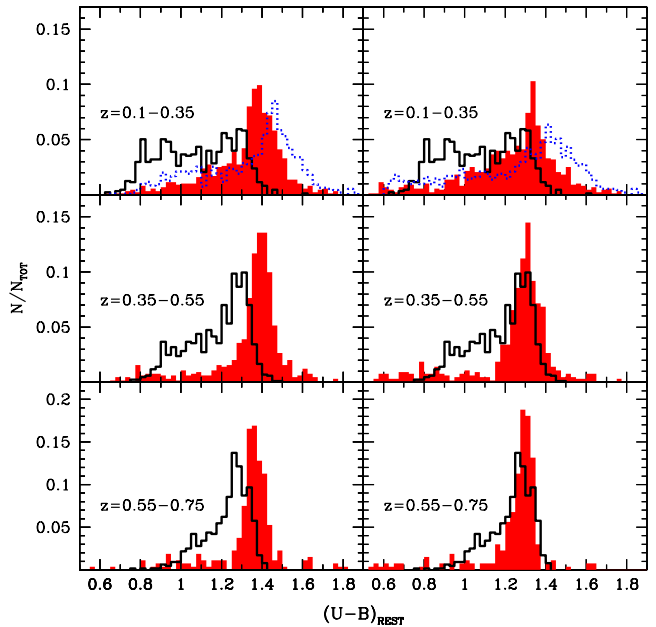}
\caption{Rest-frame $U-B$ colour distribution in a mass-limited sample 
(using the most conservative limit, see text)
in different redshift bins, as observed (solid black histogram for the total population) and predicted 
using the ``SFH-evolution" from the previous higher redshift bin (red filled histogram).
The dotted blue histogram in the top panels represent the expected $z=0$ colour distribution  
by applying the ``SFH-evolution" to the galaxies in the first redshift bin.
{\it Left panels}: assuming no evolution in the dust content, 
i.e., the same dust content derived from the SED fitting at the observed redshift; 
{\it right panels:} assuming no dust content in the evolved galaxies.
}
\label{fig:UBz}
\end{figure}

First of all, we note a modest evolution from $z\simeq 0.25$ to $z=0$,
in particular for low-to-intermediate mass galaxies (\logM$<10$)
with a shift towards higher masses being related to the SFHs. 
The evolved \GSMF\ approaches that observed locally
(Baldry et al. 2008) with which it is rather consistent for \logM$>9$.

For the zCOSMOS data set, we find reasonable agreement at each redshift between the 
``SFH-evolved \MF" and that observed at the same redshift. Close agreement at $z>0.3$ was also 
found by Bell et al. (2007) for COMBO-17 by assuming a constant SFH.
Our result suggests that the
\MF\ evolution is, therefore, driven mostly by smooth and decreasing SFHs, because of
the progressive exhaustion of the gas reservoir, for example from cold gas accretion,   
rather than by merger events or major bursts.
Very similar results
of a large contribution to the mass growth rate by the less massive and star-forming galaxies 
were obtained independently by Vergani et al. (2008) in the VVDS.
Figure \ref{fig:MFz0} (lower panel) shows the ratio of the evolved to
the observed \MFs\ in the first three 
redshift bins for \Mstar$>$\Mbias\ of evolved galaxies. 
The agreement is good particularly of intermediate-to-low masses (\logM$<10.6$), 
where most of the \MF\ evolution 
occurs and could therefore be explained mainly by the growth in mass driven by the SFHs 
in intermediate-to-low mass galaxies. 
At higher masses, galaxies have lower \SSFRs\ and therefore the SFHs are unable to explain the evolution 
observed in the MF, which in any case is quite small (see Fig. \ref{fig:NDall}). 
In the next section, we explore the role of mergers in the evolution of the \GSMF.

At all masses, we find differences at most of $\sim40$\%,
and even lower than 20\% for \logM$<10.3$, between the evolved and observed \MFs.
These residual differences may be related to different processes. 
Among them, one possibility is an additional SFR, i.e., either
more prolonged SFHs than inferred from the SED fitting, 
a different SFH functional form (Renzini 2009), or secondary bursts. 
From the SED fitting, we find that for the blue galaxies
the timescale ($\tau$) distribution for the exponential SFHs has 
a median value of $1$ Gyr, 
even if extended to higher values. This median timescale
is quite short relative to the value inferred from the global SFHs 
(the compilation of Hopkins \& Beacom 2006 is consistent with $\tau\sim3.5$ Gyr at $z<1$).
On the other hand, the residual differences could also be explained by merger events (see next section) 
or other dynamical processes that are able to induce a growth in stellar mass
unrelated to the SFH.

\begin{figure}[h!]
\centering
\includegraphics[width=0.99\hsize]{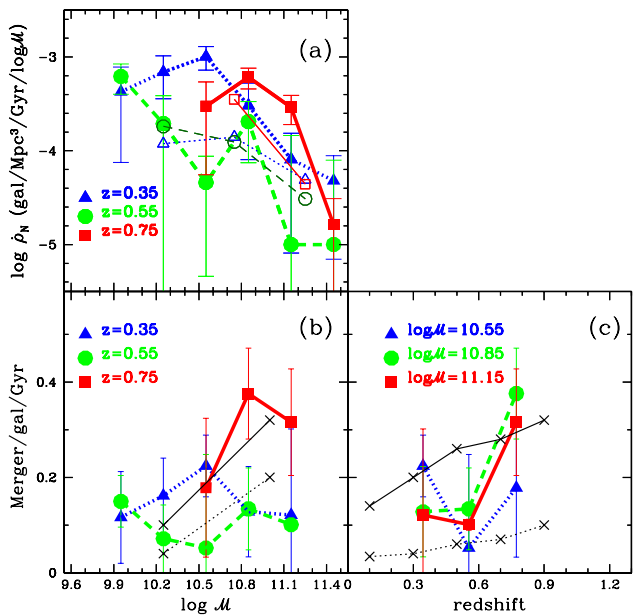}
\caption{{\it Panel (a):} Growth rate in number density after accounting for star formation evolution. 
Different thick lines and filled colour points represent estimates at different redshifts 
(dotted, dashed and solid for $z\sim 0.35, 0.55$, and $0.75$, respectively). 
For comparison, thin lines and empty points are the major merger rate estimates by de Ravel et al. (2010).
{\it Panel (b):} Merger per Gyr per galaxy as a function of mass. Same lines and points as in panel (a). 
Predictions based on cosmological simulations from Stewart et al. (2009), for mergers with mass 
ratio $m/{\cal M}>0.1$, are plotted as thin lines and black cross for the lowest and highest redshift bin.
{\it Panel (c)} Merger per Gyr per galaxy as a function of redshift. Thick lines and points represent various 
mass ranges (dotted, dashed and solid for \logM$\sim 10.55,10.85,11.15$, respectively).
Predictions from Stewart et al. (2009) for $m/{\cal M}>0.1$ are plotted as thin lines and 
black cross for \logM$\sim10.3$ (dotted line) and \logM$>11$ (solid line).
}
\label{fig:totmerge}
\end{figure}

Here, we also checked that the predicted colour distributions for the 
``SFH-evolved galaxies" were consistent with the observed colour distributions at different redshifts.
Figure \ref{fig:UBz} compares the expected colour distributions to the observed ones at different 
redshifts for two extreme hypotheses 
about the dust content: (A) (left panel): no evolution in the dust content, i.e., 
assuming the same dust content at all redshifts as derived from the SED fitting at the observed redshift, 
or (B) (right panel) no dust content in the evolved galaxies. In both panels, we 
compare the colour distribution for galaxies in a mass-limited sample using the most conservative 
limit \Mbias($z_{\rm sup}$) for evolved masses. 
Model A predicts too red colours at all redshifts, while only the extreme case of model B, 
i.e., without any dust content, is consistent with the observed colour distributions, at least at $z>0.35$,
while, in any case, it predicts too red colours at lower redshift. 
This could be related (again) to a too rapid decline in decreasing (quenching) of the SFRs of each galaxy as 
inferred by the SED fitting technique.
We therefore conclude that more extended SFHs or secondary bursts appear to be necessary to 
more accurately reproduce the observed colours, in particular at low-redshift.
One possibility would be to directly model 
the observed decrease/quenching of the \SFR\ at later cosmic times for progressively lower masses 
(see also Sect. \ref{sec:types}). 
We postpone the discussion of this possibility to a future paper.

\subsection{A limit to the galaxy merging}

The small amount ofevolution observed in the \GSMF\ argues against a dominant
contribution of galaxy mergers to galaxy evolution.
If merger events, indeed, were efficient in forming the most massive galaxies at $z<1$ 
(as predicted by hierarchical models),
we should observe instead a detectable increase in the number density of massive galaxies. 
In particular, the negligible evolution in the \GSMF\ for galaxies with \logM$>11$ up to
$z\simeq 1$ suggests that these galaxies formed and assembled their stellar mass at higher
redshifts, ruling out a major role for mergers at $z<1$.

\begin{figure*}
\centering
\includegraphics[width=0.49\hsize]{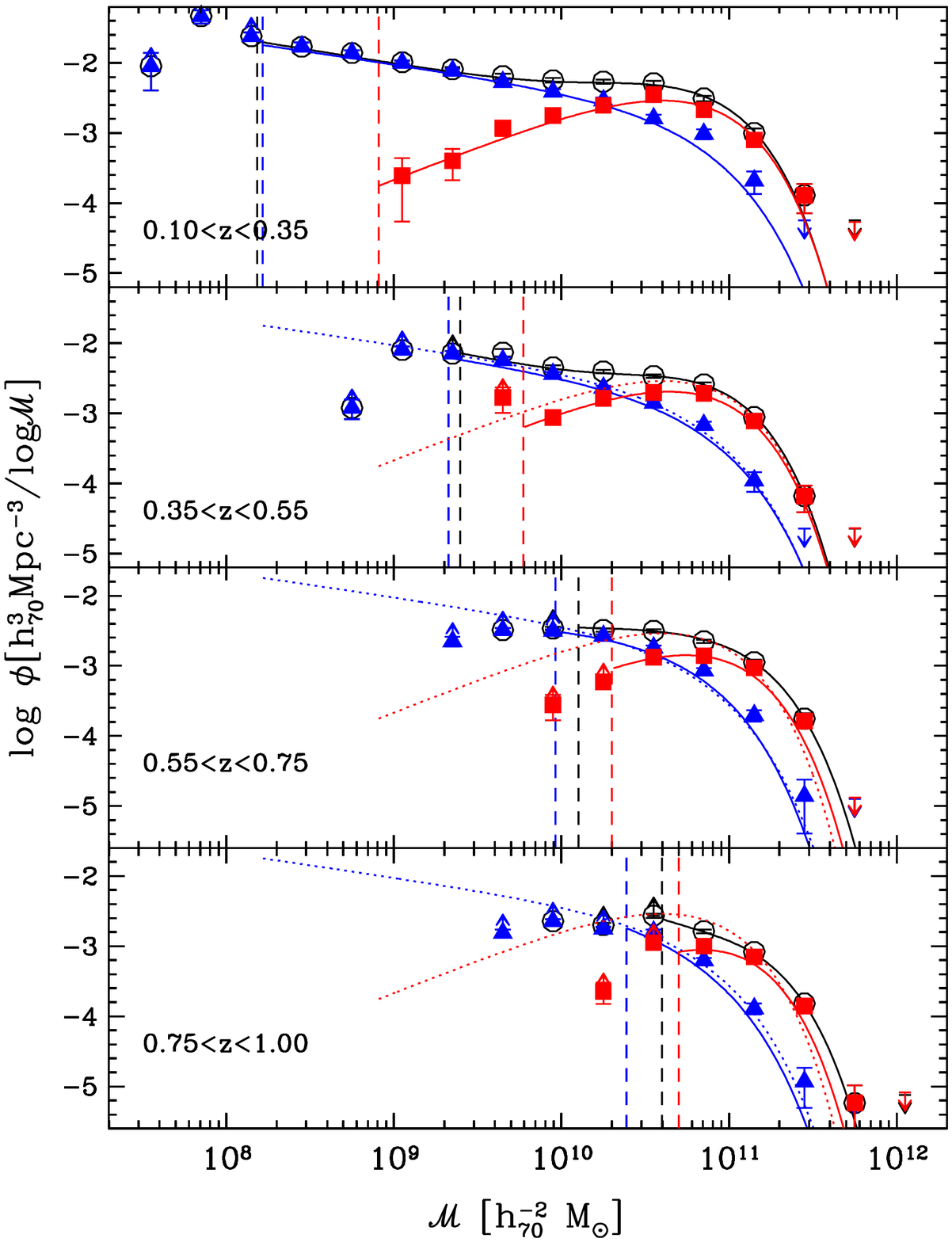}
\includegraphics[width=0.49\hsize]{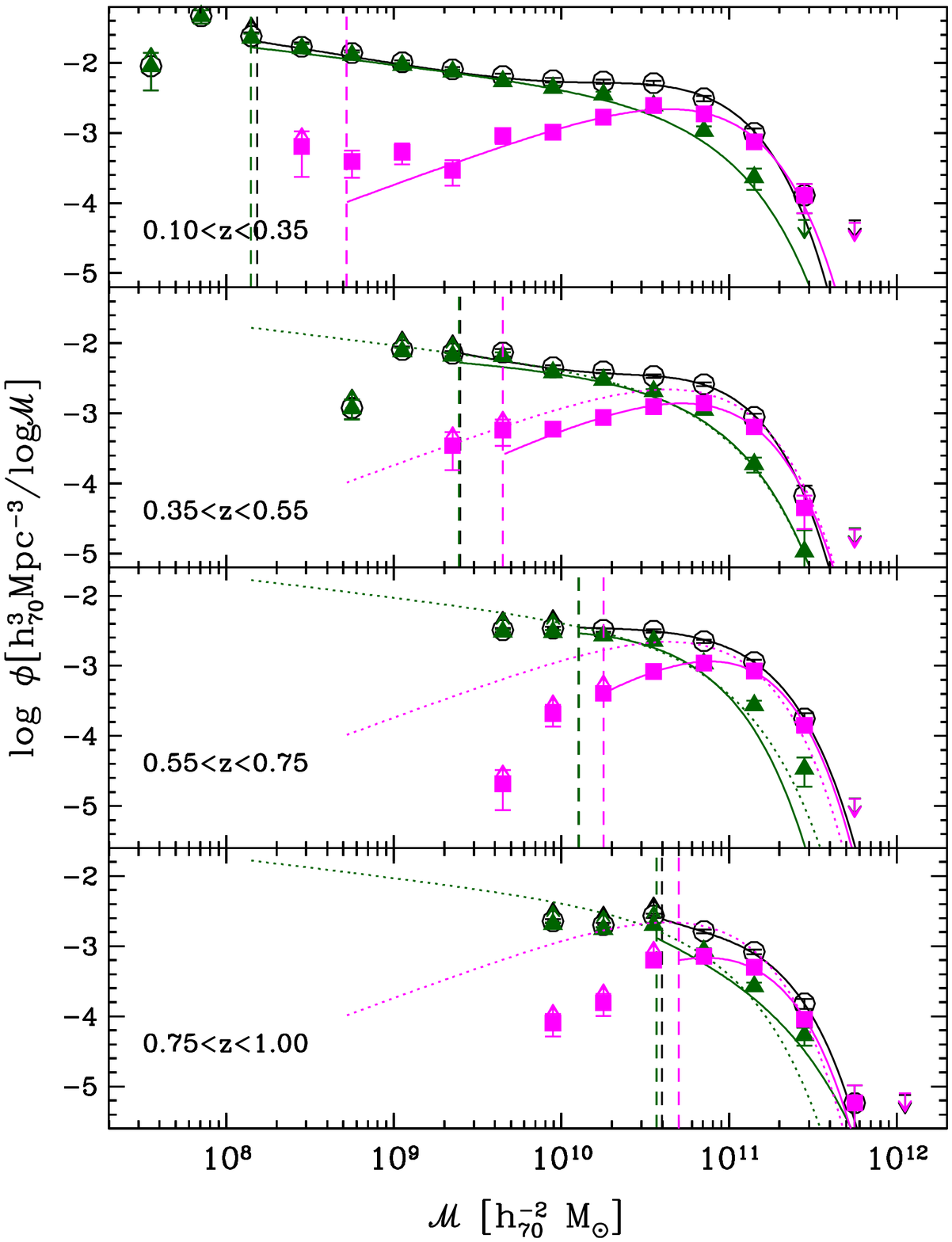}
\caption{Galaxy stellar mass function by galaxy types.
{\it Left panel:} \MFs\ by {\it photometric types} (\PT=1 as red squares, \PT=2+3+4 as blue triangles, 
total population as empty black circles).
{\it Right panel:} 
\MFs\ for {\it morphological types} (\ZT=1,2.0 as magenta squares and \ZT$>$2.0 as dark green triangles)
In both panels, points represent the \Vmax\ determination, while continuous lines the Schechter fits. 
Dotted lines reported in each panel, as a reference, are the Schechter fits to the first redshift bin.
Dashed vertical lines represent the
mass limit for the corresponding redshift bin (\Mbias). 
}
\label{fig:MFtypes_SSFR}
\end{figure*}

We derived the contribution of mergers to the evolution of the \GSMF.
We first estimated the growth rate of the number densities for different \Mstar,  
after accounting for the growth in mass produced by ongoing  star formation (see previous section).
The residual growth rate of the number density was calculated to be
\begin{equation}
\dot \rho_{N} ({\cal M},z)=\frac{[\rho^{obs}_N({\cal M},z)-\rho^{evol}_N({\cal M},z_{sup})]}{\Delta t (z,z_{sup})},
\label{eqn:growthrate}
\end{equation}
where $\rho_N$ was evaluated using the \Vmax\ method. 
The result is shown in Fig. \ref{fig:totmerge} ({\it panel a}) as a function of \Mstar\ in the various 
redshift bins. We compare these values with the 
major-merger rate ($\dot \rho_{mm}$) derived from galaxy pairs by de Ravel et al. (2010) for the zCOSMOS sample.
It is evident that major-merger rate is always below the growth rate by a factor of 2 or more, except
in the intermediate redshift range ($z\sim 0.5$).
This suggests that major mergers do not contribute significantly to the residual evolution.
We, therefore, derived  the merger rate (major+minor) that could explain the residual evolution.
We estimated the effective increase in number densities at any given \Mstar\ by accounting also for
galaxies that at the same time leave the current mass bin due to major merging. {\bf This estimate represents the
number of galaxies produced by mergers (major+minor) per Gyr per Mpc$^3$ to explain the \GSMF\ growth at 
that mass. Finally, assuming that a galaxy is the product of the last single merger
(within galaxy pairs of any mass ratio $m/{\cal M}$),
we derived the number of mergers per galaxy per Gyr dividing by the number densities of galaxies.
We note, indeed, that the time interval between two adjacent redshift bins is not long 
(1-1.5 Gyrs) relative to the typical merger timescale used in SAMs (De Lucia et al. 2010) 
or predicted by numerical simulations (Boylan-Kolchin et al. 2008)
that prevent multiple minor mergers predominating.
We, therefore,  derive the merging rate per galaxy per Gyr ($f_{\rm M}$) 
in the following way:}

\begin{equation}
f_{\rm M}({\cal M})=\frac{[\dot \rho_{N} ({\cal M},z) + \dot \rho_{mm}({\cal M},z_{sup})]}{\rho_N ({\cal M},z)} 
~~ merger/gal/Gyr.
\label{eqn:fmerge}
\end{equation}

The results are shown in Fig. \ref{fig:totmerge} as a function of mass ({\it panel b}) and redshift 
({\it panel c}). 
At high redshift, we find for massive galaxies (\logM$>10.6$ at $z>0.5$) that the merger rate 
is quite high, $\sim 0.2-0.4$ merger Gyr$^{-1}$ gal$^{-1}$, but 
decreases rapidly to $\sim 0.1$ with cosmic time, being only marginally consistent with an increase with mass.
At lower redshifts, the merger rate does not show any clear dependence on mass and always remains below 
$0.2$ at all masses. On average, we measure 0.16 merger Gyr$^{-1}$ gal$^{-1}$ since $z=1$.
The major merger rate per galaxy derived by De Ravel et al. (2010) is always below $0.1$ at all 
redshifts and masses explored ($\sim 0.06$ on average).
Integrated over cosmic time (from $z\simeq1$ to $z\simeq0.1$), massive galaxies (\logM$\sim10.6$) have 
experienced about 0.7  merger gal$^{-1}$ ($<0.3$ below $z\sim0.7$), of which less than 0.2 are major mergers. 
In Fig. \ref{fig:totmerge} we compared our estimates with the model predictions of Stewart et al. (2009) 
for mergers within galaxy pairs 
of mass ratio greater than 0.1 ($m/{\cal M}>0.1$). Galaxy mergers in a hierarchical $\Lambda$CDM scenario 
increase with stellar mass, redshift, and decreasing $m/{\cal M}$ ratio.
Minor mergers contribute considerably more than major mergers, 
dominating in all but the most massive galaxies (Parry et al. 2009).
Indeed, even in hierarchical formation models, most of ellipticals  and 
spiral bulges acquire their stellar mass through minor mergers or disc instabilities.
Given our uncertainties, we cannot draw any firm conclusions about either the redshift or 
mass dependences, even if our data appear to exclude a strong dependence on mass, at odds with hierarchical
predictions. There is, instead, some evidence of an increase with redshift, 
qualitatively in agreement with models.

We therefore conclude that
the role of major merging events is not dominant and cannot completely explain the evolution of the 
\GSMF, even after accounting for the growth in mass produced by ongoing star formation. 
Mergers in general (major+minor)
at a rate of between 0.1 and 0.4 Gyr$^{-1}$ gal$^{-1}$ may account for the residual evolution,
these rates possibly decreasing with cosmic time but not having a clear dependence on mass.
Oesch et al. (2010) discuss in greater detail the role of merging in the evolution of massive galaxies 
within the COSMOS survey.
In a future paper on the final zCOSMOS sample, we will explore in detail the role of both major 
and minor merging on the evolution of the \GSMF.

\section{Mass functions by galaxy types}\label{sec:MFtypes}

To explore the bimodality observed 
in the \MF\ and how it depends on both time and galaxy-type evolution,
we derived a type-dependent \MF\ using the different classifications described in Sect. \ref{sec:types}.

Figure \ref{fig:MFtypes_SSFR} (left panel) shows the \MF\ divided into different {\it photometric types},
red (\PT=1) versus blue (\PT=2,3,4)  galaxies; the  \MFs\ obtained by using morphological criteria to 
differentiate between ETGs and LTGs are shown in the right panel of the same figure.
We find that the red/spheroidal and blue/disc+irr populations have very different \MFs,
which can be linked to the bimodality of the global \MF. 
The \MFs\ of both ETGs and LTGs are reproduced quite well by a single Schechter function.
The LTG \MFs\ exhibit a steep low-mass end,
which in constrast with is flat for the ETGs at all redshifts. 
However, we find a hint of a small upturn in both the \MFs\ of ETGs for the morpho-ETGs 
(see Fig. \ref{fig:MFtypes_SSFR}) and in the second redshift bin of our other ETG  
classifications (see Fig. \ref{fig:MFearly}).
Drory et al. (2009), instead,  
found that neither their red (passive) nor their blue (star-forming) galaxy stellar mass functions 
could be fitted well with a single Schechter function, but show an upturn at low masses. 
In general, for all the criteria adopted, we find that the massive end of the mass function 
(\Mstar$>10^{10.5}$ \Msun) is mainly dominated by ETGs (red/spheroidal/passive) up to $z=1$, 
while LTGs (blue/disc+irr/active) mostly contribute to the 
intermediate/low-mass part (\Mstar$<$\Mstar$^*$) of the mass function at all redshifts.
For \logM$>11$, the number density of blue galaxies is always below $\sim 2\times 10^{-4}$ gal/Mpc$^3$/log\Mstar.

\begin{figure}[h!]
\centering
\includegraphics[angle=-90, width=0.99\hsize]{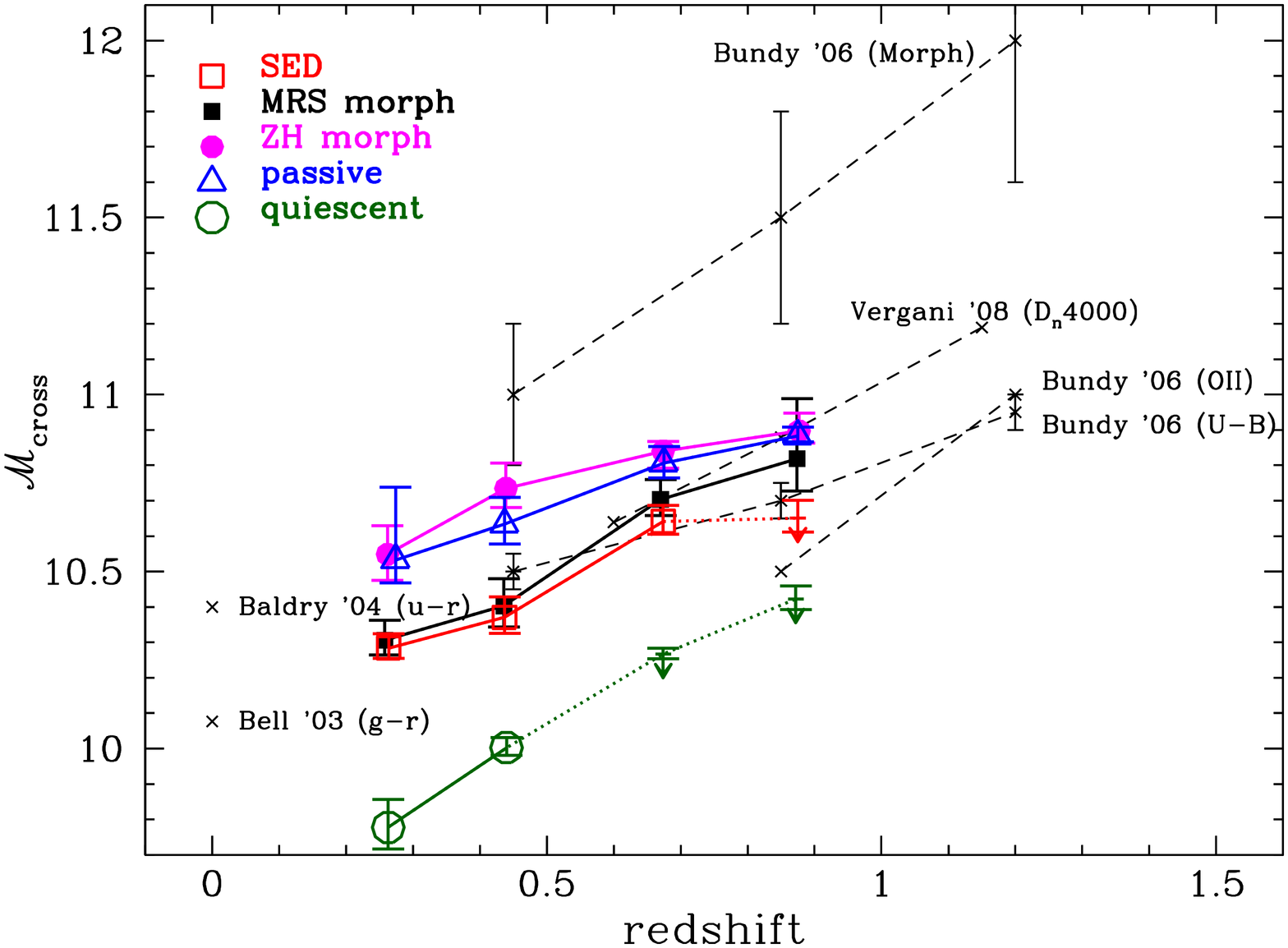}
\caption{The intersection of the \MFs\ (\Mcross) of the two populations (ETGs and LTGs) as a function of 
redshift. Values were derived from the \Vmax\ estimate of the \MFs.
Different symbols (and colours) refer to the different classification schemes adopted (see figure legend),
upper limits are shown where \Mcross\ is lower than \Mbias(z).
Black crosses show data from the literature 
(references are marked, along with the respective classification criteria).
}
\label{fig:Mcrossz}
\end{figure}

Using the \Vmax\ data points of the \MFs\ of ETGs and LTGs, 
we derived the intersection of the two populations (\Mstar$_{\rm cross}$). 
For the photometric classification, \Mcross\
is at $\sim 1.9 (\pm 0.2) \times10^{10}$ \Msun\ in the first redshift bin and evolves with redshift, increasing
 by about a factor of 2
(see red circles in Fig. \ref{fig:Mcrossz}) by $z=1$. 
Although for somewhat higher values of \Mcross,
a similar trend with redshift was  
found using the {\it morphological classifications} to divide the sample into 
spheroids and disc+irregular galaxies, for both classification schemes explored in this paper.
We determined similar values of \Mcross\ by dividing the sample in terms of 
the star formation activity
(active versus passive with \logSSFR$=-2$), while \Mcross\ is quite low when
quiescient galaxies are considered (\logSSFR$=-1$).
Our results are consistent with previous 
determinations of \Mcross\ in other deep surveys, such as VVDS (Vergani et al. 2008) and 
DEEP2 (Bundy et al. 2006), the only exception being the value of \Mcross\ derived by 
Bundy et al. (2006) using morphological types (the highest line in Fig. \ref{fig:Mcrossz}). 
In this case, the difference may be caused by a different and more extreme definition of morphological ETGs
by Bundy et al. (2006).
The extrapolation of our \Mcross\ to $z=0$ is also consistent
with local results (Baldry et al. 2004; Bell et al. 2003).

The evolution of \Mcross\ is indicative of an 
increase/decrease in the fraction of early/late galaxies with cosmic time at intermediate masses,
as already noted in previous surveys
(Fontana et al. 2004, Bundy et al. 2006, Arnouts et al. 2007, Scarlata et al. 2007b, Vergani et al. 2008). 
Oesch et al. (2010) found similar results within the COSMOS survey using extremely robust photometric redshifts.
Bolzonella et al. (2009)  found a more rapid evolution of \Mcross\ in high density environments
(see also Iovino et al. 2010 and Kova\v{c} et al. 2009 for groups).
However, previous surveys have not clearly established 
the precise evolution in absolute number densities of the two populations,
in particular for late-type galaxies. 
By comparing the zCOSMOS \MFs\ for different redshift bins, we find that the evolution of \Mcross\ 
is caused mainly by a clear increase with cosmic time in the number 
density of red/spheroids of intermediate mass 
(around \Mstars\ and below $10^{11}$ \Msun), coupled with only a marginal evolution in the number 
density of blue/spiral galaxies at the same masses.

\begin{figure}[h!]
\centering
\includegraphics[angle=-90, width=0.99\hsize]{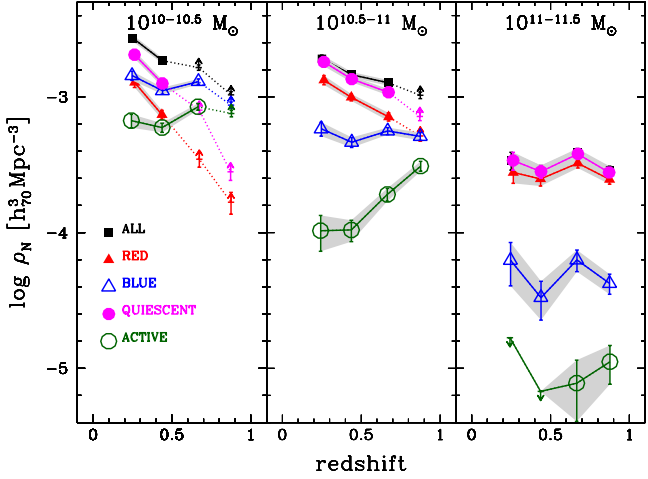}
\caption{
Number density evolution in different mass ranges. Different symbols and colours refer to different population 
classes:
black squares refers to  the total population, empty red and filled blue triangles to SED-ETGs and SED-LTGs 
using photometric classification,
 while filled magenta and empty green circles refer to quiescent and active populations, respectively. 
Hatched regions refer to \MF\ complete in the mass range and redshift considered, 
while dotted lines and lower limits are plotted at redshifts where the \MF\ is not complete in the 
considered mass range.
}
\label{fig:NDtypes_SSFR}
\end{figure}

These trends become even more distinctive when plotting
the number densities, derived using the \Vmax\ technique,
 versus redshift  in different mass ranges (\logM$=10 - 11.5$),
as shown in  Fig. \ref{fig:NDtypes_SSFR}.
We find that the number density of massive (\logM$>11$) red types is almost constant up to $z=1$ 
(see  also Pozzetti et al. 2003, Scarlata et al. 2007b, Ilbert et al. 2010) 
and dominates the total number density at these masses,
while the number density of
intermediate-mass (10$^{10}-10^{11}$ \Msun) red types increases with cosmic time. This increase  
is steeper at lower masses. Between $z\simeq 0.45$ and $z\simeq 0.25$, it is
$0.05\pm0.08, 0.14\pm0.04$, and $0.24\pm0.05$ dex 
for the three considered mass ranges of \logM$=11, 10.5$, and $10$, respectively.
An independent study (Brown et al. 2008), using halo occupation distributions, shows that while very massive 
halos often double in mass over the past 7 Gyr, the stellar masses of 
their central galaxies typically grow by only 30\%.
Our result is indicative of a {\it mass-assembly downsizing} as already noted for the global population and 
even more clearly in the build-up of the red sequence
(Cimatti et al. 2006).

\begin{figure}[h!]
\centering
\includegraphics[width=0.99\hsize]{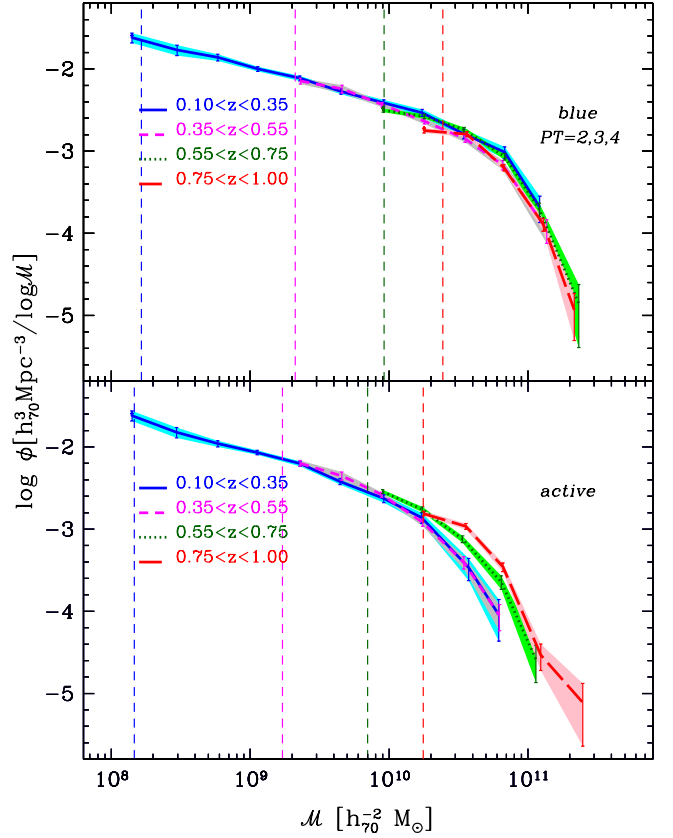}
\caption{\MFs\ for LTGs using two classification schemes. {\it Upper panel:}
blue galaxies (\PT=1), and {\it lower panel:} active galaxies (\logSSFR$>-1$), as defined in 
Sect. \ref{sec:types}. 
Only \Vmax\ determinations have been plotted, with their uncertainties.
Different lines and colours refer to different redshift range (see legend).
}
\label{fig:MFlate}
\end{figure}

In contrast, we find that the number density
of late blue types with \Mstar$>10^{10}$ \Msun\ remains approximately constant with cosmic time from $z=1$.
Their evolution in number density is less than $10-20$\% over the entire redshift and mass range
(Fig. \ref{fig:MFlate} upper panel).
These results are consistent with those of Arnouts et al. (2007), 
who found that the integrated stellar mass density of the active 
population shows only a modest mass growth rate, in contrast to an increase by a factor of 2 for the 
quiescent population.
Bundy et al. (2006) found that the abundance of blue galaxies declines by $0.1-0.2$ dex
from $z=0.75-1.0$ to $z=0.4-0.7$  at \logM$=10.6-11.3$ (see their Fig. 6).
A small evolution 
of the ``blue cloud \MF" was also detected by Bell et al. (2007) in the COMBO-17 survey out to $z\simeq0.9$. 

How is this small, possibly negligible evolution in the blue \MFs\ related to the 
global decrease with cosmic time in the \SFR\ density since $z\sim1$  (see Hopkins et al. 2006) and the
downsizing in the \SFR\ (Cowie et al. 1996)? 
To answer these questions, we analysed the \MFs\ of galaxies with different \SFR\ activities. 
In particular, we derived the \MFs\ for the most extreme 
star-forming population, i.e., those with high \SSFR.

Using estimates of the \SFR\ and \sSFR\ derived from the [OII] lines (Moustakas et al. 2006 calibration), 
Maier et al. (2009)
find that very few zCOSMOS galaxies 
with masses above \logM$>10.8$ have $1/$\sSFRs\ below the age of the universe
at $0.5 < z < 0.7$, while 
there exist several dozens of galaxies above the same mass that are strongly 
star-forming at $0.7 < z < 0.9$
(see Fig. 2 in Maier et al. 2009).
The same trend is observed using the \SFR\ and \sSFR\ derived from the fit to the  multi-band 
photometry with synthetic models.
Using an SED fitting determination, we note, first of all, that 
the median \SFR\ and \sSFR\ decreases with cosmic time 
in massive blue galaxies with \logM$>10.5$ 
(for example: the 
median \logSSFR\ is -1.0 at $0.7<z<1.0$,  while it is
-1.4 at $z<0.3$).

We therefore derived the \MFs\ using the \sSFR\ derived from the SED fitting and by dividing the sample 
as described in Sect. \ref{sec:types}, i.e., as those defining active and quiescent galaxies
with \logSSFR\ above or below $-1$. Figure \ref{fig:MFlate} (lower panel) shows indeed a decrease 
with cosmic time in the \MFs\ of the high-\sSFR\ galaxies, in particular for \logM$>10.3$. 
Their \MF\ shows, indeed, a mass-dependent evolution that is stronger at the massive end.
In the highest redshift bin 
($0.75<z<1$), the \GSMF\ of the high-\sSFR\ galaxies increases in number density and approaches that 
previously derived for the sample of blue (SED-LTG) galaxies.  

By comparing the \MFs\ of the blue galaxies with those of high-\sSFR\ galaxies, we find that,
while the number density of blue galaxies overall remains approximately constant, it
decreases significantly with cosmic time for  
high-\sSFR\ massive objects (see Fig. \ref{fig:MFlate}). From $z\sim 0.85$ to $z\sim0.25$,
the number density of high-\SSFR\ active galaxies with \logM$>10.5$ decreases by
a factor of $\sim 3$ ($-0.49\pm0.12$ dex).
Therefore, the decrease with cosmic time in the number of active massive galaxies 
is balanced by the constancy in intermediate-activity blue galaxies
and the increase of intermediate-mass quiescent and red galaxies.
Studying the $B$-band luminosity function in zCOSMOS,
Zucca et al. (2009) found results consistent with 
a scenario in which some blue galaxies are transformed into
red galaxies with increasing cosmic time.

To summarize, these data suggest that with cosmic time we are witnessing a transformation from active 
to passive galaxies and a corresponding
decrease (increase) in the fraction of late (early) types. These changes in galaxy fractions are caused mainly by
the clear increase with cosmic time in the number density of intermediate-mass 
(\Mstar$\sim 10^{10} - 10^{11}$ \Msun)  early-type galaxies, while the density of
intermediate-mass blue or morpho-LTGs remains almost constant, 
being continually replenished by blue active galaxies of even lower masses.
The median \sSFR\ of blue massive galaxies decreases with cosmic time.
Therefore, blue highly star-forming (of high \sSFR) galaxies of intermediate mass (\logM$=10-11$) 
increase in mass but decrease in either \SFR\ or \sSFR\ with cosmic time, 
transforming into low-\sSFR\ blue objects and, after the quenching of their \SFR, 
into red passive objects of intermediate mass. 
Less-massive, blue, active objects (\logM$<$10) increase in mass
to replace in the \MFs\ the blue intermediate-mass objects, whose density consequently remains almost 
constant with time. In agreement with our results, Bell et al. (2007) found that, if there were
a growth in stellar mass with a constant \SFR\ equal to
the instantaneous one (after subtracting 45\% of the return fraction),
the \MFs\ of blue galaxies would be dramatically overproduced. By only assuming that  
all the  growth in stellar mass is added to the red sequence,  
they reproduced the evolution of the blue 
and red stellar mass functions with remarkable accuracy.

\begin{figure}[h!]
\centering
\includegraphics[width=0.99\hsize]{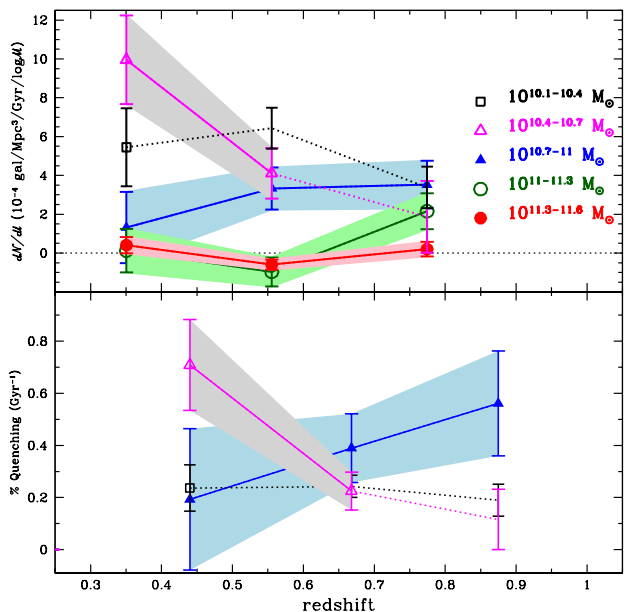}
\caption{ 
Evolution of the
growth rate in number density \dNdt\ (top panel)
as a function 
of redshift for red galaxies (\PT=1) in different mass ranges (represented by different symbols and colours). 
Values have been plotted 
at the intermediate redshift between two adjacent redshift bins.
In the bottom panel, we show the evolution with redshift of the fraction of blue galaxies (\PT=2,3,4) 
that are transforming into red galaxies per Gyr (see text).
In all the panels, hatched regions refer to \MFs\ that are complete in the mass range and redshift considered, 
while dotted lines are connecting points at redshifts where the \MF\ is not complete in the 
considered mass range.
}
\label{fig:dNdTz}
\end{figure}

\subsection{The build-up of the red sequence}

We attempt to quantify the growth rate (in terms of both number and mass density) 
of red galaxies as a function of redshift in different mass ranges.
This rate can be interpreted as the flux required to describe the migration of galaxies
with cosmic time from the blue cloud to the red sequence.

Figure \ref{fig:dNdTz} shows the evolution with redshift for the red population (\PT=1), 
in different mass ranges, of 
the growth rate in number density (\dNdt, upper panel), estimated using \Vmax\ data points.  
For \logM$<11$, the growth rate is 
in the range of $10^{-4}$ - $10^{-3}$ gal/Mpc$^3$/Gyr/log\Mstar\ and is even lower
approaching zero for higher masses.
The growth rate in mass density is 
\dMdt$\simeq$ $10^6$ - $10^7$ \Msun/Mpc$^3$/Gyr/log\Mstar.
Integrating above \logM=9.8, we find that
\dNdt=$6.8 (\pm 1.2) \times 10^{-4}$ gal/Mpc$^3$/Gyr   
and \dMdt$=1.8 (\pm 0.4) \times 10^7$ \Msun/Mpc$^3$/Gyr
between the second and the first redshift bin ($z\sim0.34$).
Despite the large errors, we find a clear  
increase with cosmic time in the growth rate of galaxies with \logM=10.4-10.7, 
but a hint of a decrease with cosmic time at higher \Mstar\ (\logM=10.7-11).
For \logM$>11$, the growth rate in number density is consistent with zero for almost all redshift bins.
These trends can be interpreted as a {\it mass-assembly downsizing} signal, 
i.e., most massive red galaxies assembled their mass earlier than lower mass red galaxies. 
For comparison, Arnouts et al. (2007) for the VVDS found that the median mass growth rate 
(integrated over the entire mass range \logM=$8-13$)
between $z\sim2$ and $z=0$ is $1.7 (\pm 0.4)\times 10^7$ \Msun/Mpc$^3$/Gyr, 
which is consistent with our values for \logM$>9.8$.
Walcher et al. (2008) derived a somewhat higher value of \dMdt$=6.5-10\times 10^7$ \Msun/Mpc$^3$/Gyr 
from $z\sim1$ to $z\sim0.5$ integrated over the whole mass range. 
The difference from our results, indeed, could be due to their more extended mass range.

In principle, by comparing this growth rate with
the number density of blue galaxies we should be able to estimate the fraction of blue galaxies 
that, at any given time, are transforming into red galaxies ($f_{Q}$) after the quenching of their \SFR.
We therefore derived the fraction rate of quenching
\begin{equation}
f_{\rm Q}({\cal M})=\frac{\dot \rho_N^{red}({\cal M})}{\rho_N^{blue}({\cal M})} ~{\rm Gyr}^{-1},
\end{equation}
using the \Vmax\ estimation for $\rho_{N}$, as 
shown in the bottom panel of Fig. \ref{fig:dNdTz}
as a function of redshift and mass.
For \logM$>11$,  the rates were unconstrained because of the low number density of blue galaxies. 
The values obtained for \logM$<11$ have a median value of $\sim 25$\%
Gyr$^{-1}$.
As for the growth rate, we find that with cosmic time this fraction rate decreases for high-mass galaxies, 
but increases
for intermediate-mass galaxies, at  least for $z<0.7$. 
At $z<0.4$, we are unable to explore the transformation from  blue to red for the
low mass regime (\logM$<10.4$) and to verify whether the trend of the blue fraction increasing
with cosmic time for low-mass galaxies is shifted to even lower
redshift, as expected by the downsizing scenario.
We note however that all fraction values remain very uncertain and their uncertainties increase 
yet further if we include the effects of cosmic variance.

Vergani et al. (2009) explore the properties and number densities of the post-starburst (PSB) galaxy population, 
spectroscopically identified  in the 
zCOSMOS sample,
as the possible link population in the transition phase between the blue cloud and the red-sequence. 
They found that 
this galaxy population, which is affected by a sudden quenching of its star-formation activity, may increase 
the stellar mass density of the red-sequence by up to a non-negligible level of 10\%. 

\begin{figure}[h!]
\centering
\includegraphics[width=0.99\hsize]{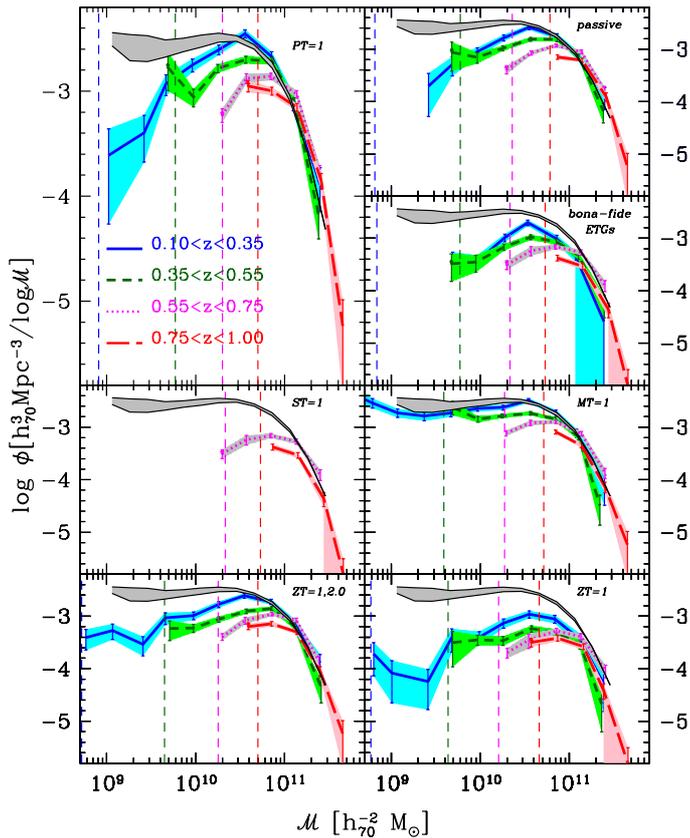}
\caption{\MFs\ for ETGs using various classification schemes:
red galaxies (\PT=1), passive (\logSSFR$<-2$), spheroidals
(\ZT=1+2.0, \ZT=1, \MT=1), absorption-line galaxies (\ST=1),
and ``bona-fide ETGs" as defined in Sect. \ref{sec:types}. 
Only \Vmax\ determinations have been plotted, with their uncertainties.
Different lines and colours refer to different redshift ranges (see upper left panel).
Grey hatched region, bounded by solid black lines, 
refer to local \GSMF\ for ETGs derived from Baldry et al. (2008) and Baldry et al. (2006) 
(see text).
}
\label{fig:MFearly}
\end{figure}

\subsection{Elliptical/early type galaxy evolution }\label{sec:ETGs}

Hierarchical semi-analytical models (De Lucia et al. 2006) predict that stars in more 
massive elliptical galaxies, relative to their lower-mass counterparts,
are {\it older} but assemble their mass by means of mergers {\it later} in cosmic time.
For example, 
50\% of ellipticals with \logM$>11$ form most (80\%) of their stars at $z_{\rm form}\simeq 1.6$, but
assemble their mass by \zassembly$\simeq 0.2$, which is defined as the redshift at which 
most (80\%) of the stars that make up the galaxy at redshift zero are already 
assembled into a single object.
Galaxies with \logM$>9.6$, instead, have a median $z_{\rm form}\simeq 1.3$ and median $z_{\rm assembly}\simeq 0.9$
(see Figs. 4  and 5 in De Lucia et al. 2006).
As pointed out by Cimatti et al. (2006), this trend is the opposite of that of the 
{\it mass-assembly downsizing} found for red galaxies in the luminosity function,
confirmed by Scarlata et al. (2007b),
i.e., most massive red galaxies assembled their mass earlier than lower-mass red ones. 
In the following, we explore 
in detail for the zCOSMOS data set the evolution of elliptical and all early-type galaxies, using 
the different classification schemes described in Sect. \ref{sec:types} based on colours, morphology, 
spectral features, or a combination of them.

Figure \ref{fig:MFearly}
of the \MFs\ for the different populations of ETGs: red galaxies (\PT=1),
passive galaxies (\logSSFR$<-2$), 
 spheroidals (\MT=1, \ZT=1+2.0, \ZT=1), absorption-line galaxies (\ST=1),
and ``bona-fide ETGs" as defined in Sect. \ref{sec:types}. 

First of all, we note that there are significant differences in the normalization of the \MFs\ at all \Mstar, 
the pure ellipticals classified by ZEST (\ZT=1) having the lowest value at all redshifts 
and mass ranges explored. 
In contrast, the shapes of the various \MFs\ are all similar, 
there being a decline in number density for \Mstar$<$\Mstars\ for all \MFs\ that is described well 
by a single Schechter function, except for
the \MF\ of the  spheroids with \MT=1, which shows an excess at low-mass (\logM$<9.5$) in the first 
redshift bin.

\begin{figure}[h!]
\centering
\includegraphics[angle=-90,width=0.99\hsize]{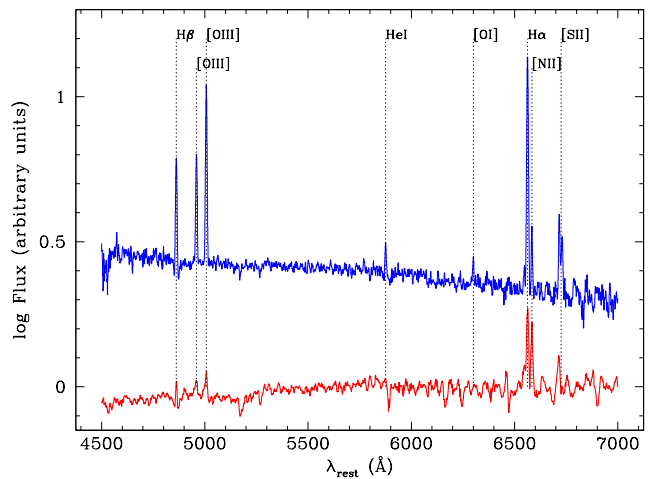}
\caption{Stacked spectra for galaxies with MRS elliptical morphology (\MT=1)
and blue colours (\PT=2+3+4)
at low-mass (\logM$<9$ upper spectrum in blue) and 
high-mass (\logM$>10.5$ lower spectrum in red).
}
\label{fig:blueE}
\end{figure}

We then studied in detail this population of low-mass spheroids (\MT=1), which dominates the \GSMF\
and exhibits blue colours (\PT$>=2$,  i.e., $(U-B)_{rest}<1$).
We found 89 of these objects with \logM$<9$  at low redshift ($z<0.5$).
They have quite compact sizes, with a small median effective radius ($r_e \simeq 0.6$ kpc).
We also identified a contribution from blue MRS spheroids (\PT$>=2$, \MT$=1$) also at high mass 
(\logM$>10.5$) at all redshifts (182 objects). 
This latter population of massive blue MRS spheroids
does not dominate the high-mass end of the \GSMF\
($\sim 20$\%), as the former does the low-mass end. Therefore, this 
 suggests that the shape of the \MF\ for this population of blue spheroids differs 
from that for red spheroids, at least when using MRS morphologies. 
This is consistent with the conclusions of Zucca et al. (2009) about the contribution of this population to 
the $B$-band luminosity function.
However, for these objects, in particular at low mass, 
  we note that MRS morphologies
do not always agree with ZEST ones, 
which implies that there are significant
uncertainties in deriving morphologies for faint low-mass objects.
We, therefore, verified their morphologies visually (i.e., PC, LT, LP). We found
that most of the blue low-mass MRS spheroids are bulge-dominated spirals (Sa).  
Most of the massive ones, even if their morphological classifications are in agreement between the two methods
(ZEST and MRS), 
appear to be spirals, between Sa and Sc, although they have morphological parameters 
(asymmetry and concentration) 
that are more similar to those of elliptical galaxies.
Composite spectra of this population were generated (following Mignoli et al. 2009) by 
averaging all their spectra at $z<0.5$ and dividing them into low-mass and high-mass objects.
In Fig. \ref{fig:blueE}, the average spectra are plotted: 
low-mass blue spheroids show a blue continuum and emission lines typical of star-forming galaxies,
indicating that these galaxies are experiencing significant star-formation,
while the high-mass blue spheroids have on average a red spectrum with absorption lines
that are largely indistinguishable 
from those of the purely passive galaxies (see Mignoli et al. 2009),
but in addition exhibit emission lines that
could be indicative of some nuclear activity contamination (i.e.,  [NII]/H$_\alpha$ and [OIII]/H$_\beta$ 
are consistent with LINERs).
An AGN component may also explain their typically high concentration index
obtained from their HST images.
These AGN might play a role in quenching the \SFR\
in this population of blue spheroidal massive galaxies,
which are possibly in the transition phase between the blue cloud and the red-sequence.

Finally, we studied the evolution with cosmic time of the \MFs\ of
each of the classes of 
red, passive, ellipticals, spheroids, absorption-line, and ``bona-fide ETG" galaxies.
In Fig. \ref{fig:MFearly}, we show the \MFs\ of these various ETG classes at increasing redshift.
We also show, as reference, the local \MF\ of ETGs, which we derived 
using the global \MF\ by Baldry et al. (2008)
and the fraction of red galaxies estimated by Baldry et al. (2006) for their 4 central environment bins.
This is indicated in Fig. \ref{fig:MFearly} by the grey shaded region. 
We note very similar evolutions in
the ETG population for all the different means of classifying an ETG.
As claimed for red galaxies, we find  a clear ``mass-assembly downsizing" evolution, 
with no evidence of an increase with cosmic time in the number density of the 
most massive ETGs since $z\sim1$ and a progressive 
increase in the \MFs\ at intermediate mass (\logM$<11$) out to $z=0$.
Using M05 models to estimate the stellar masses, we also find that 
 the same trend is visible in our data, and that the use of these models does not
strongly affect our main conclusions. 
This is because in the redshift range studied here ($0.1<z<1.0$), ETGs have on average 
old stellar populations that are 
not dominated by TP-AGB stars particularly in massive objects.

For massive ETGs, similar results were first noted in the near-IR ($K$-band) luminosity function by 
Pozzetti et al. (2003) and the mass function by Fontana et al. (2004) in 
the K20 survey, using spectral classification, 
and later confirmed by larger optical and near-IR surveys,
such as the VVDS, using colours or spectra to define ETGs (Zucca et al. 2007; Arnouts et al. 2007;
Vergani et al. 2008).  Scarlata et al. (2007b) found that in COSMOS
both morphologically and photometrically selected subsamples of ETGs show no evolution in 
their number density at the bright end of the B-band luminosity function ($L>2.5L^*$) out to $z\sim0.7$, 
and there is a deficit of a factor of about $2-3$ of fainter ETGs over the 
same cosmic period, as also confirmed by Zucca et al. (2009) in the zCOSMOS sample.
Ilbert et al. (2010), selecting ETGs (by morphology or colour) in SCOSMOS (Sanders et al. 2007)
with accurate  photometric redshifts (Ilbert et al. 2009) up to $z_{\rm photo}=2$, 
found similar results for massive ETGs at $z<1$ (evolution $<0.2$ dex for \logM$>11$) 
and a more rapid evolution at higher redshifts (by a factor of 15-20 between $z=1.5-2$ and $z=0.8-1$)
or lower masses (increasing by a factor of 4.5  between $z=0.8-1$ and $z=0.2-0.4$ for \logM$\sim10$).

\subsection{The building redshift of ETGs}

Our results suggest that lower mass ETGs (regardless of the method used to classify them) 
assembled their mass later than higher-mass ETGs. 
We can, indeed, estimate the redshift at which 50\% of ETGs have already built-up their mass
 (\zbuilding), as a function of stellar mass,
i.e., at which redshift their number density has decreased by a factor of 2 from $z=0$
($\phi_z/\phi_0=0.5$).
Since we consider the number densities of ETGs only above $z>0.25$, 
we can derive only upper limits to \zbuilding, as defined above, at least in the low mass range.
For this reason, we used as a local reference the \GSMF\ for ETGs, described in the previous section and 
shown in Fig. \ref{fig:MFearly} as a grey shaded region. 

\begin{figure}[h!]
\centering
\includegraphics[angle=-90,width=0.99\hsize]{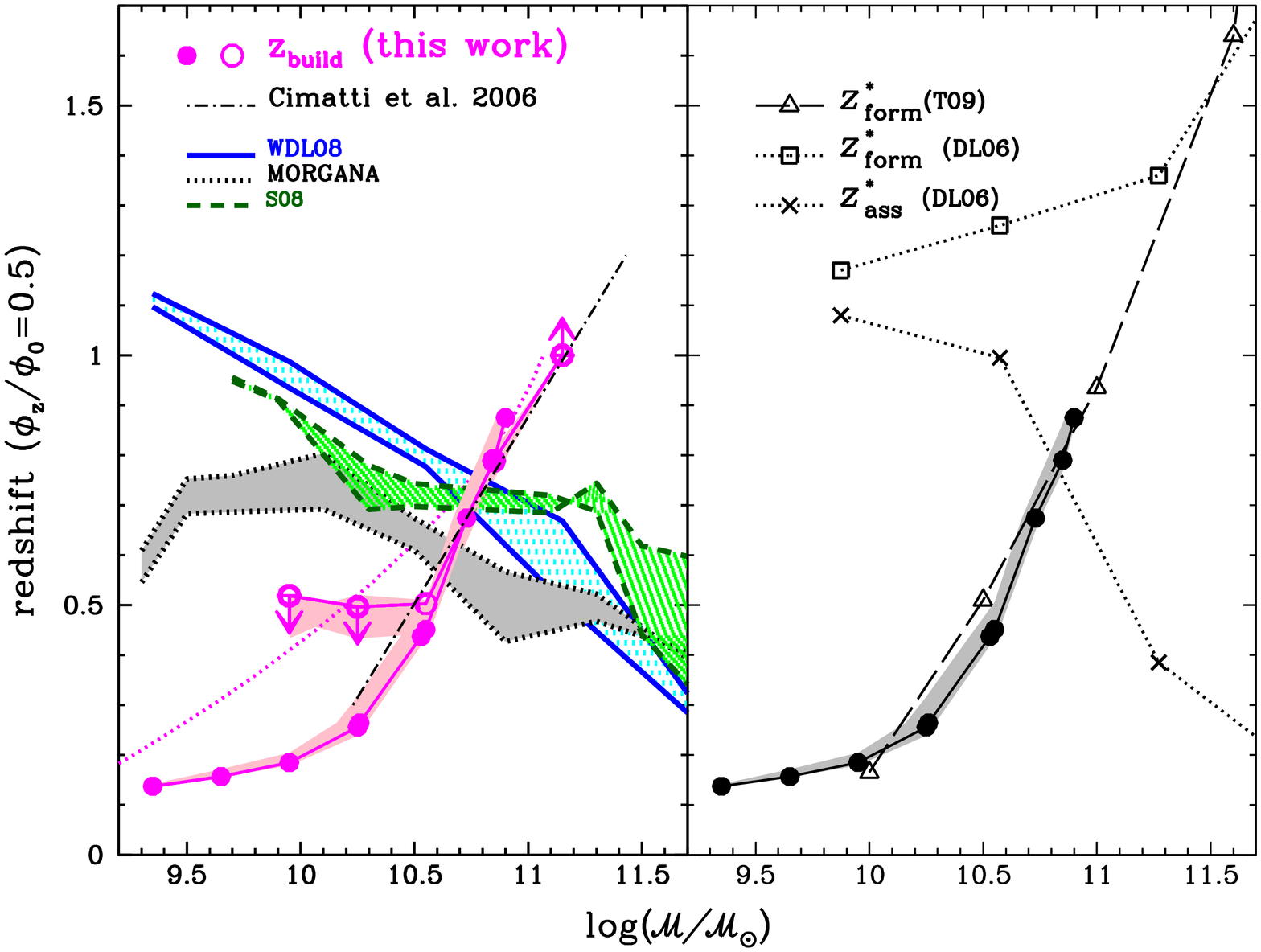}
\caption{Redshift at which the ETG \GSMF\ has decreased by a factor 2 ($\phi_z/\phi_0=0.5$, i.e., \zbuilding, 
see text) as a function of \Mstar.
{\it Left panel:} Circles magenta points refer to data derived within zCOSMOS for passive galaxies 
(\logSSFR$<-2$):
open circles have been derived completely from zCOSMOS dataset, using as a reference the \GSMF\ in the first 
redshift bin ($\phi_0=\phi(z\sim0.24$), along with the region covered using various ETG
classification criteria;
filled circles are \zbuilding\ using the local ETG \GSMF\ (see text), along with the spread in its 
shape related to their enviroment. 
Dotted magenta curve shows the \Mbias\ above which we are formally complete.
Also plotted are zbuilding\ from SAMs: WDL08 (shaded region bounded by solid blue lines), 
S08 (shaded region bounded by dotted green lines), and MORGANA (shaded region bounded by dotted black lines),
derived using the original \GSMF\ or convolved with 0.25 dex in mass uncertainties (see Fontanot et al. 2009).
 {\it Right panel:} \zbuilding\ for zCOSMOS data points are compared with the trend with stellar mass of
the median \zassembly\ and \zform\ (80\% of star assembled and formed, respectively) from 
De Lucia et al. (2006, DL06) and with the redshift at which the 
stellar formation ends (80\% of star formed) in local ETGs estimated by Thomas et al (2009, T09).
}
\label{fig:z50}
\end{figure}

Within the zCOSMOS redshift range, we find 
for ETG galaxies that \zbuilding\ increases with \Mstar\, regardless of classification method, with
\zbuilding$\sim 0.4, 0.5$, and $0.8$ for \logM$\sim10,10.4$, and $10.8$, respectively (Fig. \ref{fig:z50}).
At higher mass, the number density of ETGs is almost constant up to redshift $\sim 0.7$ 
and decreases by less than a factor of 2 after $z\sim1$, thus we set $1$ as the lower limit to \zbuilding\
for \logM$>11$.
Our estimates of \zbuilding\ for passive galaxies (\logSSFR$<-2$) are shown in Fig. \ref{fig:z50}:
the open circle symbolis  were derived completely from the zCOSMOS dataset, using the \GSMF\ in the first 
redshift bin ($\phi_0=\phi(z\sim0.24$) as reference, 
and are shown along with the region covered using various ETG 
classification criteria. 
Most of the points have \Mstar$>$\Mbias, above which we are formally complete.
Since the local ETG \GSMF, derived by Baldry et al. (2006, 2008) as described previously, approaches
the observed zCOSMOS \GSMF\ in the first redshift bin for \logM$>10.5$ for passive galaxies, below this mass
the values of \zbuilding\ should be considered upper limits.
We also show \zbuilding\ derived by assuming the local \GSMF\ for ETGs,
along with the region covered using the different fractions of ETGs 
in the 4 central environment bins in Baldry et al. (2006). In addition,
systematic uncertainties caused by the different classification methods
could affect our estimates of \zbuilding,
as for the zCOSMOS data.

First of all, we note 
that the ``downsizing" trend with mass in terms of ETG \zbuilding\  is clearly visible.
Our data are in very good agreement with the values derived by Cimatti et al. (2006), who 
used the de-evolved luminosity function for galaxies on the red-sequence.
The values of \zbuilding\ are in quite close agreement with the end of the star 
formation (80\% of the star formed)
histories estimated by Thomas et al. (2009) for early-type 
galaxies as a function of stellar mass on the basis of their spectral properties 
(T09, in the right panel of Fig. \ref{fig:z50}).
This downsizing trend with stellar mass 
conflicts instead with the prediction of semi-analytical models for $z_{\rm assembly}$ (e.g., by  
De Lucia et al. 2006, DL06, reproduced here in the right panel of Fig. \ref{fig:z50}). 
Given the increase 
with mass of the redshift at which the stars are predicted to form in this model 
(\zform, see Fig. \ref{fig:z50} right panel),
the decreasing trend with mass of \zassembly\ is caused by the increasing relevance of merging processes 
(in particular minor merging) 
with increasing mass (Wang \& Kauffmann 2008; Stewart et al. 2009; Parry et al. 2009) in the SAMs.

To perform a more accurate comparison  with semi-analytical models (SAM), we used
\zbuilding\ derived according to exactly the same
definition applied to the data. We adopt the \GSMF\ predictions 
for the three prescriptions used in Fontanot et al. (2009):
the most recent implementation of the Munich SAM (De Lucia \& Blaizot 2007, Wang et al. 2008a, hereafter WDL08),
the MORGANA model (Monaco et al. 2007, Lo Faro et al. 2009), and the fiducial model presented by Somerville 
(2008, hereafter S08).
We used the same classification criteria for passive galaxies (\logSSFR$<-2$) as used in the models.
The predicted \MFs\ for each model are shown in Fig. 6 of Fontanot et al. (2009).
In Fig. \ref{fig:z50}, we 
compare \zbuilding\ derived from \GSMF\ SAM predictions (provided by Fontanot) and 
zCOSMOS data using exactly the same definition. 
For each model, we show the region covered by \zbuilding\ using the original SAM \MFs\ and
those that take into account the 0.25 dex of uncertainties in the mass derivation.
In contrast to the data, the semi-analytical models show a reverse  trend with mass
of \zbuilding, which is similar to the trend of \zassembly.
Residual small differences, for example between the determination of mass or SFR for the data and that of 
the models (e.g., the presence of secondary bursts in the model SFH) 
or in the ETG classification criteria are unlikely to reconcile the opposite trend 
with mass.

Taken at face value, the coincidence of \zform\ estimated by Thomas et al. (2009) and \zbuilding\ 
argues against the dominant contribution of mergers (major+minor) at $z<1$ in building up ETG galaxies,
as instead predicted by semi-analytical models (Wang \& Kauffmann 2008, Stewart et al. 2009).
We therefore conclude that the build-up of ETG galaxies follows the same ``downsizing" trend in mass as
the formation of their stars, which disagrees with the ``upsizing" trend predicted in SAMs.

\subsection{Timescales for the quenching of the star formation and morphological transformation}

\begin{figure}[h!]
\centering
\includegraphics[angle=-90, width=0.99\hsize]{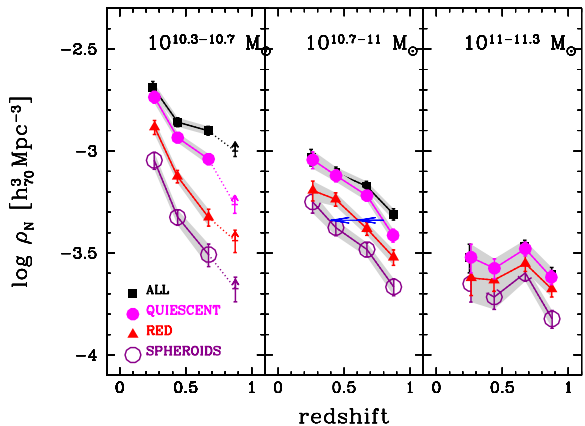}
\caption{
Number density evolution for ETGs in different mass ranges: red galaxies (red filled  triangles for \PT=1), 
spheroidals (dark magenta empty circles for \ZT=1+2.0) and low-\sSFR\ (magenta filled circles for 
log(\sSFR)$<-1$). Black filled squares refer to the global population.
Lines have the same meaning as in Fig. \ref{fig:MFtypes_SSFR}.
In the second panel, we also indicate by 2 arrows the evolutionary time sequence from 
quiescent to red to spheroidal galaxies (see text).
}
\label{fig:NDearly}
\end{figure}

The evolution with redshift in the galaxy number density in different mass ranges for 
three different samples of ETG (red (\PT=1), spheroidal (\ZT=1,2.0), and quiescent (\logSSFR$<-1$)) galaxies
is shown in Fig. \ref{fig:NDearly}.
We find that they all show a very similar trend (but a different normalization),
there being an increase in the number densities of intermediate-mass 
ETGs with cosmic time from $z=1$ to the local Universe. 
This trend
is steep for low-to-intermediate mass galaxies and is flat and almost negligible only for \logM$>11$
($<0.1$ dex between $z=0.85$ and $z=0.25$ for \logM=11-11.5). 
The similar evolutions of red and spheroidal galaxies was already noted by
Arnouts et al. (2007), who commented: ``If not by chance, this coincidence could suggest that the build-up of 
the quiescent sequence is closely followed or preceded by a morphological transformation."
In principle, if all the various classification methods here adopted are ``perfect", 
that the number density 
of SED-ETGs is higher than that of the morpho-ETGs, at a given mass, suggests that 
the colour transformation (from blue to red) precedes or has a shorter timescale than
 the morphological transformation. 
From Fig. \ref{fig:NDearly}, the delay time between the colour and the morphological transformation
could be estimated, at a fixed $\rho_N$, to be about 1-2 Gyr (as shown by the arrow for \logM=10.7-11).
In addition, since the number density of the low-\sSFR\ population is higher than that of the 
red galaxies (SED-ETGs), we estimate that the time
elapsed for quiescent galaxies to completely switch-off their SFR and become red 
is of the same order (about 1-2 Gyr, shown as an arrow in Fig. \ref{fig:NDearly}).
Wolf et al. (2008) reach a similar conclusion  in the cluster A901/2 at $z\sim0.17$:
the rich red-spiral population 
at intermediate mass (\logM$=[10, 11]$) 
is more accurately explained if quenching is a 
slow process and morphological transformation is delayed even more (see also Skibba et al. 2009).
The small contribution to the assembly of the red-sequence by the
class of post-starburst galaxies (most of which are assumed to quench 
their star formation rapidly) provides additional support to a scenario of delayed/slow
quenching (Vergani et al. 2009).

However, we are aware that
the colour-selected red galaxies may be contaminated by dusty starbursts.
We recall here that, given our 24 $\mu$m flux limit of $\sim 0.3$ mJy,  only 5\% 
of red galaxies (SED-ETGs) were detected at 24 $\mu$m (all having \logSSFR$>-2$), and  about $23\%$ have
$-2<$\logSSFR$<$-1, i.e., they are not completely ``dead".
{\bf Bundy et al. 2009, through  MIPS stacking analysis, show that the implied \SFRs\ of red galaxies 
are on average at least an order of magnitude below the star-forming population.}
On the other hand, there is a non-negligible blue population ($\sim 20\%$) among elliptical galaxies,
for which therefore the morphological transformation could precede or is faster than the 
colour transformation (see below for a discussion).
Therefore, we also checked that the number density of red passive galaxies (\PT=1 and  \logSSFR$<-2$), 
i.e., those that are ``red \& dead'', is higher at any given mass and redshift than the number density of 
passive elliptical 
galaxies (\ZT=1,2.0 and \logSSFR$<-2$), and that at a given number density and mass they are delayed by 1-2 Gyr.
We conclude that, at least statistically, the morphological transformation takes a longer time 
than for a given galaxy to become ``red and dead".

\begin{figure*}
\centering
\includegraphics[width=0.49\hsize]{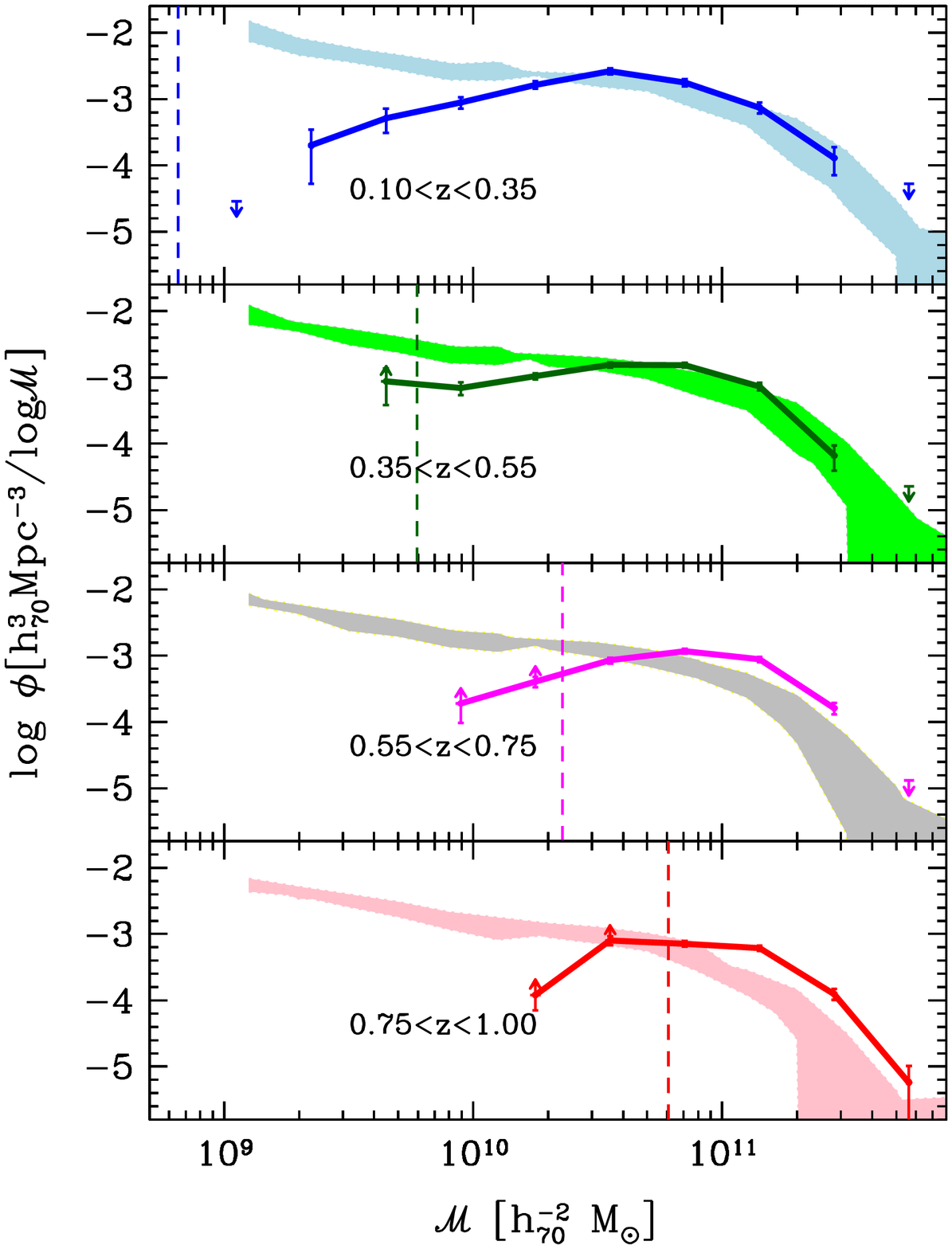}
\includegraphics[width=0.49\hsize]{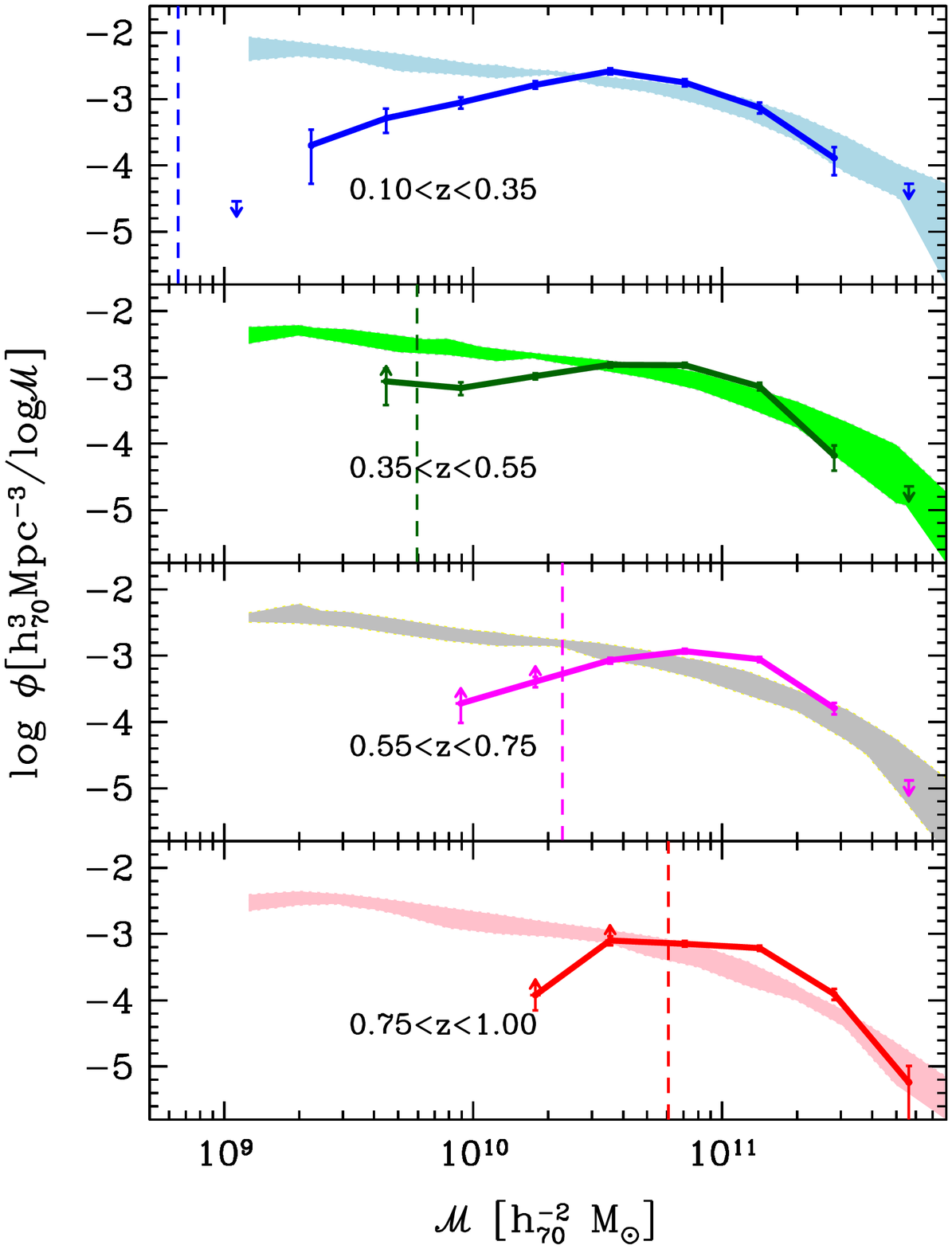}
\caption{Comparison between observed \MFs\ (lines and points) with the regions covered by the predictions 
(shaded colour regions) of three semi-analytical 
models (Morgana, WDL08, and S08 from Fontanot et al. 2009), at increasing redshifts 
(from top to bottom panels). {\it Left panels:} original \MFs\ from SAM. 
{\it Right panels:} SAM \MFs\ take into account 0.25 dex of uncertainties in the \Mstar\ derivation.
}
\label{fig:MF_SAM}
\end{figure*}

But which physical processes induce these transformations?
All the properties \SSFR, colours, and photometric types depend on the star-formation history, 
while morphological types mainly reflect the dynamical history of a galaxy.
The transformation from active blue galaxies to red passive galaxies,
as well as the transformation from spiral disc-dominated to bulge-dominated systems,
could be driven by a variety of processes such as the ageing of their 
stellar population or the fading of the disc,
because of the exhaustion of its gas reservoir, gas stripping,
or the quenching of the SFHs, for example by AGN feedback (Menci et al. 2006; Monaco et al. 2007;
Bower et al. 2006), 
or by truncating the gas accretion from infall or from satellite galaxies.
In addition, ``starvation" (or ``strangulation") is expected to affect a galaxy star formation history on 
a quite long timescale and therefore to cause a slowly declining activity:
when 
a galaxy is accreted onto a larger structure, the gas supply can no longer be 
replenished by cooling, which is suppressed (Larson et al. 1980). 
This process is an important element of semi-analytic models of galaxy 
formation.
Some of the aforementioned processes are also obviously environment-dependent.

On the other hand, the strong increase in the number density 
of intermediate-mass pure morphological ellipticals (\ZT=1) or ``bona-fide ETGs" (red passive spheroidals) 
provides support for dynamical processes, such as the merging between two low-mass galaxies 
(Toomre \& Toomre 1972, Barnes 1992).
Mergers are included in standard semi-analytic models of galaxy formation and represent indeed the main channel 
of formation for bulges.
Menci et al. (2002) also included in their hierarchical semi-analytical models binary 
aggregations of satellite galaxies inside a common halo.
However, if merger events were responsible for morphological transformations,
 we would expect to observe a rapid evolution from blue to red
after the formation of the bulge/elliptical, 
i.e., after the burst triggered by the merger (Mihos \& Hernquist 1996). 
We find, instead, that the colour transformation precedes the morphological one, and 
that only a population ($\sim 20$\%) of massive blue spheroidal galaxies exists,
which could
be the end-product of merger events. The massive tail of this population
has some signature of an AGN component in its stacked spectrum (see Fig. \ref{fig:blueE}), which
may be responsible for quenching the \SFR,
allowing their transition to red ellipticals with cosmic time.
The study of this particular population is beyond the aim of this paper and will be 
the subject of future zCOSMOS papers.

Our results, therefore, suggest that dynamical processes, other than mergers, are the main mechanisms 
responsible for the morphological transformation that follows the colour transformation.
For example, bulge growth through disc instability could be efficient (Carollo et al. 2001, Bower et al. 2006, Dekel et al. 2009).
Disc instability is also a possible formation channel for bulges in SAMs 
(De Lucia et al. 2006, Bower et al. 2006, Parry et al. 2009).
Dynamical observations of massive galaxies at $z\simeq2$ with SINFONI (Genzel et al. 2006, 2008)
also suggest that disc instability can produce
massive bulges, even in the absence of major mergers. 
Indeed, dynamical friction and viscous processes at $z\simeq2$ proceed on a 
timescale of $<1$ Gyr, 
which is at least one order of magnitude faster than in $z\sim 0$ disc galaxies. 
In overdense environments, such as clusters, ram pressure may also play a role, 
demolishing the star-forming disc (Cortese  et al. 2006). 
Recently, Bundy et al. (2009) demonstrated that the colour and morphological transformations 
should proceed through several separate stages and 
explored the strengths and weaknesses of several more sophisticated 
explanations, including environmental effects, internal stabilization, and disc regrowth by means of gas-rich 
mergers.

\section {Comparison with models}\label{sec:SAM}

To establish the cause of the ``age-downsizing" of elliptical galaxies,
according to which the stars in more massive galaxies formed earlier and over a shorter 
period than those in less massive galaxies, 
Cattaneo et al. (2008) 
discuss a model in which star formation shutdowns in dark 
matter haloes of mass above a critical value of  $\sim 10^{12}$ \Msun.
However, this model predicts an evolution in the massive tail of the \GSMF\ of red galaxies 
that is more rapid than that of intermediate-mass red galaxies 
from $z=1$ to $z=0$, which is in contrast, at least qualitatively, with our results
(see Ilbert et al. 2010 for a comparison between model and data).

The same behaviour is clearly visible in Fig. 6 of Fontanot et al. (2009), which shows that the 
evolution of the \MF\ for passive galaxies with redshift follows 
exactly the opposite trends with mass and cosmic time observed in the data 
(compare with our Fig. \ref{fig:MFearly}).  
They also conclude that a robust prediction of SAM models seems to be that the evolution of less 
massive galaxies is slower than for more massive ones, 
i.e., the models do not predict ``mass-downsizing" but rather the opposite behaviour 
(sometimes called ``upsizing").
Here we recall that the three semi-analytical model renditions considered in Fontanot et al. (2009) 
are indeed unable to reproduce the ``mass-assembly downsizing'' evolution of observed ETG \GSMF, 
as parametrized for example by the median building redshift described in the previous section
(Fig. \ref{fig:z50}).

In Fig. \ref{fig:MF_SAM}, we also show the direct comparison of zCOSMOS ETG \GSMF\ with the SAM model 
predictions for passive galaxies (\logSSFR$<-1$).
Hatched region show the range covered by the three models from Fontanot et al. (2009) for the original \MFs\ 
(left panel) and convolved with 0.25 dex uncertainties in the mass estimate (right panel).
A more detailed and quantitative comparison between each model and data is beyond the aim of this paper.
We first note that the shape of the predicted \MFs\ differs from that of the observed \MFs\ in all 
the mass and redshift ranges explored. In particular, the models are unable to recover the steep 
decline with decreasing mass in the number densities of passive ETGs, but instead 
overproduce their number densities
increasingly with decreasing mass
for \logM$<10.3$. This is because in the models, intermediate-mass galaxies form too early 
and are too passive and red at late times.
In contrast, at the massive mass end (\logM$>11$) the original models have the well-known problem 
of underestimating the number density of massive galaxies at high redshift (for $z>0.6$ in zCOSMOS dataset). 
We show that the convolution with stellar mass uncertainties (0.25 dex) only partially solve the problem at 
high redshift and high masses (\logM$>11.3$), while models continue to 
underpredict \MFs\ around \logM$\sim11-11.3$,
where they fail to reproduce the \GSMF\ shape,  and 
overproduce massive galaxies at low redshifts, as already pointed out by Fontanot et al. (2009).
To qualitatively reconcile data and models at the massive-end, the mass uncertainties would have to
 increase rapidly 
with redshift, which is not the case for most current surveys with accurate and multi-band photometry 
that extends to near-IR rest-frame wavelengths.

We conclude that the three semi-analytical model explored here (WDL08, MORGANA, and S08) are unable to 
completely reproduce either the observed shape of the \MFs\ of passive galaxies or the evolutionary trend.
In particular, SAMs overpredict the low-mass ETG number densities and, even if the uncertainties in \Mstar\ 
determination partially assount for the underprediction at high masses and high redshifts, 
they predict an evolution with cosmic time that is of an opposite trend with mass to that observed 
for the data, 
i.e., they are unable to reproduce the later formation and more rapid evolution in number densities of the 
lower mass galaxies (``mass-assembly downsizing'').

\section{Summary and conclusions}\label{sec:concl}

We have investigated the evolution of the galaxy stellar mass function
up to $z=1$ in the zCOSMOS 10k bright spectroscopic survey using VIMOS spectroscopy
($\sim 8500$ galaxies with $15.0<I<22.5$ over 1.4 deg$^2$) and multiband photometry (from UV to near-IR).
Our main observational results can be summarized as follows:

\begin{itemize}
\item {\it Unveiling bimodality in the \MF:}
The shape of the \MF\ is more accurately reproduced by two Schechter functions at least up to $z\simeq0.55$,
the maximum redshift at which we can explore the \MF\ upturn due to our mass limit.
This shape is linked to the bimodality in the galaxy properties of the ETG and LTG populations, 
which dominate the \GSMF, respectively, at high masses (\logM$>10.5$) 
and at low/intermediate-masses up to the highest redshift explored ($z\simeq1$).

\item {\it Mass-assembly downsizing: }
We find a continuous, rapid increase with cosmic time from $z\simeq1$
to $z=0$ in the global \MF\ for \logM$<11$ (by a factor of $\sim 2$ for
\logM$\sim10.5$) and a more slow increase ($<15$\%) for \logM$>11$.
A similar but even stronger trend is inferred for the ETG population, i.e.,
massive galaxies build-up their stars and mass earlier than lower mass galaxies.
We show that the \zbuilding\ of ETGs increases rapidly with \Mstar\ from \zbuilding$\sim0.2$ to $\sim0.8$ for
\logM$=10, 10.8$, respectively,  and is significantly higher ($z\simgt 1$) 
for massive ETGs.

\item {\it ``SFH-evolved \MF" and mergers:}
Assuming only a growth in stellar mass driven by exponentially decreasing star-formation 
rates, we find that the predicted ``SFH-evolved \MFs" agree with the observed \MFs\
with differences of at most 20-40\%, even if they predict 
redder colours than observed, in particular at low redshift.
Therefore, mass assembly is controlled primarily by SFH, which
can explain most of the evolution of the \GSMF\ at intermediate-low \Mstar\ (\logM$<10.6$).
The low residual evolution of the \GSMF\ is consistent with 0.16 merger per galaxy per Gyr, on average,
with a hint of a decrease with cosmic time but no clear dependence on mass.
Major merging events contribute in terms of fewer than 0.1 merger/gal/Gyr ($\sim 0.06$ on average) and are, 
if not marginal, not the dominant evolutionary process.

\item{\it \Mcross\ and the evolution of the LTGs:}
The intersection (\Mcross) between the \MFs\ of the ETG and LTG populations decreases 
with cosmic time.
This is due mainly to a clear increase with cosmic time in the number 
density of ETGs, regardless of their classification method, for \logM$<11$.
In contrast, for LTGs the number density of blue or disc+irregular galaxies shows
only a mild or negligible evolution, while 
the most extreme population of star-forming galaxies (those with high specific star
formation) decreases rapidly in number
density with cosmic time, in particular at high mass (\logM$>10.3$).

\item {\it Flow to build-up the red population:}
We have quantified the flow rate from blue to red population at different redshifts and masses.
We have found a growth rate in number and mass density of the red galaxies
of about a few $10^{-4}$ gal/Mpc$^3$/Gyr/log\Mstar\   and a few $10^6 - 10^7$ \Msun/Mpc$^3$/Gyr/log\Mstar\
for \logM$<11$, and lower in value above this mass.
The corresponding fraction of blue galaxies
that, at any given time, are transforming into red galaxies per Gyr 
is on average $\sim 25$\% for \logM$<11$.
This fraction and the growth rate for red galaxies increases with cosmic time 
at \logM$=10.4-10.7$ at least for $z<0.7$, while it seems to  
decrease with cosmic time at higher \Mstar\ (\logM$=10.7-11$). 
 
 \item {\it ETG evolutionary sequence:}
Using different methods to classify ETGs, we have found that 
the  number density of quiescent galaxies is higher than that of red galaxies, which is even higher than that 
of spheroidals for \logM$<11$.
This behaviour suggests that a transformation occurs from quiescent to red objects, 
as inferred mainly from SFHs, before or on a shorter timescales than the 
morphological tranformation, which is most likely primarily driven by dynamical processes.

\item{\it Comparison with SAM:} Semi-analytical models are unable to fully account for \GSMF\ evolution, 
in particular for ETGs. 
They predict a different shape at any mass and redshift explored, there being a strong excess of low-mass 
passive galaxies and a deficit of high-mass galaxies at high-$z$, 
disagreement that can be only partially resolved by the bias introduced by 
uncertainties in the mass determination. 
Furthermore, they predict an ``upsizing" evolution in number density  with cosmic time, 
i.e., mass assemply occurring slower and 
earlier in low-mass galaxies, which is the opposite of the ``downsizing" trend observed in zCOSMOS data 
for the \GSMF\ and quantified by \zbuilding.

\end{itemize}

We attempt to summarise our results within a galaxy evolutionary scenario that emerges from
several observational studies.
It is almost established that there is an age-downsizing of galaxy stellar populations,
according to which less massive galaxies contain younger stars,
which is also applicable to similar spectral/morphological types
both among ETGs (Nelan et al. 2005; Thomas et al. 2005; Graves et al. 2007; Fontana et al. 2004)
and LTGs (Noeske et al. 2007a,b). 
Furthermore, we find that downsizing is also applicable to the mass-assembly,  
i.e., massive galaxies assemble their mass in a single object earlier than
low-mass galaxies, both for the global population and for ETGs, regardless of
their definition (in terms of their colours, morphology, or SFR activity).

We therefore propose a scenario where intermediate-mass galaxies (\logM$=10-11$)
decrease their star formation activity to intermediate values
gradually (with an e-folding time $\tau=1-3$ Gyr) 
and finally to that of quiescent galaxies, by means of the exhaustion of either the gas reservoir 
or cold gas accretion, or quenching due to AGN feedback. 
Noeske et al. (2007a,b) found that the e-folding time is mass-dependent, 
being longer for less massive galaxies, which also have a later onset of the SFH. 
Therefore, if no other mechanism intervenes to again switch on the star-formation 
activity (e.g., re-juvenation hypothesis, Hasinger et al. 2008), 
they evolve to have red colours on a timescale of 1-2 Gyr and finally undergo a 
dynamical morphological transformation into spheroidal galaxies. 

In this scenario,
massive galaxies (\logM$>11$), in particular ETGs, are already in place
at $z=1$, and their \MF\ exhibit negligible evolution thereafter.
In contrast, the number density of intermediate-mass ETGs continues to increase  
with cosmic time (by a factor $\sim4-2$ from $z=0.7$ to $z=0.2$ for \logM$=10.3-10.7$) 
with a growth rate of a few $10^{-4}$ gal/Mpc$^3$/Gyr/log\Mstar\ and a few $10^{7}$ \Msun/Mpc$^3$/Gyr/log\Mstar.
We have found that the most likely evolutionary process to explain our results 
is the transformation from blue to red galaxies by means of
the smooth decrease or the quenching of the SFR (as described above), 
followed by a morphological transformation probably by means of  internal dynamical processes, 
such as disc instabilities. 
Major merger events are unlikely to be able to explain why
colour transformation precedes the morphological one.
In contrast, blue galaxies at low mass (\logM$<10$) continue to grow in stellar mass, because of 
their high \SSFR,
and replace intermediate-mass LTGs, whose density remains almost constant in time. 
We have found that only a small fraction  of low-mass (\logM$\sim10$)  blue galaxies 
need to transform into red galaxies to explain the observed ETG growth rate at least at $z\ge 0.4$. 
Merger events (minor and major) contribute with a rate of $\sim 0.16$ merger/gal/Gyr on average
(of which fewer than $0.1$ are major) to form galaxies at redshifts lower than $z\sim1$.

\begin{figure}[h!]
\centering
\includegraphics[angle=-90,width=0.99\hsize]{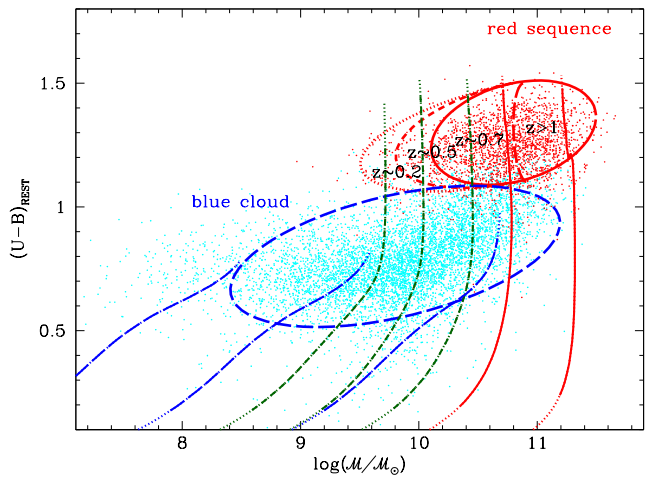}
\caption{
Schematic scenario for galaxy evolution.  Different ellipses refer to different redshifts and galaxy populations,
while different evolutionary tracks are plotted between $0.1$ and $10$ Gyr and 
refer to SFHs with different timescales  and
normalized to the total final mass (($\tau$; \logMtot)=
($0.1; 11.5$), ($0.3;11.0$) [solid red curves];
($0.6; 10.7$), ($1; 10.3$), ($2; 10.0$) [short dashed green curves];
($5; 11$), ($10; 10$), ($15; 9$) [long dashed blue curves].
}
\label{fig:scenario}
\end{figure}

Following a schematic diagram, as originally proposed by Faber et al. (2007), this scenario is presented in
Fig. \ref{fig:scenario}. The dominant processes are the transformation 
with cosmic time of blue star-forming galaxies of progressively lower mass 
to red passive galaxies according to their SFH. 
Different evolutionary tracks are plotted between $0.1$ and $10$ 
Gyr and 
refer to constraints on their inferred SFHs with different timescales  (longer for less massive galaxies) and
normalized to the total final mass (\Mstar$_{\rm tot}$). 
As examples that only qualitatively reproduce our data,
we show tracks for models with ($\tau$; \logMtot)=
($0.1; 11.5$), ($0.3;11.0$) [solid red curves];
($0.6; 10.7$), ($1; 10.3$), ($2; 10.0$) [short dashed green curves];
($5; 11$), ($10; 10$), ($15; 9$) [long dashed blue curves].

We conclude that the build-up of galaxies, and ETGs in particular, follows the same ``downsizing" trend 
with mass 
(i.e., occurring earlier in high-mass galaxies) of the formation of their stars, which is the
opposite of that predicted by current SAM models.
{\bf The SAM LCDM models, with rapid galaxy growth, possibly overestimate the merger rate within halos 
(and underestimate the time-scale for merging).
Our analysis also suggests that morphological transformation should occur on longer 
timescales than colour transformation.}
Merging processes are not marginal but their frequency ($\sim 0.16$ merger/gal/Gyr of which $<0.1$ are major) 
appears to decrease with redshift, while no strong dependence on mass has been found, in
disagreement with
SAM models (Wang \& Kauffmann 2008, Stewart et al. 2009).  
The prevailing galaxy formation and evolutionary scenario 
is therefore moving towards one 
in which a smoother evolution in mass growth and star formation (due to accretion of cold gas) 
and minor merger play a more important role than major merging events. 
In this scenario, however, it appears quite difficult to explain the ``fine-tuning" of the 
transformation of low-mass into intermediate-mass blue galaxies required to ensure that  
their number density remains almost constant at any given mass.
In this scenario, we expect negligible evolution of the galaxy baryonic mass function (\GBMF) 
for the global population at all masses and a decrease with cosmic time in the  \GBMF\ for the 
blue galaxy population at intermediate-to-high masses.

\begin{acknowledgements}
LP wish to thank Fabio Fontanot and Gabriella De Lucia for the
stimulating discussions and to have provided SAM models and derived properties in electronic form. 
LP thanks Micheal Brown and Ramin Skibba for comments.
We acknowledge support from an INAF contract PRIN-2007/1.06.10.08 and
an ASI grant ASI/COFIS/WP3110 I/026/07/0. 
\end{acknowledgements}

\clearpage


\begin{thebibliography}{}

\bibitem[]{} 
	Abraham, R. G., van den Bergh, S., Nair, P., 2003, ApJ, 588, 218
\bibitem[2007]{arn07} 
    Arnouts, S., Walcher, C. J., Le F\`evre, O., et al. 2007, 476, 137
\bibitem[]{} 
    Baldry, I.K., Glazebrook, K., Brinkmann, J., et al. 2004, ApJ, 600, 681
\bibitem[]{}
    Baldry, I.K., Balogh, M. L., Bower, R. G., et al. 2006, MNRAS, 373, 469
\bibitem[]{} 
    Baldry, I.K., Glazebrook, K., Driver, S. P., et al. 2008, MNRAS, 388, 945
\bibitem[]{} 
	Balogh, M. L., Baldry, I. K., Nichol, R., et al. 2004, ApJ, 615, L101
\bibitem[]{} 
	Barnes, J. E. 1992, \apj, 393, 484
\bibitem[]{} 
    Bauer, A. E., Drory, N., Hill, G. J., Feulner, G., et al. 2005, \apj, 621, L89
\bibitem[2003]{bel03}%
    Bell, E. F., McIntosh, D. H., Katz, N., \& Weinberg, M. D. 2003, \apj, 149, 289 
\bibitem[2004]{bel04} 
    Bell, E. F., Wolf, C., Meisenheimer, K., et al. 2004, \apj, 608, 752
\bibitem[2005]{bel06} %
    Bell, E.F., Phleps, S., Somerville, R. S., et al. 2006, \apj, 652, 270
\bibitem[2007]{bel07} %
	Bell, E. F., Zheng, X. Z., Papovich, C., et al.  2007, ApJ, 663, 834	
\bibitem[]{} 
	Bertoldi, F., Carilli, C., Aravena, M., Schinnerer, E., et al. 2007, ApJS, 172, 132
\bibitem[]{} 
    Blanton, M.R., Hogg, D.W., Bahcall, N.A.  et al. 2003, \apj, 594, 186
\bibitem[]{} 
	Blanton, M. R., Lupton, R. H., Schlegel, D. J., et al., 2005, ApJ, 631, 208	
\bibitem[2000]{bol00}%
    Bolzonella, M., Miralles, J.-M., Pell\'o, R.  2000, A\&A, 363, 476
\bibitem[]{} 
	Bolzonella, M.,  Kova\v{c}, K., Pozzetti, L. et al. 2009, A\&A submitted (arXiv:0907.0013)
\bibitem[2006]{bor06} 
    Borch, A., Meisenheimer, K., Bell, E. F., et al. 2006, \aap, 453, 869
\bibitem[2006]{bow06}%
	Bower, R. G., Benson, A. J., Malbon, R., et al. 2006, MNRAS, 370, 645
\bibitem[2000]{bri00} 
    Brinchmann, J., Ellis, R. S. 2000, \apj, 536, L77
\bibitem[2004]{bri04}%
    Brinchmann, J., Charlot, S., White, S. D. M., et al. 2004, MNRAS, 351. 1151
\bibitem[2008]{brown08}%
	Brown, M.J.I., Zheng, Z., White, M., Dey, A., Jannuzi, B.T., et al., 
	2008, ApJ, 682, 937
\bibitem[2003]{bru03} 
    Bruzual, G., \& Charlot, S. 2003, \mnras, 344, 1000
\bibitem[2006]{bun06} 
    Bundy, K., Ellis, R.S., Conselice, C.J., et al. 2006, \apj, 651, 120
\bibitem[2009]{bun09} 
	Bundy, K., Scarlata, C., Carollo, C. M., Ellis, R. S., Drory, N., et al., 2009,
	Apj, submitted (arXiv0912.1077)
\bibitem[2000]{cal00} 
    Calzetti, D., Armus, L., Bohlin, R. C., et al. 2000, \apj, 533, 682
\bibitem[]{} 
    Capak, P., Aussel, H., Ajiki, M., et al. 2007, ApJS, 172, 99
\bibitem[2006]{cap06}%
    Caputi, K. I., McLure, R. J., Dunlop, J. S., et al. 2006, MNRAS, 366, 609Q
\bibitem[]{} 
	Carollo, C. M., Stiavelli, M., de Zeeuw, P. T., Seigar, M., Dejonghe, H., 2001, ApJ, 546, 216	
\bibitem[]{} 
	Cassata, P., Guzzo, L., Franceschini, A., et al. 2007, ApJS, 172, 270
\bibitem[]{} 
    Cassata, P., Cimatti, A., Kurk, J., et al. 2008, A\&A, 483, L39
\bibitem[]{} 
	Cattaneo, A., Dekel, A., Faber, S. M., Guiderdoni, B., 2008, MNRAS, 389, 567
\bibitem[2003]{cha03} 
    Chabrier, G. 2003, \pasp, 115, 763
\bibitem[]{}  Charlot \& Bruzual 2007
\bibitem[2006]{cim06} 
    Cimatti, A., Daddi, E., Renzini, A., 2006, A\&A, 453, L29
\bibitem[]{} 
    Cirasuolo, M., McLure, R. J., Dunlop, J. S., et al. 2007, MNRAS, 380, 585 
\bibitem[2000]{col00}%
	Cole, S., Lacey, C. G., Baugh, C. M., Frenk, C. S., 2000, MNRAS, 319, 168	
\bibitem[2001]{col01}%
    Cole, S., Norberg, P., Baugh, C. M.,  et al. 2001, MNRAS, 326, 255
\bibitem[]{}
    Coleman, G. D., Wu, C.-C., Weedman, D. W. 1980, ApJS, 43, 393
\bibitem[]{}
	Cortese, L., Gavazzi, G., Boselli, A., et al., 2006, A\&A, 453, 847	
\bibitem[1996]{cow96} 
    Cowie, L. L., Songaila, A., Hu, E. M., Cohen, J. G. 1996, AJ, 112, 839
\bibitem[]{} 
    Cucciati, O., Iovino, A., Marinoni, C., et al. 2006, A\&A, 458, 39
\bibitem[]{} 
	Dekel, A., Birnboim, Y., Engel, G., et al., 2009, Nature, 457, 451
\bibitem[2006]{del06}%
    De Lucia, G., Springel, V., White, S. D. M., Croton, D., Kauffmann, G. 2006, MNRAS, 366, 499
\bibitem[2007]{del07} 
	De Lucia, G., Poggianti, B. M., Arag\'on-Salamanca, A., et al. 2007, MNRAS, 374, 809
\bibitem[2007]{del07} 
	De Lucia, G., Blaizot, J., 2007, MNRAS, 375, 2
\bibitem[]{} 
	de Ravel, L., Le F\`evre, O., Tresse, L.,  et al., 2009, A\&A, 498, 379	
\bibitem[]{} 
	Dickinson, M., Papovich, C., Ferguson, H. C., Budav\'ari, T., 2003, ApJ, 587, 25	
\bibitem[]{} 
	Driver, S. P., Phillipps, S., Davies, J. I., et al., 1994, MNRAS, 268, 393	
\bibitem[2005]{dro05} 
	Drory, N., Salvato, M., Gabasch, A., et al. 2005, \apj, 619, L131
\bibitem[2009]{dro09} 
	Drory, N., Bundy, K., Leauthaud, A., Scoville, N., Capak, P., et al., 2009, ApJ, 707, 1595	
\bibitem[]{} Elvis, M., Civano, F., Vignali, C., et al., 2009, ApJS, 184, 158
\bibitem[2007]{fab07}%
	Faber, S. M., Willmer, C. N. A., Wolf, C., et al., 2007, ApJ, 665, 265	
\bibitem[]{} 
	Feldmann, R., Carollo, C. M., Porciani, C., et al. 2006, MNRAS, 372, 565
\bibitem[1976]{fel76}%
    Felten, J. E. 1976, \apj, 207, 700
\bibitem[]{} 
	Feulner, G., Gabasch, A., Salvato, M., et al. 2005a, \apj, 633, L9
\bibitem[]{} 
	Feulner, G., Goranova, Y., Drory, N., et al. 2005b, MNRAS, 358, L1
\bibitem[2004]{fon04} 
	Fontana, A., Pozzetti, L., Donnarumma, I., et al. 2004, A\&A, 424, 23, F04
\bibitem[2006]{fon06} 
	Fontana, A., Salimbeni, S., Grazian, A., et al. 2006, A\&A, 459, 745
\bibitem[2009]{fon09} 
	Fontanot, F., De Lucia, G., Monaco, P., Somerville, R. S., Santini, P.  2009, MNRAS, 397, 1776
\bibitem[2004]{franz07} 
	Franzetti, P., Scodeggio, M., Garilli, B., et al. 2007, A\&A, 465, 711
\bibitem[]{} 
	Gavazzi, G., Scodeggio, M., 1996, A\&A, 312, L29
\bibitem[]{}
	Gehrels, N. 1986, ApJ, 303, 336
\bibitem[]{} 
	Genzel, R., Tacconi, L. J., Eisenhauer, F., et al., 2006, Nature, 442, 786	
\bibitem[]{} 
	Genzel, R., Burkert, A., Bouch\'e, N., et al., 2008, ApJ, 687, 59	
\bibitem[]{} 
	Giallongo, E., Salimbeni, S., Menci, N, et al. 2005, ApJ, 622, 116
\bibitem[]{} 
	Graves, G. J., Faber, S. M., Schiavon, R. P., Yan, R., 2007, ApJ, 671, 243	
\bibitem[]{} 
	Gwyn, S.D.J., Hartwick, F.D.A., 2005, AJ, 130, 1337
\bibitem[]{} 
	Juneau, S., Glazebrook K., Crampton, D., et al. 2005
\bibitem[]{} 
	Hasinger, G., Cappelluti, N., Brunner, H., et al. 2007, ApJS, 172, 29
\bibitem[]{} 
	Hogg, D.W., Blanton, M., Strateva, I.,  et al. 2002, AJ, 124, 646
\bibitem[]{} 
	Hopkins, A. M., Beacom, J. F., 2006, \apj, 652, 864
\bibitem[2006]{ilb06} 
	Ilbert, O., Arnouts, S., McCracken, H. J., et al. 2006, A\&A, 457, 841
\bibitem[2009]{ilb09} 
	Ilbert, O., Capak, P., Salvato, M., Aussel, H., McCracken, H. J.,  et al.,
	2009, ApJ, 690, 1236
\bibitem[2009]{ilb10} 
	Ilbert, O., Salvato, M., Le Floc'h, E., et al. 2010, ApJ, 709, 644
\bibitem[2009]{iovino08}
   Iovino, A., Cucciati, O., Scodeggio, M., Knobel, C., Kova\v{c}, K., et al. 2010, A\&A, 509, 40 
\bibitem[2003]{kauff03}%
    Kauffmann, G., Heckman, T. M., White, S. D. M., et al. 2003, MNRAS, 341, 33
\bibitem[1998]{ke98}%
    Kennicutt, R. C. Jr., 1998, ARA\&A, 36, 189	
\bibitem[2003]{kw07}%
    Kitzbichler, M. G., White, S. D. M., 2007, MNRAS, 376, 2
\bibitem[]{}
	Kinney, A. L., Calzetti, D., Bohlin, R. C., et al. 1996, ApJ, 467, 38
\bibitem[]{} 
	Kodama, T., Yamada, T., Akiyama, M., et al. 2004, MNRAS, 350, 1005
\bibitem[]{}
	Koekemoer, A. M., Aussel, H., Calzetti, D., et al., 2007, ApJS, 172, 196	
\bibitem[2009a]{kovac08}
   Kova\v{c}, K., Lilly, S.J., Cucciati, O., et al. 2010, ApJ, 708, 505
\bibitem[2009a]{kovac09}
	Kova\v{c}, K., Lilly, S. J., Knobel, C., Bolzonella, M., Iovino, A., et al., 2009, ApJ submitted (arXiv:0909.2032)
\bibitem[]{} 
	Larson, R. B., Tinsley, B. M., Caldwell, C. N., 1980, ApJ, 237, 692	
\bibitem[2003a]{vimos} 
	Le~F\`evre, O., Saisse, M., Mancini, D., et al. 2003, SPIE, vol. 4841, p.1670 
\bibitem[]{} 
	Lilly, S. J., Le Fevre, O., Hammer, F., Crampton, D., 1996, ApJ, 460, L1	
\bibitem[]{} 
	Lilly, S.J., Le F\`evre, O., Renzini, A., et al. 2007, ApJS, 172, 70
\bibitem[2009]{lilly08}
   Lilly, S.J., Le Brun, V., Maier, C., et al. 2009, ApJS, 184, 218 
\bibitem[2004]{lin2004}%
    Lin, L., Koo, D. C., Willmer, C. N. A., et al.,  2004, \apj, 617, 9L
\bibitem[]{} Lo Faro B., Monaco P., Fontanot F. et al. 2009, MNRAS, 399, 827L
\bibitem[1996]{mad96}%
	Madau, P., Ferguson, H. C., Dickinson, M. E., Giavalisco, M., Steidel, C. C., Fruchter, A.,
	1996, MNRAS, 283, 1388	
\bibitem[1998]{mad98}%
    Madau, P., Pozzetti, L., \& Dickinson, M. 1998, \apj, 498, 106
\bibitem[]{} 
	Maier, C., Lilly, S. J., Zamorani, G., et al., 2009, ApJ, 694, 1099	
\bibitem[2006]{ma06}%
    Maraston, C., Daddi, E., Renzini, A., et al. 2006, \apj, 652, 85
\bibitem[2005]{ma05} 
	Maraston, C., 2005, MNRAS, 362, 799
\bibitem[]{} 
	Marchesini, D., van Dokkum, P. G., Forster Schreiber, N. M., Franx, M., Labbe', I., Wuyts, S., 
        2009, ApJ, 701, 1765
\bibitem[]{} 
	Menci, N., Fontana, A., Giallongo, E., et al. 2005, ApJ, 632, 49 
\bibitem[2006]{menci2006}
    Menci, N., Fontana, A., Giallongo, E., et al. 2006, \apj, 647, 753
\bibitem[2009]{meneux09}
	Meneux, B., Guzzo, L., de la Torre, S., Porciani, C., Zamorani, G., et al., 2009, A\&A, 505, 463
\bibitem[2009]{hjmcc09}
   McCracken, H.J., Capak, P., Salvato, M., et al. 2010, ApJ, 708, 202
\bibitem[2009]{mignoli08}
   Mignoli, M., Zamorani, G., Scodeggio, M., et al. 2009, A\&A, 493, 39
\bibitem[]{} 
	Mihos, J. C., Hernquist, L., 1996, ApJ, 464, 641	
\bibitem[2006]{monaco2006}
	Monaco, P., Murante, G., Borgani, S., Fontanot, F.,  2006, ApJ, 652, L89	
\bibitem[2007]{monaco2007}
   Monaco, P, Fontanot, F., Taffoni, G., 2007, MNRAS, 375, 1189	
\bibitem[2009]{moresco09}
   Moresco, M., Pozzetti, L., Cimatti, A., et al. 2010, A\&A, submitted
\bibitem[]{Moustakasetal2006} 
	Moustakas, J., Kennicutt, R.C., Jr., Tremonti, C. A. 2006, ApJ, 642, 775	
\bibitem[]{} 
	Nelan, J. E., Smith, R. J., Hudson, M. J., et al., 2005, ApJ, 632, 137	
\bibitem[]{} 
	Noeske, K. G., Faber, S. M., Weiner, B. J., et al. 2007a, ApJ, 660, L47	
\bibitem[]{} 
	Noeske, K. G., Weiner, B. J., Faber, S. M., et al. 2007b, ApJ, 660, L43	
\bibitem[]{} 
	Norberg, P., Cole, S., Baugh, C. M., et al. 2002, MNRAS, 336, 907
\bibitem[]{} 
	Oesch, P. A., Carollo, C. M., Feldmann, R., Hahn, O., Lilly, S. J., et al. 2010, ApJ, 714, L47
\bibitem[]{} 
	Parry, O. H., Eke, V. R., Frenk, C. S., 2009, MNRAS, 396, 1972	
\bibitem[]{} 
	Popesso, P., Biviano, A., Bohringer, H.,  Romaniello, M. 2006, A\&A, 445, 29	
\bibitem[2000]{pozzetti00} 
	Pozzetti, L., Mannucci, F., 2000, MNRAS, 317, L17	
\bibitem[2003]{pozzetti03} 
	Pozzetti, L., Cimatti, A., Zamorani, G., et al. 2003, A\&A, 402, 837
\bibitem[2007]{pozzetti07} 
	Pozzetti, L., Bolzonella, M., Lamareille, F., et al. 2007, A\&A, 474, 443
\bibitem[2007]{ren07}%
    Renzini, A. 2007, in 'At the Edge of the Universe: Latest Results from the Deepest Astronomical Surveys', 
	ed.  by J. Afonso,  ASP Conf. Ser. 380, 309
\bibitem[2009]{ren09}%
	Renzini, A., 2009, MNRAS accepted
\bibitem[1955]{sal55} 
	Salpeter, E. E. 1955, \apj, 121, 161
\bibitem[1979]{san79} 
	Sandage, A., Tammann, G.A., \& Yahil, A., 1979, \apj, 232, 352
\bibitem[]{} 
	Sanders, D. B., Salvato, M., Aussel, H., et al. 2007, ApJS, 172,  86
\bibitem[]{} 
	Sargent, M. T., Carollo, C. M., Lilly, S. J., et al. 2007, ApJS, 172, 434
\bibitem[]{} 
	Scarlata, C., Carollo, C.M., Lilly, S.J., et al. 2007a, ApJS, 172, 406
\bibitem[]{} 
	Scarlata, C., Carollo, C.M., Lilly, S.J., et al. 2007b, ApJS, 172, 494
\bibitem[1976]{schechter76}%
    Schechter, P. 1976, \apj, 203, 297
\bibitem[]{} 
	Schinnerer, E., Smol\v{c}i\'c, V., Carilli, C. L., et al. 2007, ApJS, 172, 46
\bibitem[2005]{vipgi} 
	Scodeggio, M., Franzetti, P., Garilli, B., et al. 2005, PASP, 117, 1284
\bibitem[]{} 
	Scoville, N., Aussel, H., Brusa, M., et al. 2007, ApJS, 172, 1 
\bibitem[1959]{sch59}%
    Schmidt, M. 1959, \apj, 129, 243
\bibitem[1968]{sch68}%
    Schmidt, M. 1968, \apj, 151, 393
\bibitem[]{} 
	Skibba, R. A., Bamford, S. P., Nichol, R. C., et al., 2009, MNRAS, 399, 966
\bibitem[]{} 
	Somerville R. S., Hopkins P. F., Cox T. J., Robertson B. E., Hernquist L., 2008, MNRAS, 391, 481 
\bibitem[]{} 
	Stewart, K. R., Bullock, J. S., Barton, E. J., Wechsler, R. H.,  2009, ApJ, 702, 1005
\bibitem[]{} 
	Strateva, I., Ivezic, Z., Knapp, G.R., et al. 2001, AJ, 122, 1861
\bibitem[]{} 
\bibitem[]{} 
	Taniguchi, Y., Scoville, N., Murayama, T., et al. 2007, ApJS, 172, 9
\bibitem[2009]{tasca08}
   Tasca, L., Kneib, J.P., Iovino, A., et al. 2009, A\&A, 503, 379 
\bibitem[]{} 
	Thomas, D., Maraston, C., Bender, R., Mendes de Oliveira, C., 2005, ApJ, 621, 673
\bibitem[]{} 
	Thomas, D., Maraston, C., Schawinski, K., Sarzi, M., Silk, J., 2009, MNRAS, 404, 1775
\bibitem[]{} 
	Toomre, A., Toomre, J., 1972, ApJ, 178, 623	
\bibitem[2005]{van05}%
    van Dokkum, P. G.  2005, AJ, 130, 2647
\bibitem[]{} 
	Vergani, D., Scodeggio, M., Pozzetti, L.,  et al. 2008, A\&A, 487, 89
\bibitem[]{} 
	Vergani, D., Zamorani, G., Lilly, S., Lamareille, F., Halliday, C., et al. (2009), A\&A, 509, 42
\bibitem[]{} 
	Walcher, C. J., Lamareille, F., Vergani, D., et al., 2008, A\&A, 491, 713	
\bibitem[]{} 
	Wang, L., Kauffmann, G., 2008, MNRAS, 391, 785	
\bibitem[]{} 
	Wang J., De Lucia G., Kitzbichler M. G., White S. D. M., 2008, MNRAS, 384, 1301
\bibitem[]{} 
	Weiner, B.J., Phillips, A.C., Faber, S.M., et al. 2005, ApJ, 620, 595
\bibitem[]{} 
	Williams, R. J., Quadri, R. F., Franx, M., van Dokkum, P., Labb\'e, I., ApJ, 691, 1879
\bibitem[]{} 
	Zamojski, M. A., Schiminovich, D., Rich, R. M., et al. 2007, ApJS, 172, 468
\bibitem[]{} 
	Zucca, E., Zamorani, G., Vettolani, G., et al. 1997, A\&A, 326, 477
\bibitem[2006]{zuc06} 
	Zucca, E., Ilbert, O., Bardelli, S., et al. 2006, A\&A, 455, 879
\bibitem[]{} Zucca, E., Bolzonella, M., Bardelli, S.,  et al. 2009, A\&A, 508, 1217

\end{thebibliography}
\end{document}